\documentclass[runningheads]{llncs}
\usepackage[export]{adjustbox}

\usepackage{microtype}%if unwanted, comment out or use option "draft"
\usepackage[rawfloats=true]{floatrow} 
\restylefloat{figure}    
\DeclareMathSizes{4}{4}{4}{4}
\usepackage{varwidth}
\usepackage{lmodern}
\usepackage{amsfonts}
\usepackage[fleqn]{amsmath}
\usepackage{mathtools}
\usepackage{amssymb}
\usepackage{hyperref}
\usepackage{proof}
\usepackage{stmaryrd}
\usepackage{xspace}
\usepackage{listings}
\usepackage{comment}
\usepackage{longtable}
\usepackage{float}
\usepackage[section]{placeins}
\usepackage[toc,page]{appendix}
\usepackage[export]{adjustbox}
\usepackage{amssymb}
\usepackage{graphics,graphicx}
\usepackage{mathpartir}
\usepackage{array}
\usepackage{tikz}
\usetikzlibrary{shapes.geometric}
\usetikzlibrary{arrows.meta,arrows}
\usetikzlibrary{positioning}
\usetikzlibrary{calc, shapes, backgrounds}
\usepackage{verbatim}
\usepackage{tikz-qtree}
\usepackage{pgfplots}
\usepackage{caption,capt-of}
\usepackage{color}
\usepackage{soul}
\setul{1ex}{0.8ex}
\definecolor{orange}{rgb}{1,0.5,0}
\setulcolor{orange}
\usepackage{subcaption}
\usepackage{lipsum}
\usepackage{afterpage}
\newcommand{\specline}[1]{{\color{blue}\left\{#1\right\}}}
\newcommand{\assertion}[1]{\textsf{assert} {\color{red}\left\{#1\right\}}}

\newcommand{\proofoutstepnr}[5][1pt/3pt]{\small \left.
\begin{array}{@{}l@{}}
\hdashline[#1]
\ensuremath{\specline{#2}}\\
\begin{array}{@{\;\;\;}l}
#3
\end{array}\\
\ensuremath{\specline{#5}}\\
\hdashline[1pt/3pt]
\end{array}
\color{green}\right)\hspace{-6pt}{\color{green}-}
\begin{array}{@{}l@{}}
#4
\end{array}
}
\newcommand{\proofoutstep}[5][1pt/3pt]{
\proofoutstepnr[#1]{#2}
{#3}
{\rotatebox[origin=c]{90}{$\begin{array}{@{}c@{}}{}#4\end{array}$}}
{#5}
}

\newcommand{\NULL}{\texttt{NULL}}
\newcommand{\NEW}{\texttt{new}}
\newcommand{\SKIP}{\texttt{skip}}

\newcommand{\INT}{\texttt{int}}
\newcommand{\bool}{\texttt{bool}}

\newcommand{\db}[1]{\llbracket#1\rrbracket}

\newcommand{\subt}{\prec:}
 \newcommand{\fname}{\textsf{FName}}

\newcommand{\udef}{\textsf{undef}}
\newcommand{\unlinked}{\textsf{unlinked}}

\newcommand{\var}{\textsf{Var}}

\newcommand{\fmapfieldemp}{\N_{f,\emptyset} = \N \setminus \{f \rightharpoonup \_ \}}
\newcommand{\fmapname}{\N = \{ f_0|\ldots|f_n\rightharpoonup  \{y\} \mid f_{i} \in \fname \land \; 0\leq i\leq n  \land \; (y \in \var \lor y \in \{null\}) \}}
\newcommand{\fmapextend}{  \N(f\rightharpoonup x \setminus y) = \N \setminus \{f \rightharpoonup x\} \cup \{f \rightharpoonup y\} }
\newcommand{\fmapextendn}{  \N(\cup_{f\rightharpoonup y}) = \N \cup \{f \rightharpoonup y\} }
\newcommand{\fmapdiff}{\N(\setminus_{f\rightharpoonup y}) = \N - \{f\rightharpoonup y\}}
\newcommand{\fmapwithf}{\N([f\rightharpoonup  y]) = \N \text{   where  } f \rightharpoonup y \in \N }
\newcommand{\fmapempty}{\N_{\emptyset} =  \{ \}}

\newcommand{\rcunf}[3]{#1 : \textsf{rcuFresh} \;\; \N(\cup_{ #2 \rightharpoonup #3}) } %x, f , y
\newcommand{\rcunextfresh}[1]{\textsf{rcuFresh} \;\; \N #1 }

\newcommand{\rcuitertype}[4]{\textsf{rcuItr}(#1,#2)\;#3\;#4}
\newcommand{\rcuit}[4]{#1:\textsf{rcuItr}\;\; #3 \;\; #4 }
\newcommand{\rcuitrT}[5]{#1:\textsf{rcuItr}\;\; \rho \;\; \N }

\newcommand{\result}[1]{\mathsf{y = \mathsf{result}}}

\newcommand{\N}{\ensuremath{\mathcal{N}}}

\newcommand{\alternatingpath}[1]{f^{*} \; = \; f_1 \mid f_2  \; \text{where} \;  (f=f_1) \;\oplus \; ( f=f_2 )  }

\newcommand{\assert}[1]{\{#1\}}

\definecolor{framed}{gray}{0.50}

\usepackage{todonotes}

\newcommand{\mypar}[1]{\textbf{#1}~\xspace}

\newcommand{\grammar}{
\[
\begin{array}{rclcl}
\alpha &::= & \mathsf{skip} \mid \mathsf{x.f = y} \mid \mathsf{y=x} \mid \mathsf{y =x.f}  \mid \mathsf{y =new} \mid  \mathsf{Free}\mathsf{(\textsf{x})} \mid \mathsf{Sync}                   
\end{array}\quad
\textsf{Sync} \overset{\Delta}{=} \texttt{\textsf{SyncStart};\textsf{SyncStop}}
\]
}

\usepackage{lipsum} % for dummy text

\renewcommand{\and}{\\}

\usepackage{microtype}%if unwanted, comment out or use option "draft"
\begin{document}
\title{Safe Deferred Memory Reclamation with Types}
%
%\titlerunning{RCU Types}
\author{Ismail Kuru \orcidID{0000-0002-5796-2150} \and
Colin S. Gordon \orcidID{0000-0002-9012-4490}}
%%%Third Author\inst{3}\orcidID{2222--3333-4444-5555}}
%
\authorrunning{I. Kuru et al.}
\institute{Drexel University \\ %\and
%%%Springer Heidelberg, Tiergartenstr. 17, 69121 Heidelberg, Germany
\email{\{ik335,csgordon\}@drexel.edu}}%\\
%%%\url{http://www.springer.com/gp/computer-science/lncs} \and
%%%ABC Institute, Rupert-Karls-University Heidelberg, Heidelberg, Germany\\
%%%\email{\{abc,lncs\}@uni-heidelberg.de}}
%
\maketitle              % typeset the header of the contribution

\begin{abstract}
Memory management in lock-free data structures remains a major challenge in concurrent programming.  Design techniques including read-copy-update (RCU) and hazard pointers provide workable solutions, and are widely used to great effect.  These techniques rely on the concept of a grace period: nodes that should be freed are not deallocated immediately, and all threads obey a protocol to ensure that the deallocating thread can detect when all possible readers have completed their use of the object. This provides an approach to safe deallocation, but only when these subtle protocols are implemented correctly.

We present a static type system to ensure correct use of RCU memory management: that nodes removed from a data structure are always scheduled for subsequent deallocation, and that nodes are scheduled for deallocation at most once. As part of our soundness proof, we give an abstract semantics for RCU memory management primitives which captures the fundamental properties of RCU. 
Our type system allows us to give the first proofs of memory safety for RCU linked list and binary search tree implementations without requiring full verification.
\end{abstract}\raggedbottom
\section{Introduction}
For many workloads, lock-based synchronization -- even fine-grained locking -- has unsatisfactory performance.  Often lock-free algorithms yield better performance, at the cost of more complex implementation and additional difficulty reasoning about the code.  Much of this complexity is due to memory management: 
developers must reason about not only other threads violating local assumptions, but whether other threads are \emph{finished accessing} nodes to deallocate. 
At the time a node is unlinked from a data structure, an unknown number of additional threads may have already been using the node, having read a pointer to it before it was unlinked in the heap.

A key insight for manageable solutions to this challenge is to recognize that just as in traditional garbage collection, the unlinked nodes need not be reclaimed immediately, but can instead be reclaimed later after some protocol finishes running.
Hazard pointers~\cite{Michael:2004:HPS:987524.987595} are the classic example: all threads actively collaborate on bookkeeping data structures to track who is using a certain reference.
For structures with read-biased workloads, Read-Copy-Update (RCU)~\cite{Mckenney:2004:EDD:1048173} provides an appealing alternative. The programming style resembles a combination of reader-writer locks and lock-free programming. 
Multiple concurrent readers perform minimal bookkeeping -- often nothing they wouldn't already do.  A single writer at a time runs in parallel with readers, performing additional work to track which readers may have observed a node they wish to deallocate.
There are now RCU implementations of many common tree data structures~\cite{urcu_ieee,Triplett:2011:RSC:2002181.2002192,DBLP:conf/asplos/ClementsKZ12,mc_report,Arbel:2014:CUR:2611462.2611471,Kung:1980:CMB:320613.320619}, and
RCU plays a key role in Linux kernel memory management~\cite{Mckenney01read-copyupdate}.
%There are now RCU variants of linked lists~\cite{urcu_ieee}, hash tables~\cite{Triplett:2011:RSC:2002181.2002192}, and balanced trees~\cite{DBLP:conf/asplos/ClementsKZ12,mc_report,Arbel:2014:CUR:2611462.2611471,Kung:1980:CMB:320613.320619,mc_report}, and RCU plays a central role in portions of the Linux kernel~\cite{Mckenney01read-copyupdate}.

However, RCU primitives remain non-trivial to use correctly: developers must ensure they release each node exactly once, from exactly one thread, \emph{after} ensuring other threads are finished with the node in question.
Model checking can be used to validate correctness of implementations for a mock client~\cite{LiangMKM16,Desnoyers:2013:MSM:2506164.2506174,Kokologiannakis:2017:SMC:3092282.3092287,DBLP:conf/cav/AlglaveKT13}, but this does not guarantee correctness of arbitrary client code.
Sophisticated verification logics can prove correctness of the RCU primitives and clients~\cite{Gotsman:2013:VCM:2450268.2450289,fu2010reasoning,verrcu,Mandrykin:2016:TDV:3001219.3001297}.
But these techniques require significant verification expertise to apply, and are specialized to individual data structures or implementations.
One important reason for the sophistication in these logics stems from the complexity of the underlying memory reclamation model. However, Meyer and Wolff~\cite{myr} show that a suitable abstraction enables separating verifying \textit{correctness} of concurrent data structures from its underlying reclamation model under the assumption of \textit{memory safety}, and study proofs of correctness assuming memory safety.
% We propose the first type system to guarantee memory safety for clients of an RCU implementation, by providing suitable abstractions of the RCU model. 

We propose a type system to ensure that RCU client code uses the RCU primitives safely, ensuring memory safety for concurrent data structures using RCU memory management.
We do this in a general way, not assuming the client implements any specific data structure, only one satisfying some basic properties common to RCU data structures (such as having a \textit{tree} memory footprint).  In order to do this, we must also give a formal operational model of the RCU primitives that abstracts many implementations, without assuming a particular implementation of the RCU primitives.
We describe our RCU semantics and type system, prove our type system sound against the model (which ensures memory is reclaimed correctly), and show the type system in action on two important RCU data structures.
% Specifically what we have done is :
% \begin{itemize}
% \item \textbf{\textsf{Formalising RCU Semantics}}, We formalise the RCU programming model in section \textsf{3.1}.  capturing the core semantics of RCU.  We define a core language including \textsf{Read} and \textsf{Write} RCU block constructs in section \textsf{3} and operational semantics for this language in section \textsf{3.1}
% \item \textbf{\textsf{An Ownership Type System}}, We got inspired by "Multiple Aggregate Entry Points For Ownership Types" paper when deciding the fundemental theory of our type system as \textsf{Ownership-Types}. We propose a type system based on \textsf{Ownership-Types} ~\cite{Ostlund:2012:MAE:2367163.2367175} to enforce memory-safe usage of RCU primitives. To our best knowledge this is the first use of ownership types for weak models.
% \item \textbf{\textsf{Formalism, Proofs}} We formalised the type system in section \textsf{3.2}. We run type-system on Linked List and Binary Serach Tree example. We prove the soundness of the types system using \textsf{Views-Framework} ~\cite{views}.
% \end{itemize}

Our contributions include:
\begin{itemize}
\item A general (abstract) operational model for RCU-based memory management
\item A type system that ensures code uses RCU memory management correctly, which is signifiantly simpler than full-blown verification logics
\item Demonstration of the type system on two examples: a linked-list based bag and a binary search tree  
\item A proof that the type system guarantees memory safety when using RCU primitives.
\end{itemize}
\raggedbottom
\section{Background \& Motivation}
In this section, we recall the general concepts of read-copy-update concurrency.
We use the RCU linked-list-based bag~\cite{McKenney2015SomeEO} from Figure \ref{fig:rculist} as a running example.  It includes annotations for our type system, which will be explained in Section \ref{subsection:type-action}.
%One of the most well-known and frequently used linked data structures following RCU programming pattern, singly acyclic linked list. Table ~\ref{tab:rculist} shows how we implement a RCU linked list with our types. We are going to explain the RCU setting with running code in Table ~\ref{tab:rculist} along with pictorial explanations to crucial interleavings in following subsections.
\begin{figure*}[t!]
\begin{tabular}{p{0.47\textwidth}p{0.47\textwidth}}
\begin{lstlisting}[basicstyle=\scriptsize\ttfamily]
struct BagNode{
  int data;
  BagNode<rcuItr> Next;
}
BagNode<rcuRoot> head;
void add(int toAdd){
WriteBegin;
BagNode nw = new;
$\assert{nw\mathsf{:rcuFresh}  \{\}}$
nw.data = toAdd;
$\assert{head\mathsf{:rcuRoot},par\mathsf{:undef}, cur\mathsf{:undef}}$
BagNode<rcuItr> par,cur = head;
$\assert{head\mathsf{:rcuRoot},par\mathsf{:rcuItr} \epsilon  \{\}}$
$\assert{cur\mathsf{:rcuItr} \epsilon  \{\}}$
cur = par.Next;
$\assert{cur\mathsf{:rcuItr} Next \{\}}$
$\assert{par\mathsf{:rcuItr} \epsilon \{Next\mapsto cur\}}$
while(cur.Next != null){
  $\assert{cur\mathsf{:rcuItr} (Next)^{k}.Next \{\}}$
  $\assert{par\mathsf{:rcuItr}  (Next)^{k} \{Next \mapsto cur\}}$
  par = cur;
  cur = par.Next;
  $\assert{cur\mathsf{:rcuItr} (Next)^{k}.Next.Next \{\}}$
  $\assert{par\mathsf{:rcuItr}(Next)^{k}.Next \{Next \mapsto cur\}}$
}
$\assert{nw\mathsf{:rcuFresh}  \{\}}$
$\assert{cur\mathsf{:rcuItr} (Next)^{k}.Next \{Next\mapsto null\}}$
$\assert{par\mathsf{:rcuItr}  (Next)^{k} \{Next \mapsto cur\}}$
nw.Next= null;
$\assert{nw\mathsf{:rcuFresh}  \{Next\mapsto null\}}$
$\assert{cur\mathsf{:rcuItr} (Next)^{k}.Next \{Next \mapsto null\} }$
cur.Next=nw;
$\assert{nw\mathsf{:rcuItr}  (Next)^{k}.Next.Next  \{Next\mapsto null\}}$
$\assert{cur\mathsf{:rcuItr} (Next)^{k}.Next \{Next \mapsto nw\} }$
WriteEnd;
}
\end{lstlisting}&
\begin{lstlisting}[basicstyle=\scriptsize\ttfamily]
void remove(int toDel){
WriteBegin;
$\assert{head\mathsf{:rcuRoot},par:\mathsf{undef}, cur\mathsf{:undef}}$
BagNode<rcuItr> par,cur = head;
$\assert{head\mathsf{:rcuRoot},par\mathsf{:rcuItr}\epsilon  \{\},cur\mathsf{:rcuItr} \epsilon  \{\}}$
cur = par.Next;
$\assert{cur\mathsf{:rcuItr} Next \{\}}$
$\assert{par\mathsf{:rcuItr} \epsilon \{Next\mapsto cur\}}$
while(cur.Next != null&&cur.data != toDel)
{
  $\assert{cur\mathsf{:rcuItr} (Next)^{k}.Next \{\}}$
  $\assert{par\mathsf{:rcuItr}  (Next)^{k} \{Next \mapsto cur\}}$
  par = cur;
  cur = par.Next;
  $\assert{cur\mathsf{:rcuItr} (Next)^{k}.Next.Next \{\}}$
  $\assert{par\mathsf{:rcuItr}  (Next)^{k}.Next \{Next \mapsto cur\}}$
}
$\assert{nw\mathsf{:rcuFresh} \{\}}$
$\assert{par\mathsf{:rcuItr}  (Next)^{k} \{Next \mapsto cur\}}$
$\assert{cur\mathsf{:rcuItr} (Next)^{k}.Next \{\}}$
BagNode<rcuItr> curl = cur.Next;
$\assert{cur\mathsf{:rcuItr} (Next)^{k}.Next \{Next \mapsto curl\}}$
$\assert{curl\mathsf{:rcuItr} (Next)^{k}.Next.Next \{\}}$
par.Next = curl;
$\assert{par\mathsf{:rcuItr}  (Next)^{k} \{Next \mapsto curl\}}$
$\assert{cur\mathsf{:unlinked}}$
$\assert{cur\mathsf{:rcuItr} (Next)^{k}.Next \{\}}$
SyncStart;
SyncStop;
$\assert{cur\mathsf{:freeable}}$
Free(cur);
$\assert{cur\mathsf{:undef}}$
WriteEnd;
}
\end{lstlisting}
%\end{minipage}
%\end{tabular*}
\end{tabular}
\vspace{-2em}
\caption{RCU client: singly linked list based bag implementation.}
\label{fig:rculist}
\end{figure*}

As with concrete RCU implementations, we assume threads operating on a structure are either performing read-only traversals of the structure --- \emph{reader threads} --- or are performing an update --- \emph{writer threads} --- similar to the use of many-reader single-writer reader-writer locks.\footnote{RCU implementations supporting multiple concurrent writers exist~\cite{Arbel:2014:CUR:2611462.2611471}, but are the minority.}  It differs, however, in that readers may execute concurrently with the (single) writer.

This distinction, and some runtime bookkeeping associated with the read- and write-side critical sections, allow this model to determine at modest cost when a node unlinked by the writer can safely be reclaimed.

Figure \ref{fig:rculist} gives the code for adding and removing nodes from a bag. Type checking for all code, including membership queries for bag, can be found in our technical report~\cite{isotek}. Algorithmically, this code is nearly the same as any sequential implementation. There are only two differences.
First, the read-side critical section in \texttt{member} is indicated by the use of \lstinline|ReadBegin| and \lstinline|ReadEnd|; the write-side critical section is between \lstinline|WriteBegin| and \lstinline|WriteEnd|.
Second, rather than immediately reclaiming the memory for the unlinked node, \texttt{remove} calls \lstinline|SyncStart| to begin a \emph{grace period} --- a wait for reader threads that may still hold references to unlinked nodes to finish their critical sections.
 %with saving the readers on the detached node as bounding readers. 
 \lstinline|SyncStop| blocks execution of the writer thread until these readers exit their read critical section (via \lstinline|ReadEnd|). These are the essential primitives for the implementation of an RCU data structure.

These six primitives together track a critical piece of information: which reader threads' critical sections overlapped the writer's.
%When a reader thread executes \lstinline|ReadBegin|, some bookkeping data structures are updated to indicate that thread is an active reader; \lstinline|ReadEnd| atomically revokes that status.
Implementing them efficiently is challenging~\cite{urcu_ieee}, but possible.
The Linux kernel for example finds ways to reuse existing task switch mechanisms for this tracking, so readers incur no additional overhead.
The reader primitives are semantically straightforward -- they atomically record the start, or completion, of a read-side critical section. % for that thread.

The more interesting primitives are the write-side primitives and memory reclamation.
\lstinline|WriteBegin| performs a (semantically) standard mutual exclusion with regard to other writers, so only one writer thread may modify the structure \emph{or the writer structures used for grace periods}. % It also initializes a
%\texttt{Free}, as the name suggests, reclaims the memory pointed by its argument when active readers are done with the memory that is to be reclaimed.

\lstinline|SyncStart| and \lstinline|SyncStop| implement \emph{grace periods}~\cite{grc}: a mechanism to wait for readers to finish with any nodes the writer may have unlinked.  A grace period begins when a writer requests one, and finishes when all reader threads active \emph{at the start of the grace period} have finished their current critical section.
Any nodes a writer unlinks before a grace period are physically unlinked, but not logically unlinked until after one grace period.

An attentive reader might already realize that our usage of logical/physical unlinking is different than the one used in data-structures literature where typically a \textit{logical deletion} (e.g., marking) is followed by a \textit{physical deletion} (unlinking).
Because all threads are forbidden from holding an interior reference into the data structure after leaving their critical sections, waiting for active readers to finish their critical sections ensures they are no longer using any nodes the writer unlinked prior to the grace period.
This makes actually freeing an unlinked node after a grace period safe.
%a period of time where the writer waits for all readers active at the that after a node is logically removed from a data structure, the freeing thread must wait for some time to ensure threads that may have read the pointer before it was unlinked are done using the reference.  Typically, the notion of a reader thread being ``done'' with a reference is tied to its read-side critical section: reader threads are forbidden to use interior pointers to inside an RCU structure after leaving their critical section --- even if the use would occur in a subsequent critical section, since the referenced memory may have been reclaimed.

\lstinline|SyncStart| conceptually takes a snapshot of all readers active when it is run.
\lstinline|SyncStop| then blocks until all those threads in the snapshot have finished at least one critical section.
%is a blocking operation.  Intuitively, it waits for all readers whose critical sections overlapped with \lstinline|SyncStart-SyncStop| calls to finish their critical section.  Once those threads have finished, no other threads retain usable references to unlinked nodes, and the writer thread may continue.
\lstinline|SyncStop| does not wait for \emph{all} readers to finish, and does not wait for all overlapping readers to simultaneously be out of critical sections.

To date, every description of RCU semantics, most centered around the notion of a grace period, has been given algorithmically, as a specific (efficient) implementation.  While the implementation aspects are essential to real use, the lack of an abstract characterization makes judging the correctness of these implementations -- or clients -- difficult in general. In Section \ref{sec:semantics} we give formal \emph{abstract}, \emph{operational} semantics for RCU implementations -- inefficient if implemented directly, but correct from a memory-safety and programming model perspective, and not tied to specific low-level RCU implementation details.
%We handle the management of grace periods (and in particular, distinguishing multiple overlapping read-side critical sections by the same thread) in intuitive ways, which would be terribly inefficient to implement directly, but express the intended semantics of more sophisticated implementations.
%For these semantics to correctly manage memory reclamation in an RCU data structure implementation, a number of assumptions must be satisfied:
To use these semantics or a concrete implementation correctly, client code must ensure:
\begin{itemize}
\item Reader threads never modify the structure
\item No thread holds an interior pointer into the RCU structure across critical sections
\item Unlinked nodes are always freed by the unlinking thread \emph{after} the unlinking, \emph{after} a grace period, and \emph{inside} the critical section
\item Nodes are freed at most once
\end{itemize}
In practice, RCU data structures typically ensure additional invariants to simplify the above, e.g.:
\begin{itemize}
\item The data structure is always a tree
\item A writer thread unlinks or replaces only one node at a time. 
\end{itemize}
and our type system in Section \ref{sec:tslbl} guarantees these invariants.
\raggedbottom
\section{Semantics}
\label{sec:semantics}
In this section, we outline the details of an abstract semantics for RCU implementations. It captures the core client-visible semantics of most RCU primitives, but not the implementation details required for efficiency~\cite{Mckenney01read-copyupdate}.
%RCU allows a single mutator thread and  multiple  reader threads to traverse a data structure. When a writer mutates some state, it waits  until all concurrent reads are completed to free the mutated state. To represent this in our semantics, we augment the  machine state,$\textsf{MState}$, ranged over by $\sigma$, consists of:
In our semantics, shown in Figure \ref{fig:operationalsemrcu}, an abstract machine state, \textsf{MState}, contains:\\

%\noindent
%\begin{minipage}{0.5\textwidth}
\begin{itemize}
\item A stack $s$, of type $\textsf{Var} \times \textsf{TID} \rightharpoonup \textsf{Loc}$
\item A heap, $h$, of type $\textsf{Loc} \times \textsf{FName} \rightharpoonup \textsf{Val}$
\item A lock, $l$, of type $\textsf{TID} \uplus \{\textsf{unlocked}\}$
\item A root location $rt$ of type $\textsf{Loc}$
%\end{itemize}
%\end{minipage}
%\begin{minipage}{0.5\textwidth}
%\begin{itemize}
\item A read set, $R$, of type $\mathcal{P}(\textsf{TID})$ and 
\item A bounding set, $B$, of type $\mathcal{P}(\textsf{TID})$ 
\end{itemize}
%\end{minipage}\\

The lock $l$ enforces mutual exclusion between write-side critical sections.
The root location $rt$ is the root of  an \textsf{RCU} data structure. We model only a single global RCU data structure; the generalization to multiple structures is straightforward but complicates formal development later in the paper.
The reader set $R$ tracks the thread IDs (TIDs) of all threads currently executing a read block. 
The bounding set $B$ tracks which threads the writer is \emph{actively} waiting for during a grace period --- it is empty if the writer is not waiting.
\begin{figure*}\scriptsize%%%%%%%%%%%%OPdERATIONAL SEMANTICS%%%%%%%%%%%%%%
  \grammar
  %$\desugarsync$ \\
%  $\desugarwrite$\\
%  $\desugarread$
\[\begin{array}{c@{\;}rl@{\Downarrow_{\mathit{tid}}}ll} 
(\textsc{RCU-WBegin}) & \llbracket\texttt{WriteBegin}\rrbracket & (s,h,\mathsf{unlocked},rt,R,B) &(s,h,l,rt,R,B) & \\
(\textsc{RCU-WEnd}) & \llbracket\texttt{WriteEnd}\rrbracket & (s,h,l,rt,R,B) & (s,h,\mathsf{unlocked},rt,R,B) & \\
(\textsc{RCU-RBegin}) & \llbracket\texttt{ReadBegin}\rrbracket & (s,h,tid,rt,R,B) & (s,h,tid,rt,R\uplus \{tid\},B) \qquad \mathit{tid} \neq l &  \\
(\textsc{RCU-REnd}) & \llbracket\texttt{ReadEnd}\rrbracket & (s,h,tid,rt,R\uplus\{tid\},B) & (s,h,l,rt,R,B\setminus \{tid\}) \qquad  \mathit{tid} \neq l& \\
(\textsc{RCU-SStart}) & \llbracket\texttt{SyncStart}\rrbracket & (s,h,l,rt,R,\emptyset) & (s,h,l,rt,R,R) &\\  %\forall_{t\in R}\ldotp t \notin r \\
(\textsc{RCU-SStop}) & \llbracket\texttt{SyncStop}\rrbracket & (s,h,l,rt,R,\emptyset) & (s,h,l,rt,R,\emptyset) &  \\
(\textsc{Free}) & {\llbracket\texttt{Free}(x)\rrbracket} & (s,h,l,rt,R,\emptyset) & (s,h',l,rt,R,\emptyset) &
\end{array}\]
$\textrm{provided}~ \forall_{f,o'}\ldotp rt \neq s(x,tid) \textrm{ and } o' \neq s(x,tid) \implies h(o',f) = h'(o',f) \textrm{ and } \forall_{f}\ldotp h'(o,f)=\textsf{undef}$
\[\begin{array}{c@{\;}rl@{\Downarrow_{\mathit{tid}}}ll} 
(\textsc{HUpdt}) & \llbracket\texttt{x.f=y}\rrbracket   &(s,h,l,rt,R,B)& (s,h[s(x,tid),f \mapsto s(y,tid)],l,rt,R,B) &\\
(\textsc{HRead})   & \llbracket\texttt{y=x.f}\rrbracket   &(s,h,l,rt,R,B)& (s[(y,tid) \mapsto h(s(x,tid),f)],h,l,rt,R,B) & \\
(\textsc{SUpdt}) & \llbracket\texttt{y=x}\rrbracket     &(s,h,l,rt,R,B)& (s[(y,tid) \mapsto (x,tid)],h,l,rt,R,B) &\\
(\textsc{HAlloc}) & \llbracket\texttt{y=new}\rrbracket &(s,h,l,rt,R,B)& (s,h[\ell\mapsto\mathsf{nullmap}],l,rt,R,B) & 
\end{array}\]
$  \textrm{provided}~ rt \neq s(y,tid) \textrm{ and } s[(y,tid) \mapsto \ell], \textrm{ and }$
$h[\ell \mapsto \mathsf{nullmap}] \overset{\mathrm{def}}{=} \lambda (o',f) . \textrm{ if } o=o' \textrm{ then } skip \textrm{ else } h(o',f)$
\caption{Operational semantics for \textsf{RCU}.}
\label{fig:operationalsemrcu}
\vspace{-2mm}
\end{figure*}%%%%%%%%%%%%%%%%%%%%%OPERATIONAL SEMANTICS ENDS%%%%%%%%%%%%%%%%%%

Figure \ref{fig:operationalsemrcu} gives operational semantics for \emph{atomic} actions; conditionals, loops, and sequencing all have standard semantics, and parallel composition uses sequentially-consistent interleaving semantics.

%(\textbf{\textit{Basic Actions}}) - 
The first few atomic actions, for writing and reading fields, assigning among local variables, and allocating new objects, are typical of formal semantics for heaps and mutable local variables. \lstinline|Free| is similarly standard.
%
%(\textbf{\textit{Writer Thread Actions}}) - 
A writer thread's critical section is bounded by \lstinline|WriteBegin| and \lstinline|WriteEnd|, which acquire and release the lock that enforces mutual exclusion between writers.  \lstinline|WriteBegin| only reduces (acquires) if the lock is \textsf{unlocked}.

Standard RCU APIs include a primitive \texttt{synchronize\_rcu()} to wait for a grace period for the current readers.  We decompose this here into two actions, \lstinline|SyncStart| and \lstinline|SyncStop|.
%, for reasons both pedagogical (to separate the start of the grace period from its completion), and technical (our proof framework makes this formulation most convenient).
\lstinline|SyncStart| initializes the blocking set to the current set of readers --- the threads that may have already observed any nodes the writer has unlinked.
\lstinline|SyncStop| blocks until the blocking set is emptied by completing reader threads. However, it does not wait for \emph{all} readers to finish, and does not wait for all overlapping readers to simultaneously be out of critical sections. If two reader threads $A$ and $B$ overlap some \lstinline|SyncStart|-\lstinline|SyncStop|'s critical section, it is possible that $A$ may exit and re-enter a read-side critical section before $B$ exits, and vice versa.  Implementations must distinguish subsequent read-side critical sections from earlier ones that overlapped the writer's initial request to wait: since \lstinline|SyncStart| is used \emph{after} a node is physically removed from the data structure and readers may not retain RCU references across critical sections, $A$ re-entering a fresh read-side critical section will not permit it to re-observe the node to be freed.
%requires the blocking set to be empty to run, and thus blocks when it is non-empty.  
%Reader threads are responsible for shrinking the blocking set. 
%Once the grace period ends, the unlinked/mutated node is reclaimed with \lstinline|Free|.

%(\textbf{\textit{Reader Thread Actions}}) - 
%A reader thread, like the writer thread, has a critical section, in this case 
Reader thread critical sections are
bounded by \lstinline|ReadBegin| and \lstinline|ReadEnd|.  \lstinline|ReadBegin| simply records the current thread's presence as an active reader.
\lstinline|ReadEnd| removes the current thread from the set of active readers, and also removes it (if present) from the blocking set --- if a writer was waiting for a certain reader to finish its critical section, this ensures the writer no longer waits once that reader has finished its current read-side critical section.

%(\textbf{\textit{Implementing Grace Periods}}) - 
Grace periods are implemented by the combination of \lstinline|ReadBegin|, \lstinline|ReadEnd|, \lstinline|SyncStart|, and \lstinline|SyncStop|.
% work together to implement grace periods.  
\lstinline|ReadBegin| ensures the set of active readers is known.  When a grace period is required, \lstinline|SyncStart;SyncStop;| will store (in $B$) the active readers (which may have observed nodes before they were unlinked), and wait for reader threads to record when they have completed their critical section (and implicitly, dropped any references to nodes the writer wants to free) via \lstinline|ReadEnd|.
 %of threads the writer waits for to the set of active readers when \lstinline|SyncStart| is executed (recording the readers that may have observed any nodes unlinked by the writer), and \lstinline|SyncStop| waits for that set of threads to become empty.
%\lstinline|ReadEnd| ``informs'' the writer thread of its exit from the critical section by removing itself from $B$, and once every reader that overlapped the \lstinline|SyncStart| has completed its critical section, the blocking set will be empty and the writer will resume.

These semantics do permit a reader in the blocking set to finish its read-side critical section and enter a \emph{new} read-side critical section before the writer wakes.  In this case, \emph{the writer waits only for the first critical section of that reader to complete}, since entering the new critical section adds the thread's ID back to $R$, but not $B$.
\raggedbottom
\makeatletter % allow us to mention @-commands
\def\arcr{\@arraycr}
\makeatother
\section{Type System \& Programming Language}
\label{sec:tslbl}
In this section, we present a simple imperative programming language with two block constructs for modeling \textsf{RCU}, and a type system that ensures proper (memory-safe) use of the language.
The type system ensures memory safety by enforcing these sufficient conditions:
%To ensure memory safety, we must show how our type system makes programs preserve the following properties:
\begin{itemize}
\item A heap node can only be freed if it is no longer accessible from an RCU data structure or from local variables of other threads. To achieve this we ensure the reachability and access which can be suitably restricted. We explain how our types support a delayed ownership transfer for the deallocation.
\item Local variables may not point inside an RCU data structure unless they are inside an \textsf{RCU} read or write block.
\item Heap mutations are \textit{local}: each unlinks or replaces exactly one node.
\item The RCU data structure remains a tree.  While not a fundamental constraint of RCU, it is a common constraint across known RCU data structures because it simplifies reasoning (by developers or a type system) about when a node has become unreachable in the heap.
\end{itemize}

We also demonstrate that the type system is not only sound, but useful:
we show how it types Figure \ref{fig:rculist}'s list-based bag implementation~\cite{McKenney2015SomeEO}.
We also give type checked fragments of a binary search tree to motivate advanced features of the type system; the full typing derivation can be found in our technical report~\cite{isotek} Appendix B. The BST requires type narrowing operations that refine a type based on dynamic checks (e.g., determining which of several fields links to a node).
In our system, we presume all objects contain all fields, but the number of fields is finite (and in our examples, small). This avoids additional overhead from tracking well-established aspects of the type system --- class and field types and presence, for example --- and focus on checking correct use of RCU primitives.  Essentially, we assume the code our type system applies to is already type-correct for a system like C or Java's type system.
\subsection{\textsf{RCU} Type System for \textsf{Write} Critical Section}
\label{subsection:rcu-typing}
Section \ref{subsection:rcu-typing} introduces \textsf{RCU} types and the need for subtyping. Section \ref{subsection:type-action}, shows how types describe program states, through code for Figure \ref{fig:rculist}'s list-based bag example. Section \ref{subsection:type-rules} introduces the type system itself.
\paragraph{}
\mypar{\textsf{RCU} Types}
\label{subsection:typwrt}
There are six types used in \textsf{Write} critical sections
\[
\tau ::=  \textsf{rcuItr}\;\rho\;\N \mid \textsf{rcuFresh}\;\N  \mid \textsf{unlinked} \mid \textsf{undef} \mid \textsf{freeable}
\mid \textsf{rcuRoot}\]
\paragraph{rcuItr} is the type given to references pointing into a shared RCU data structure. A \textsf{rcuItr} type can be used in either a write region or a read region (without the additional components). It indicates both that the reference points into the shared \textsf{RCU} data structure and that the heap location referenced by \textsf{rcuItr} reference is reachable by following the path $\rho$ from the root. A component $\N$ is a set of field mappings taking the field name to local variable names. Field maps are extended when the referent's fields are read.
The field map and path components track reachability from the root, and local reachability between nodes.  These are used to ensure the structure remains acyclic, and for the type system to recognize exactly when unlinking can occur.
 %The field map and path components are important for keeping track of heap locations through \textit{reached-by} and \textit{points-to} relations so that effects of mutations using references in \textsf{rcuItr} type are \textit{localized}.
%In general, we utilize path information to ensure that we preserve well shaped heap for the data structures, e.g. acyclicity. In conjuction with the path component, we employ the field mappings to assist in checking whether proper linking/mapping exists between the related nodes shown by references when unlinking occurs.

Read-side critical sections use \textsf{rcuItr} without path or field map components.  These components are both unnecessary for readers (who perform no updates) and would be invalidated by writer threads anyways.
%Similar to the ones in \textsf{Write} critical section, they  point into a shared RCU data structure. 
Under the assumption that reader threads do not hold references across critical sections, the read-side rules essentially only ensure the reader performs no writes, so we omit the reader critical section type rules. They can be found in our technical report~\cite{isotek} Appendix E.
 %together with the other standard rules (\textsc{T-Exchange}, \textsc{T-Par}, \textsc{T-Seq}, \textsc{T-Conseq} and \textsc{T-Skip}).
\paragraph{unlinked} is the type given to references to unlinked heap locations --- objects previously part of the structure, but now unreachable via the heap. A heap location referenced by an unlinked reference may still be accessed by reader threads, which may have acquired their own references before the node became unreachable. Newly-arrived readers, however, will be unable to gain access to these referents.
\paragraph{freeable} is the type given to references to an unlinked heap location that is safe to reclaim because it is known that no concurrent readers hold references to it.
Unlinked references become freeable after a writer has waited for a full grace period.
\paragraph{undef} is the type given to references where the content of the referenced location is inaccessible. A local variable of type \textsf{freeable} becomes \textsf{undef} after reclaiming that variable's referent.
\paragraph{rcuFresh} is the type given to references to freshly allocated heap locations. Similar to \textsf{rcuItr} type, it has field mappings set $\N$. We set the field mappings in the set of an existing \textsf{rcuFresh} reference to be the same as field mappings in the set of \textsf{rcuItr} reference when we replace the heap referenced by \textsf{rcuItr} with the heap referenced by \textsf{rcuFresh} for memory safe replacement.
\paragraph{rcuRoot} is the type given to the fixed reference to the root of the RCU data structure.  It may not be overwritten.
%An undefined reference can be loaded with a value -- it becomes \textsf{rcuItr} -- once we read the global unique root to it.
\paragraph{}
\mypar{Subtyping}
It is sometimes necessary to use imprecise types --- mostly for control flow joins.  Our type system performs these abstractions via subtyping on individual types and full contexts, as in Figure \ref{fig:sub-typing}.
%  We discuss subtyping before the full type rules to give a sense of what kinds of imprecision exist in our type system.

%%%%%%%%%%%%%%%%%%%%%%%%SUB TYPING JUDGEMENTS%%%%%%%%%%%%%%%%%%%%%%%%%%%%%%%%%%%%
\begin{figure}[!t]\scriptsize
\begin{mathpar}
 \fmapname \quad \fmapfieldemp \\ \fmapempty \quad \fmapextendn \quad \fmapdiff \\ \fmapwithf \quad \fmapextend %\and \typelocalmap %\and  \ftypedecl \and \rpth 
\\
\fbox{$\vdash \, \N \subt \N'$}
\and
\inferrule[\scriptsize(T-NSub3)]
{
}
{\vdash \, \N_{f,\emptyset} \subt \N([f \rightharpoonup y]) }
\and
\inferrule[\scriptsize(T-NSub4)]
{
}
{\vdash \,  \N_{\emptyset} \subt \N}
\and
\inferrule[\scriptsize(T-NSub5)]
{
}
{\vdash \, \N \subt \N}
\and
\inferrule*[left=\scriptsize(T-NSub2)]
{
}
{\vdash \, \N([f_2 \rightharpoonup y]) \subt \N([f_1|f_2 \rightharpoonup y])}
\and
\inferrule*[left=\scriptsize(T-NSub1)]
{
}
{\vdash \, \N([f_1 \rightharpoonup y]) \subt  \N([f_1|f_2 \rightharpoonup y])}
\and
\fbox{$\vdash \, \rho \subt \rho'$} \;\;
\inferrule*[left=\scriptsize(T-PSub1)]
{
}
{
 \vdash  \rho.f_1 \subt \rho.f_1|f_2
}
\and
\inferrule*[left=\scriptsize(T-PSub2)]
{
}
{
 \vdash  \rho.f_2 \subt  \rho.f_1|f_2
}
\and
\inferrule*[left=\scriptsize(T-PSub3)]
{
}
{
 \vdash  \rho \subt  \rho
}
\\
\fbox{$\vdash \, T \subt T'$} 
\and
\inferrule[\scriptsize(T-TSub2)]
{
}
{
\vdash \textsf{rcuItr}   \subt \textsf{rcuItr}
}
\and
\inferrule[\scriptsize(T-TSub)]
{
}
{
\vdash \textsf{rcuItr}\; \_   \subt \textsf{undef}
}
\and
\inferrule[\scriptsize(T-TSub1)]
{
\vdash\rho\subt \rho' \qquad \vdash \N \subt \N'
}
{
\vdash \textsf{rcuItr} \, \rho \, \N  \subt  \textsf{rcuItr} \, \rho' \, \N'
}
\and
\fbox{$\vdash \, \Gamma \subt \Gamma'$} \;\;
\inferrule*[left=\scriptsize(T-CSub1)]
{
\vdash\Gamma \subt \Gamma' \;\; \vdash \text{T}  \subt \text{T'}
}
{
\vdash \Gamma\, , x:\text{T}  \subt \Gamma'\, , x:\text{T'}
}
\and
\inferrule*[left=\scriptsize(T-CSub)]
{
}
{
 \vdash \Gamma \subt  \Gamma
}
\end{mathpar}
\vspace{-2em}
\caption{Subtyping rules.}
\label{fig:sub-typing}
\end{figure}

Figure \ref{fig:sub-typing} includes four judgments for subtyping.  The first two --- $\vdash\N\prec:\N'$ and $\vdash\rho\prec:\rho'$ --- describe relaxations of field maps and paths respectively.  $\vdash\N\prec:\N'$ is read as ``the field map $\N$ is more precise than $\N'$'' and similarly for paths.
The third judgment $\vdash T\prec:T'$ uses path and field map subtyping to give subtyping among \textsf{rcuItr} types --- one \textsf{rcuItr} is a subtype of another if its paths and field maps are similarly more precise --- and to allow \textsf{rcuItr} references to be ``forgotten'' --- this is occasionally needed to satisfy non-interference checks in the type rules.
The final judgment $\vdash\Gamma\prec:\Gamma'$ extends subtyping to all assumptions in a type context.

\begin{figure}[!t]\scriptsize
\begin{mathpar}
\fbox{$\Gamma \vdash_{M,R} C \dashv \Gamma'$} \;\;\;\;\;\;
\inferrule[\scriptsize(T-ReIndex)]
{
}
{
\Gamma \vdash C_{k} \dashv  \Gamma[\rho.f^{k}/\rho.f^{k}.f]
}
\and
\inferrule*[left=\scriptsize(T-Loop1)]
{
\Gamma(x) = \textsf{bool} \and
\Gamma \vdash C \dashv \Gamma 
}
{
\Gamma \vdash \textsf{while}(x)\{C\} \dashv  \Gamma
}
\and
\inferrule*[left=\scriptsize(T-Branch1)]
{
 \Gamma\, ,x:\textsf{rcuItr}\,\rho \, \N([f_1 \rightharpoonup  z]) \vdash C_1 \dashv\Gamma_4 \and
  \Gamma\, ,x:\textsf{rcuItr}\, \rho \, \N([f_2 \rightharpoonup z]) \vdash C_2 \dashv \Gamma_4
}
{
\Gamma\, ,x:\textsf{rcuItr}\, \rho \, \N([f_1\mid f_2 \rightharpoonup z]) \vdash \textsf{if}(x.f_1==z)  \textsf{ then } C_1 \textsf{ else } C_2 \dashv \Gamma_4
}
\and
\inferrule*[left=\scriptsize(T-Branch3)]
{
 \Gamma,\,x:\textsf{rcuItr} \, \rho \, \N([f\rightharpoonup y \setminus \textsf{null}]) \vdash C_1 \dashv \Gamma' \and \Gamma,\,x:\textsf{rcuItr} \, \rho \, \N([f\rightharpoonup y]) \vdash C_2 \dashv \Gamma'
}
{
\Gamma,\,x:\textsf{rcuItr} \, \rho \, \N([f\rightharpoonup y]) \vdash \textsf{if}(x.f == \textsf{null}) \textsf{ then } C_1  \textsf{ else } C_2 \dashv \Gamma'
}
\and
\inferrule*[left=\scriptsize(T-Loop2)]
{
\Gamma,\, x:\textsf{rcuItr}\;\rho \; \N([f\rightharpoonup \_]) \vdash C \dashv \Gamma, x:\textsf{rcuItr}\;\rho' \; \N([f\rightharpoonup \_])
}
{
\Gamma,\, x:\textsf{rcuItr}\;\rho \; \N([f\rightharpoonup \_]) \vdash \textsf{while}(x.f \neq \textsf{null})\{C\} \dashv   x:\textsf{rcuItr}\;\rho' \; \N([f\rightharpoonup \textsf{null}]) ,\,  \Gamma
}
\and
\inferrule*[left=\scriptsize(T-Branch2)]
{
\Gamma(x)= \textsf{bool} \\
 \Gamma \vdash C_1 \dashv\Gamma' \and \Gamma \vdash C_2 \dashv \Gamma'
}
{
\Gamma \vdash \textsf{if}(x) \textsf{ then } C_1  \textsf{ else } C_2 \dashv \Gamma'
}
\end{mathpar}
\vspace{-2em}
\caption{Type rules for control-flow.}
\label{fig:type-judgements-for-cf}
\end{figure}

It is often necessary to abstract the contents of field maps or paths, without simply forgetting the contents entirely.  In a binary search tree, for example, it may be the case that one node is a child of another, but \emph{which} parent field points to the child depends on which branch was followed in an earlier conditional (consider the lookup in a BST, which alternates between following left and right children).
%Since our list based bag example's objects have single field($Next$), we take small code snippet from our \textsf{BST} type checking effort in Appendix \ref{appendix:bst_del} in which objects have two fields $Left$ and $Right$. Field mappings take field names to local variable names. They are added to the field map, $\N$, when the object's fields are read. 
In Figure \ref{fig:altfield}, we see that \texttt{cur} aliases different fields of \texttt{par} -- either $Left$ or $Right$ -- in different branches of the conditional.
%We see the join of the type assertions in lines 6 and 11 to the one in the line 12. 
The types after the conditional must overapproximate this, here as $Left|Right\mapsto cur$ in \texttt{par}'s field map, and a similar path disjunction in \texttt{cur}'s path.
This is reflected in Figure \ref{fig:sub-typing}'s \textsc{T-NSub1-5} and \textsc{T-PSub1-2} -- within each branch, each type is coerced to a supertype to validate the control flow join.

Another type of control flow join is handling loop invariants -- where paths entering the loop meet the back-edge from the end of a loop back to the start for repetition.
Because our types include paths describing how they are reachable from the root, some abstraction is required to give loop invariants that work for any number of iterations -- in a loop traversing a linked list, the iterator pointer would na\"ively have different paths from the root on each iteration, so the exact path is not loop invariant. However, the paths explored by a loop are regular, so we can abstract the paths by permitting (implicitly) existentially quantified indexes on path fragments, which express the existence of \emph{some} path, without saying \emph{which} path. The use of an explicit abstract repetition allows the type system to preserve the fact that different references have common path prefixes, even after a loop.

Assertions for the \texttt{add} function in lines 19 and 20 of Figure \ref{fig:rculist} show the \textit{loop}'s effects on paths of iterator references used inside the loop, \texttt{cur} and \texttt{par}. 
On line 20, \texttt{par}'s path contains has $(Next)^{k}$.
The $k$ in the $(Next)^{k}$ abstracts the number of loop iterations run, implicitly assumed to be non-negative.
The trailing $Next$ in \texttt{cur}'s path on line 19 -- $(Next)^{k}.Next$ -- expresses the relationship between \texttt{cur} and \texttt{par}: \texttt{par} is reachable from the root by following $Next$ $k$ times, and \texttt{cur} is reachable via one additional $Next$.
The types of 19 and 20, however, are not the same as lines 23 and 24, so an additional adjustment is needed for the types to become loop-invariant.
\emph{Reindexing} (\textsc{T-ReIndex} in Figure \ref{fig:type-judgements-for-cf}) effectively increments an abstract loop counter, contracting $(Next)^k.Next$ to $Next^k$ everywhere in a type environment. This expresses the same relationship between \texttt{par} and \texttt{cur} as before the loop, but the choice of $k$ to make these paths accurate after each iteration would be one larger than the choice before.
Reindexing the type environment of lines 23--24 yields the type environment of lines 19--20, making the types loop invariant. The reindexing essentially chooses a new value for the abstract $k$. This is sound, because the uses of framing in the heap mutation related rules of the type system ensure uses of any indexing variable are never separated -- either all are reindexed, or none are.
\looseness=-1

\begin{figure}
\begin{lstlisting}[numbersep=1pt,xrightmargin=-1cm,xleftmargin=0.1\textwidth, numbers=left, numbersep=5pt,basicstyle=\scriptsize\ttfamily]
$\assert{cur:\mathsf{rcuItr}\; Left|Right \;\{\},\;\;par:\mathsf{rcuItr}\;  \epsilon \;\{Left|Right\mapsto cur\}}$
if(par.Left == cur){
  $\assert{cur:\mathsf{rcuItr}\; Left \;\{\} ,\;\; par:\mathsf{rcuItr}\;  \epsilon \;\{Left\mapsto cur\}}$
  par = cur;
  cur = par.Left;
  $\assert{cur:\mathsf{rcuItr}\; Left.Left \;\{\}, \;\; par:\mathsf{rcuItr}\;  Left \;\{Left\mapsto cur\}}$
}else{
  $\assert{cur:\mathsf{rcuItr}\; Right \;\{\}, \;\; par:\mathsf{rcuItr}\;  \epsilon \;\{Right\mapsto cur\}}$
  par = cur;
  cur = par.Right;
  $\assert{cur:\mathsf{rcuItr}\; Right.Right \;\{\}, \;\; par:\mathsf{rcuItr}\;  Right \;\{Right\mapsto cur\}}$
}
$\assert{cur:\mathsf{rcuItr}\; Left|Right.Left|Right \;\{\}, \;\; par:\mathsf{rcuItr}\;  Left|Right \;\{Left|Right\mapsto cur\}}$
\end{lstlisting}
\vspace{-1em}
\caption{Choosing fields to read.}
\label{fig:altfield}
\end{figure}

While abstraction is required to deal with control flow joins, reasoning about whether and which nodes are unlinked or replaced, and whether cycles are created, requires precision. Thus the type system also includes means (Figure \ref{fig:type-judgements-for-cf}) to refine imprecise paths and field maps.
%
%We should also note that being precise in the field names is so crucial for mutation actions. This means we need to know which fields are used to replace/unlink a heap node. Then, what is the mechanism to become more/less precise in the fields used in both path and field mappings of the \textsf{rcuItr} type? 
In Figure \ref{fig:altfield}, we see a conditional with the condition $par.Left == cur$. The type system matches this condition to the imprecise types in line 1's typing assertion, and refines the initial type assumptions in each branch accordingly (lines 2 and 7) based on whether execution reflects the truth or falsity of that check.
Similarly, it is sometimes required to check -- and later remember -- whether  a field is null, and the type system supports this.
%refines the assertions in Listing \ref{lst:altfield} line 1 to the one in the line 3 for the \textit{then} branch and to the one in the line 8 for the \textit{else} branch via making types more precise on the object fields(\textsc{T-Branch1} shown in Figure \ref{fig:type-judgements-for-cf}).
%(\textit{\textbf{Points to }}\textsf{null}) - Typing rules(\textsc{T-Branch3} and \textsc{T-Loop2}) in Figure \ref{fig:type-judgements-for-cf} refine field mappings with respect to \texttt{null} value in control flow statements.
%%%%%%%%%%%%%%%%%%%%%%%%%%%%%%%%%%%%%%%%%%%%%%%
\subsection{Types in Action}
\label{subsection:type-action}
%%%%%%%%%%%%%%%%%%%%%%TYPE JUDGEMENTS ENDS%%%%%%%%%%%%%%%%%%%%%%%%%%%
%We have typing context $\Gamma$ of the form $x_1 : T_1, \ldots x_n : T_n$.

The system has three forms of typing judgement: 
$\Gamma \vdash C$ for standard typing outside RCU critical sections; 
$\Gamma \vdash_R C \dashv \Gamma'$ for reader critical sections, and
$\Gamma \vdash_M C \dashv \Gamma'$ for writer critical sections.
The first two are straightforward, essentially preventing mutation of the data structure, and preventing nesting of a writer critical section inside a reader critical section.
The last, for writer critical sections, is flow sensitive: the types of variables may differ before and after program statements.  This is required in order to reason about local assumptions at different points in the program, such as recognizing that a certain action may unlink a node.
Our presentation here focuses exclusively on the judgment for the write-side critical sections.

Below, we explain our types through the list-based bag implementation~\cite{McKenney2015SomeEO} from Figure \ref{fig:rculist}, highlighting how the type rules handle different parts of the code.
Figure \ref{fig:rculist} is annotated with ``assertions'' -- local type environments -- in the style of a Hoare logic proof outline.  As with Hoare proof outlines, these annotations can be used to construct a proper typing derivation.
%  We aim to
%\begin{itemize}
%\item explain why we employ each of these components. How do they serve to certify memory safety in \textsf{RCU} programs? How do they represent the program?
%\item explain the \textit{flow-sensitive} nature of the type system. 
%\end{itemize}
%The code fragments we refer to below from \ref{fig:rculist} are annotated with assertions about type environments between each statement (in the style of a proof outline in Hoare Logic) to demonstrate informally how types change before and after various statements.

\mypar{Reading a Global \textsf{RCU} Root}
All \textsf{RCU} data structures have fixed roots, which we characterize with the \textsf{rcuRoot} type.  Each operation in Figure \ref{fig:rculist} begins by reading the root into a new \textsf{rcuItr} reference used to begin traversing the structure.
After each initial read (line 12 of \lstinline|add| and line 4 of \lstinline|remove|), the path of \texttt{cur} reference is the empty path ($\epsilon$) and the field map is empty ($\{\}$), because it is an alias to the root, and none of its field contents are known yet.

\mypar{Reading an Object Field and a Variable}
As expected, we explore the heap of the data structure via reading the objects' fields. 
Consider line 6 of \lstinline|remove| and its corresponding pre- and post- type environments.  Initially \texttt{par}'s field map is empty.  After the field read, its field map is updated to reflect that its $Next$ field is aliased in the local variable \texttt{cur}.  Likewise, afer the update, \texttt{cur}'s path is $Next$ ($=\epsilon\cdot Next$), extending the par node's path by the field read.  This introduces field aliasing information that can subsequently be used to reason about unlinking.
%For example, the assertion of \textsf{remove} function in line 7 says that the path to reach the heap node pointed by \texttt{current} is $Next$. The assertion in line 8 for \texttt{parent} tells that there exists a link from the referent of \textsf{parent} to the referent of \textsf{current}. As we observe in the post condition in the lines 7 and 8, reading \texttt{current} to \texttt{parent}'s $Next$ field in line 6 adds the field mapping entry $Next \mapsto \texttt{current}$ to the field mapping set of \texttt{parent} and extends the path of \texttt{current} with $Next$. We call this "properly" linked heap nodes which is a requirement for safe unlinking.

%(\textit{\textbf{Mutation - Unlinking a Heap Node}}) - When we unlink a heap node, the type of the reference to it changes from \textsf{rcuItr} to \textsf{unlinked}. We observe this change for \texttt{current} reference in the assertions of \texttt{remove} in lines 22 and 26.
\mypar{Unlinking Nodes}
Line 24 of \lstinline|remove| in Figure \ref{fig:rculist} unlinks a node.
The type annotations show that before that line \lstinline|cur| is in the structure (\textsf{rcuItr}), while afterwards its type is \textsf{unlinked}.
The type system checks that this unlink disconnects only one node: note how the types of \lstinline|par|, \lstinline|cur|, and \lstinline|curl| just before line 24 completely describe a section of the list.

%(\textit{\textbf{Mutation - Reclaiming a Heap Node after a Grace Period}}) - 
\mypar{Grace and Reclamation}
After the referent of \texttt{cur} is unlinked, concurrent readers traversing the list may still hold references.  So it is not safe to actually reclaim the memory until after a grace period.
Lines 28--29 of \lstinline|remove| initiate a grace period and wait for its completion.  At the type level, this is reflected by the change of \lstinline|cur|'s type from \textsf{unlinked} to \textsf{freeable}, reflecting the fact that the grace period extends until any reader critical sections that might have observed the node in the structure have completed.
This matches the precondition required by our rules for calling \lstinline|Free|, which further changes the type of \lstinline|cur| to \textsf{undef} reflecting that \lstinline|cur| is no longer a valid reference. The type system also ensures no local (writer) aliases exist to the freed node and understanding this enforcement is twofold. First, the type system requires that only \textsf{unlinked} heap nodes can be freed. Second, framing relations in rules related to the heap mutation ensure no local aliases still consider the node linked.
%enforce \textit{locality} on the effects of the heap update -- only one node becomes \textsf{unlinked}.

\mypar{Fresh Nodes}
Some code must also allocate new nodes, and the type system must reason about how they are incorporated into the shared data structure.
Line 8 of the \texttt{add} method allocates a new node \texttt{nw}, and lines 10 and 29 initialize its fields.
The type system gives it a \textsf{fresh} type while tracking its field contents, until line 32 inserts it into the data structure.  The type system checks that nodes previously reachable from \lstinline|cur| remain reachable: note the field maps of \lstinline|cur| and \texttt{nw} in lines 30--31 are equal (trivially, though in general the field need not be  null).
%and set its fields in lines 10 and 29. The type of the variable \texttt{nw} in line 9 represents a freshly allocated heap node, whose fields are not set yet. A \textsf{rcuFresh} reference can be used either in setting the fields of a fresh object or linking the fresh object. We will investigate  linking a fresh object in the next analysis but, first we anaylze how we set the fields of the fresh object. In  line 29, we set the fields of the object referenced by \texttt{nw} and we end up having the type in line 30 which seems similar to a type \textsf{rcuItr} with empty path. We will justify why we have separate type \textsf{rcuFresh} instead of typing the freshly allocated heap with \textsf{rcuItr} when we explain the type rules.
\subsection{Type Rules}
\label{subsection:type-rules}
\begin{figure}[!t]\scriptsize
\begin{mathpar}
\fbox{$\Gamma \vdash_{M} \alpha \dashv \Gamma'$} \;\;
\inferrule*[left=\scriptsize(T-Root)]
{ %\textsf{FV}(\Gamma) \cap \{r,y\}  =\emptyset
  y\not\in\textsf{FV}(\Gamma)
}
{\Gamma,r\textsf{:rcuRoot},y\textsf{:undef} \vdash y = r \dashv y\textsf{:rcuItr} \epsilon \N_{\emptyset},r\textsf{:rcuRoot},\Gamma}
\and
\inferrule*[left=\scriptsize(T-ReadS)]
{ %\textsf{FV}(\Gamma) \cap \{z,x\}  =\emptyset 
  z\not\in\textsf{FV}(\Gamma)
}
{
\Gamma, z:\_  ,\rcuitrT{x}{G}{k}{k+1}{\_} \vdash z=x \dashv  \rcuitrT{x}{G}{k}{k+1}{\_}, \rcuitrT{z}{G}{k}{k+1}{\_}, \Gamma
}
\and
\inferrule*[left=\scriptsize(T-Alloc)]
{
}
{
 \Gamma, x\textsf{:undef} \vdash x = \NEW \dashv x\textsf{:rcuFresh} \N_{\emptyset} ,  \Gamma
}
\;\;
\inferrule*[left=\scriptsize(T-Free)]
{
}
{
 x\textsf{:freeable} \vdash \textsf{Free}(x) \dashv x\textsf{:undef}
}
\and
\inferrule*[left=\scriptsize(T-ReadH)]
{
\rho.f=\rho' \\
 %\textsf{FV}(\Gamma) \cap \{z,x\} =\emptyset
  z\not\in\textsf{FV}(\Gamma)
}
{
 \Gamma , z:\_  ,  x\textsf{:rcuItr} \rho \N   \vdash
 	z=x.f
 \dashv  x\textsf{:rcuItr} \rho  \N([f\rightharpoonup z]),z\textsf{:rcuItr} \rho'  \N_{\emptyset}  , \Gamma
}
\and
\inferrule[\scriptsize(T-WriteFH)]
{
z:\textsf{rcuItr}  \rho.f  \_ \quad \N(f)=z \quad f \notin dom (\N')
}
{
 \Gamma, p\textsf{:rcuFresh}\N', x\textsf{:rcuItr} \rho  \N
  \vdash_{M} p.f = z \dashv
  p\textsf{:rcuFresh}\N'([f\rightharpoonup z])   , x\textsf{:rcuItr} \rho  \N([f\rightharpoonup z]),  \Gamma
}
\and
\inferrule*[left=\scriptsize(T-Sync)]
{
}
{
\Gamma \vdash \textsf{SyncStart};\textsf{SyncStop} \dashv \Gamma[\overline{x\textsf{:freeable}/x\textsf{:unlinked}}]
}
\and
{
\inferrule[\scriptsize(T-UnlinkH)]
{
\N(f_1)=z\\
\rho.f_1=\rho_1 \\
\rho_1.f_2=\rho_2 \\
\N'=\N([f_1\rightharpoonup z \setminus r])\\
\forall_{f\in dom(\N_1)} \ldotp f\neq f_2 \implies (\N_1(f) = \textsf{null})
\\
\N(f_1)=z\\
\N_1(f_2)=r\\
\begin{array}{l}
\forall_{n\in \Gamma,m,\N_3, p_3,f}\ldotp n\textsf{:rcuItr}\,\rho_3\,\N_3([f\rightharpoonup m]) \implies %\arcr
\left\{\begin{array}{l}
((\neg\mathsf{MayAlias}(\rho_3,\{\rho,\rho_1,\rho_2\})  ) \land (m\not\in\{z,r\} ) ) \arcr
\land (\forall_{\rho_4\neq \epsilon} \ldotp \neg\mathsf{MayAlias}(\rho_3, \rho_2.\rho_4) )
\end{array}\right.
\end{array}
}
{
\Gamma,x\textsf{:rcuItr} \rho  \N  ,
z\textsf{:rcuItr} \rho_1  \N_1  ,  r\textsf{:rcuItr}  \rho_2  \N_2 
\vdash x.f_1=r \dashv
z\textsf{:unlinked},
x\textsf{:rcuItr}  \rho  \N' ,
r\textsf{:rcuItr}  \rho_1  \N_2, \Gamma
}
}
\and
{
\inferrule[\scriptsize(T-Replace)]
{
\N(f)=o \\ \N'=\N([f \rightharpoonup o\setminus n])\\ \rho.f  = \rho_1  \\
\N_1 = \N_2 \and
\textsf{FV}(\Gamma) \cap \{p,o,n\}  =\emptyset \\
\forall_{x \in \Gamma, \N_3, \rho_2,f_1,y} \ldotp (x\textsf{:rcuItr}\,\rho_2\,\N_3([f_1 \rightharpoonup y])) \implies (\neg\mathsf{MayAlias}(\rho_2,\{\rho,\rho_1\}) \land (y\neq o  ))
}
{
\Gamma,
 p\textsf{:rcuItr} \rho  \N ,
  o\textsf{:rcuItr} \rho_1  \N_1  , n\textsf{:rcuFresh}  \N_2
  \vdash p.f = n \dashv
  p\textsf{:rcuItr} \rho  \N'  ,
  n\textsf{:rcuItr} \rho_1  \N_2 ,
  o\textsf{:unlinked} ,  \Gamma
}
}
\and
{
\inferrule[\scriptsize(T-Insert)]
{
\N'=\N([f \rightharpoonup o\setminus n])\\ \rho.f  = \rho_1 \\ \rho_1.f_4 = \rho_2 \\ \N(f) = \N_1(f_4) \\
\forall_{f_2\in dom(\N_1)} \ldotp f_4\neq f_2 \implies \N_1(f_2)=\textsf{null} \\
\textsf{FV}(\Gamma) \cap \{p,o,n\}  =\emptyset \\
\forall_{x \in \Gamma, \N_3, \rho_3,f_1,y} \ldotp (x:\textsf{rcuItr}\,\rho_3\,\N_3([f_1 \rightharpoonup y])) \implies
 (\forall_{\rho_4\neq \epsilon} \ldotp \neg\mathsf{MayAlias}(\rho_3, \rho.\rho_4) )
}
{
\Gamma,
 p\textsf{:rcuItr} \rho  \N ,
  o\textsf{:rcuItr} \rho_1  \N_2  , n\textsf{:rcuFresh}  \N_1
  \vdash p.f = n \dashv
  p\textsf{:rcuItr} \rho  \N'  ,
  n\textsf{:rcuItr} \rho_1  \N_1 ,
  o\textsf{:rcuItr} \rho_2 \N_2 ,  \Gamma
}
}
\and
\fbox{$\Gamma \vdash_{M} C \dashv \Gamma'$} \;\;
\inferrule*[left=\scriptsize(ToRCUWrite)]
{
  \textsf{NoFresh}(\Gamma') \and \textsf{NoUnlinked}(\Gamma')  \and \textsf{NoFreeable}(\Gamma') \\
 \Gamma , y\textsf{:rcuItr} \_ \vdash_M C \dashv \Gamma' \and \textsf{FType}(f)=\textsf{RCU} 
}
{
\Gamma \vdash \textsf{RCUWrite}\, x.f \textsf{ as }y\textsf{ in } \{ C \}
}
\end{mathpar}
\caption{Type rules for \textsf{write} side critical section.}
\label{fig:tss}
\end{figure}
Figure \ref{fig:tss} gives the primary type rules used in checking write-side critical section code as in Figure \ref{fig:rculist}.
%Our aim in this section is to investigate potential memory safety issues, discuss what memory inconsistencies they can cause and how type rules prevent it and justify why the flow of change in types serves to the correct \textsf{RCU} programming.

\textsc{T-Root} reads a root pointer into an \textsf{rcuItr} reference, and \textsc{T-ReadS} copies a local variable into another.  In both cases, the free variable condition ensures that updating the modified variable does not invalidate field maps of other variables in $\Gamma$. These free variable conditions recur throughout the type system, and we will not comment on them further.
\textsc{T-Alloc} and \textsc{T-Free} allocate and reclaim objects. These rules are relatively straightforward.
\textsc{T-ReadH} reads a field into a local variable.
As suggested earlier, this rule updates the post-environment to reflect that the overwritten variable $z$ holds the same value as $x.f$.
\textsc{T-WriteFH} updates a field of a \emph{fresh} (thread-local) object, similarly tracking the update in the fresh object's field map at the type level.
The remaining rules are a bit more involved, and form the heart of the type system.

\mypar{Grace Periods}
\textsc{T-Sync} gives pre- and post-environments to the compound statement \lstinline|SyncStart;SyncStop| implementing grace periods.  As mentioned earlier, this updates the environment afterwards to reflect that any nodes \textsf{unlinked} before the wait become \textsf{freeable} afterwards.

\mypar{Unlinking}
\textsc{T-UnlinkH} type checks heap updates that remove a node from the data structure.  The rule assumes three objects $x$, $z$, and $r$, whose identities we will conflate with the local variable names in the type rule.  The rule checks the case where $x.f_1==z$ and $z.f_2==r$ initially (reflected in the path and field map components, and a write $x.f_1=r$ removes $z$ from the data structure (we assume, and ensure, the structure is a tree).
%(\textit{\textbf{Mutation - Unlinking a Heap Node}}) - Preserving invariants of a data structure against possible mutations under \textsf{RCU} semantics is challenging. Unlinking a heap node is one way of mutating the heap and the type rule for unlinking(\textsc{T-UnlinkH}) enforces the locality on the effect of unlinking to enforce memory safety. First essential property is "proper local linkage" in between the heap nodes involved in unlinking(footpring of unlinking). The type rule asserts that the $x$, $z$ and $r$ references are linked with field mappings $\N([f_1\rightharpoonup z])$ e.g. ($Left \mapsto current$) of $x$, $\N'([f_2\rightharpoonup r])$ e.g. ($Left \mapsto currentL$) of $y$. In accordance with the field mappings, the type rule also asserts that $x$ has the path $\rho$ e.g. ($(Left)^{k}$), $y$ has the path $\rho.f_1$ e.g. ($(Left)^{k}.Left$) and $z$ has the path $\rho.f_1.f_2$ e.g. ($(Left)^{k}.Left.Left$).

The rule must also avoid unlinking multiple nodes: this is the purpose of the first (smaller) implication: it ensures that beyond the reference from $z$ to $r$, all fields of $z$ are null.

Finally, the rule must ensure that no types in $\Gamma$ are invalidated.  This could happen one of two ways: either a field map in $\Gamma$ for an alias of $x$ duplicates the assumption that $x.f_1==z$ (which is changed by this write), or $\Gamma$ contains a descendant of $r$, whose path from the root will change when its ancestor is modified.
The final assumption of \textsc{T-UnlinkH} (the implication)
checks that for every \textsf{rcuItr} reference $n$ in $\Gamma$, it is not a path alias of $x$, $z$, or $r$; no entry of its field map ($m$) refers to $r$ or $z$ (which would imply $n$ aliased $x$ or $z$ initially); and its path is not an extension of $r$ (i.e., it is not a descendant). \textsf{MayAlias} is a predicate on two paths (or a path and set of paths) which is true if it is possible that any concrete paths the arguments may abstract (e.g., via adding non-determinism through $|$ or abstracting iteration with indexing) \emph{could} be the same.  The negation of a \textsf{MayAlias} use is true only when the paths are guaranteed to refer to different locations in the heap.
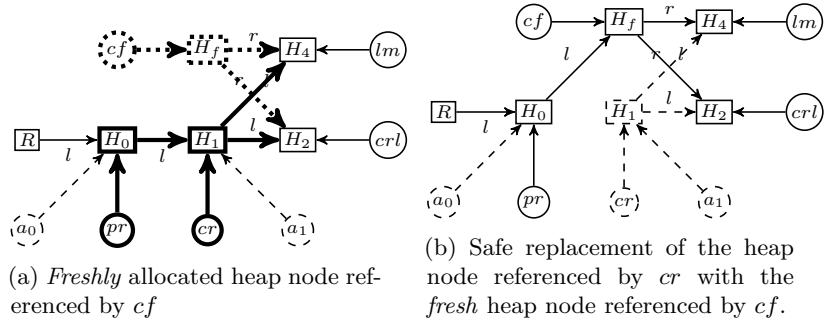
\begin{figure}[!t]\scriptsize
 \centering
\begin{subfigure}[b]{.4\linewidth}
 \centering
 \begin{tikzpicture}[>=stealth',node distance=1.2cm,semithick,auto]
 \tikzstyle{hollow node}=[circle,draw,inner sep=1]
 \tikzstyle{sub node}=[triangle,draw,inner sep=1]
 \tikzstyle{solid node}=[rectangle,draw,inner sep=1.5]
 \tikzstyle{solids node}=[rectangle,draw,ultra thick,inner sep=1.5]
  \tikzstyle{solidss node}=[dashed,rectangle,draw=red,inner sep=1.5]
  \tikzstyle{freshollow node}=[ultra thick,dotted ,circle,draw,inner sep=1]

 \tikzstyle{fresh node}=[ultra thick, dotted ,rectangle,draw,inner sep=1.5]
  
 \tikzstyle{null node}=[circle,draw=red,fill=red]
 \tikzset{
 	red node/.style={rectangle,draw=black,fill=red,inner sep=1.5},
 	blue node/.style={rectangle,draw=black,inner sep=1.5},
 	reader node/.style={circle,draw=black,inner sep=1},
        redreader node/.style={circle,draw,ultra thick,inner sep=1},
        readerr node/.style={dashed,circle,draw=black,inner sep=1},
 	writer node/.style={circle,draw=black,inner sep=1}
 }

       \node[solid node] (R) {$R$};
       \node[solids node] (0) [right of=R] {$H_0$};
       \node[solids node] (1) [right of=0] {$H_1$};
       \node[fresh node] (2) [above of=1] {$H_f$};
       \node[solid node] (3) [right of=1] {$H_2$};
       \node[solid node] (4) [above  of=3] {$H_4$};
     %         \node[solid node] (5) [above left of=0] {$H_5$};
     %  \node[solid node] (6) [above left of = 1]{$H_6$};

       \node[redreader node] (r0) [below  of= 0]  {$pr$};
       \node[redreader node] (r1) [below  of= 1]  {$cr$};
       \node[freshollow node] (f)  [left  of= 2] {$cf$};
        \node[reader node] (crl)  [right  of= 3] {$crl$};
       \node[reader node] (lm) [right  of= 4]  {$lm$};
       \node[readerr node] (r1a) [below   of= R]  {$a_0$};
       \node[readerr node] (r2a)  [below of= 3] {$a_1$};

     %  \node[reader node] (r1f) [above  of= 5]  {$a_5$};
     %  \node[reader node] (r2f)  [above  of= 6] {$a_6$};

     \path[->]  (R) edge node[below] {$l$} (0);
     \path[ultra thick,->]  (0) edge node[below] {$l$} (1);
     \path[->]  (crl) edge  node {} (3);
     \path[ultra thick,dotted,->]  (f) edge node {} (2);
     \path[ultra thick,dotted,->]  (2) edge node {$r$} (4);
     \path[ultra thick,dotted,->]  (2) edge node {$l$} (3);

    % \path[dashed,->]  (5) edge node {$l$} (0);
    % \path[dashed,->]  (6) edge node {$l$} (1);

     \path[ultra thick,->]  (1) edge node {$r$} (4);
     \path[ultra thick,->]  (1) edge node {$l$} (3);
     
     \path[ultra thick,->]  (r0) edge node {} (0);
     \path[ultra thick,->]  (r1) edge node  {}  (1);
     \path[->]  (lm) edge  node  {}   (4);

     \path[dashed,->]  (r1a) edge node {} (0);
     \path[dashed,->]  (r2a) edge node  {}  (1);
     
    % \path[->]  (r1f) edge  node  {}   (5);
    % \path[->]  (r2f) edge  node  {}   (6);

     %\path[draw=red, ->] (2) edge node[below] {$r$} (nl);
 ;
 \end{tikzpicture}

 \caption{\textit{Freshly} allocated heap node referenced by $cf$}
 \label{fig:frframeout}
 \end{subfigure} \quad\quad
 \begin{subfigure}[b]{.4\linewidth}
 \centering
 \begin{tikzpicture}[>=stealth',node distance=1.2cm,semithick,auto]
 \tikzstyle{hollow node}=[circle,draw,inner sep=1]
 \tikzstyle{sub node}=[triangle,draw,inner sep=1]
 \tikzstyle{solid node}=[rectangle,draw,inner sep=1.5]
 \tikzstyle{solids node}=[rectangle,draw,inner sep=1.5]
  \tikzstyle{solidss node}=[dashed,rectangle,draw,inner sep=1.5]
  \tikzstyle{freshollow node}=[dashed,circle,draw=green,inner sep=1]

 \tikzstyle{fresh node}=[dashed,rectangle,draw=green,inner sep=1.5]
  
 \tikzstyle{null node}=[circle,draw=red,fill=red]
 \tikzset{
 	red node/.style={rectangle,draw=black,fill=red,inner sep=1.5},
 	blue node/.style={rectangle,draw=black,inner sep=1.5},
 	reader node/.style={circle,draw=black,inner sep=1},
        redreader node/.style={circle,draw,inner sep=1},
        readerr node/.style={dashed,circle,draw=black,inner sep=1},
 	writer node/.style={circle,draw=black,inner sep=1},
               rredreader node/.style={dashed,circle,draw,inner sep=1}
 }

       \node[solid node] (R) {$R$};
       \node[solids node] (0) [right of=R] {$H_0$};
       \node[solidss node] (1) [right of=0] {$H_1$};
       \node[solid node] (2) [above of=1] {$H_f$};
       \node[solid node] (3) [right of=1] {$H_2$};
       \node[solid node] (4) [above  of=3] {$H_4$};
       %\node[solid node] (5) [above of=R] {$H_5$};
       %\node[solid node] (6) [above of = 0]{$H_6$};

       \node[redreader node] (r0) [below  of= 0]  {$pr$};
       \node[rredreader node] (r1) [below  of= 1]  {$cr$};
       \node[reader node] (f)  [left  of= 2] {$cf$};
        \node[reader node] (crl)  [ right  of= 3] {$crl$};
       \node[reader node] (lm) [right  of= 4]  {$lm$};
       \node[readerr node] (r1a) [below   of= R]  {$a_0$};
       \node[readerr node] (r2a)  [below of= 3] {$a_1$};

      %\node[reader node] (r1f) [above  of= 5]  {$a_5$};
      %\node[reader node] (r2f)  [above  of= 6] {$a_6$};

     \path[->]  (R) edge node[below] {$l$} (0);
    % \path[draw=red,->]  (0) edge node[below] {$l$} (1);
     \path[->]  (crl) edge  node {} (3);
     \path[->]  (f) edge node {} (2);
     \path[->]  (2) edge node {$r$} (4);
     \path[->]  (2) edge node {$l$} (3);
     \path[dashed,->]  (1) edge node {$r$} (4);
     \path[dashed,->]  (1) edge node {$l$} (3);

     \path[->]  (r0) edge node {} (0);
     \path[dashed,->]  (r1) edge node  {}  (1);
     \path[->]  (lm) edge  node  {}   (4);

     \path[dashed,->]  (r1a) edge node {} (0);
     \path[dashed,->]  (r2a) edge node  {}  (1);
     %     \path[dashed,->]  (5) edge node {$l$} (0);
     %\path[dashed,->]  (6) edge node {$l$} (1);

     %\path[dashed,->]  (r2a) edge  node  {}   (4);

     \path[->]  (0) edge node  {$l$}  (2);

    % \path[->]  (r1f) edge  node  {}   (5);
    % \path[->]  (r2f) edge  node  {}   (6);

 ;
 \end{tikzpicture}

 \caption{Safe replacement of the heap node referenced by $cr$ with the \textit{fresh} heap node referenced by $cf$.}
 \label{fig:freshframeout}
 \end{subfigure}
 \caption{Replacing \textit{existing} heap nodes with \textit{fresh} ones. Type rule \textsc{T-Replace}.}
 \label{fig:alloc}\vspace{-2mm}
 \end{figure}
%\textit{\textbf{Creating a Fresh Node}} - We deffered answering the question: "Why do we need a separate type \textsf{rcuFresh} instead of using \textsf{rcuItr} for typing a freshly allocated heap node in \textsf{Write} critical section ?". To understand and answer the question easily, we illustrate the setting for creating a fresh \textsf{BST} heap node and linking it in Figures \ref{fig:frframeout} and \ref{fig:freshframeout}.

\mypar{Replacing with a Fresh Node}
Replacing with a \textsf{rcuFresh} reference faces the same aliasing complications as direct unlinking.
We illustrate these challenges in Figures \ref{fig:frframeout} and \ref{fig:freshframeout}. Our technical report~\cite{isotek} also includes Figures 32a and 32b in Appendix D to illustrate complexities in unlinking.
The square $R$ nodes are root nodes, and $H$ nodes are general heap nodes.
All resources in thick straight lines and dotted lines form the memory foot print of a node replacement. The hollow thick circular nodes -- $pr$ and $cr$ -- point to the nodes involved in replacing $H_1$ (referenced by \texttt{cr}) wih $H_f$ (referenced by $cf$) in the structure.
We may have $a_0$ and $a_1$ which are aliases with $pr$ and $cr$ respectively. They are \textit{path-aliases} as they share the same path from root to the node that they reference. 
Edge labels $l$ and $r$ are abbreviations for the $Left$ and $Right$ fields of a binary search tree.
The thick dotted $H_f$ denotes the freshly allocated heap node referenced by thick dotted  $cf$. The thick dotted field $l$ is set to point to the referent of $cl$ and the thick dotted field $r$ is set to point to the referent of the heap node referenced by $lm$.

$H_f$ initially (Figure \ref{fig:frframeout}) is not part of the shared structure.  If it was, it would violate the tree shape requirement imposed by the type system.  This is why we highlight it separately in thick dotts --- its static type would be \textsf{rcuFresh}.
Note that we cannot duplicate a \textsf{rcuFresh} variable, nor read a field of an object it points to.
This restriction localizes our reasoning about the effects of replacing with a fresh node to just one fresh reference and the object it points to.
Otherwise another mechanism would be required to ensure that once a fresh reference was linked into the heap, there were no aliases still typed as fresh --- since that would have risked linking the same reference into the heap in two locations.

The transition from the Figure \ref{fig:frframeout} to \ref{fig:freshframeout} illustrates the effects of the heap mutation (replacing with a fresh node). The reasoning in the type system for replacing with a fresh node is nearly the same as for unlinking an existing node, with one exception.
In replacing with a fresh node, there is no need to consider the paths of nodes deeper in the tree than the point of mutation.  In the unlinking case, those nodes' static paths would become invalid.  In the case of replacing with a fresh node, those descendants' paths are preserved.
Our type rule for ensuring safe replacement (\textsc{T-Replace}) prevents path aliasing (representing the nonexistence of $a_0$ and $a_1$ via dashed lines and circles) by negating a \textsf{MayAlias} query and prevents field mapping aliasing (nonexistence of any object field from any other context pointing to $cr$) via asserting $(y\neq o)$. It is important to note that objects($H_4,H_2$) in the field mappings of the $cr$ whose referent is to be unlinked captured by the heap node's field mappings referenced by $cf$ in \textsf{rcuFresh}. This is part of enforcing locality on the heap mutation and captured by assertion $\N = \N'$ in the type rule(\textsc{T-Replace}). 
%(\textit{\textbf{Mutation - Reclaiming a Heap Node after a Grace Period}}) - The reason why we have type \textsf{freeable} after the grace period is to prevent the ability to call \texttt{Free} before the grace period ends. This means that we cannot give ownership of a heap node to heap right after unlinking. Our type rule(\textsc{T-Sync}) ensures this via enforcing that we call \texttt{Free} only with a reference in type \textsf{freeable}. To get \textsf{freeable} reference, first we wait the grace period \texttt{SyncStart};\texttt{SyncStop} for unlinked heap nodes to end. The \texttt{SyncStart} does not have any effect on type of the variables.

\mypar{Inserting a Fresh Node}
\textsc{T-Insert} type checks heap updates that link a fresh node into a linked data structure. Inserting a \textsf{rcuFresh} reference also faces some of the aliasing complications that we have already discussed for direct unlinking and replacing a node. Unlike the replacement case, the path to the last heap node (the referent of $o$) from the root is \emph{extended} by $f$, which risks falsifying the paths for aliases and descendants of $o$. The final assumption(the implication) of \textsc{T-Insert} checks for this inconsistency.

There is also another rule, \textsc{T-LinkF-Null}, not shown in Figure \ref{fig:tss}, which handles the case where the fields of the fresh node are not object references, but instead all contain null (e.g., for appending to the end of a linked list or inserting a leaf node in a tree).

\mypar{Critical Sections}
(\textit{\textbf{Referencing inside \textsf{RCU} Blocks}})
We introduce the \textit{syntactic sugaring} $\textsf{RCUWrite } x.f \textsf{ as } y \textsf{ in } \{C\}$ for write-side critical sections where the analogous syntactic sugaring can be found for read-side critical sections in Appendix E of the technical report~\cite{isotek}.

The type system ensures \textsf{unlinked} and \textsf{freeable} references are handled linearly, as they cannot be dropped -- coerced to \textsf{undef}. The top-level rule \textsc{ToRCUWrite} in Figure \ref{fig:tss} ensures \textsf{unlinked} references have been freed by forbidding them in the critical section's post-type environment. Our technical report~\cite{isotek} also includes the analogous rule \textsc{ToRCURead} for the read critical section in Figure 33 of Appendix E.

Preventing the reuse of \textsf{rcuItr} references across critical sections is subtler: the non-critical section system is not flow-sensitive, and does not include \textsf{rcuItr}. Therefore, the initial environment lacks \textsf{rcuItr} references, and trailing \textsf{rcuItr} references may not escape.
\raggedbottom
\makeatletter % allow us to mention @-commands
\def\arcr{\@arraycr}
\makeatother
\section{Evaluation}
\label{sec:eval}
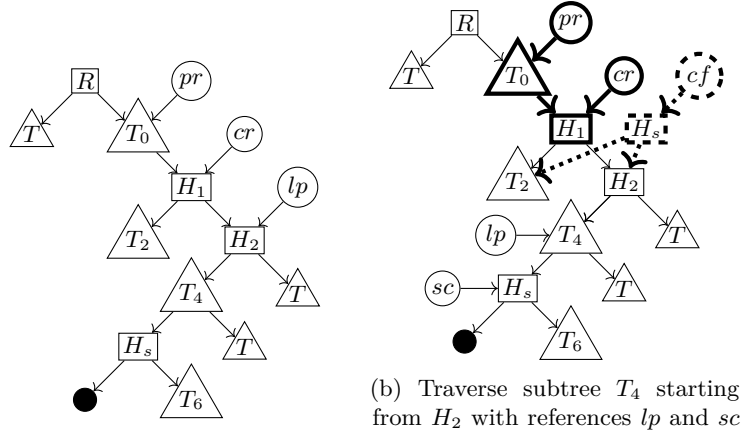
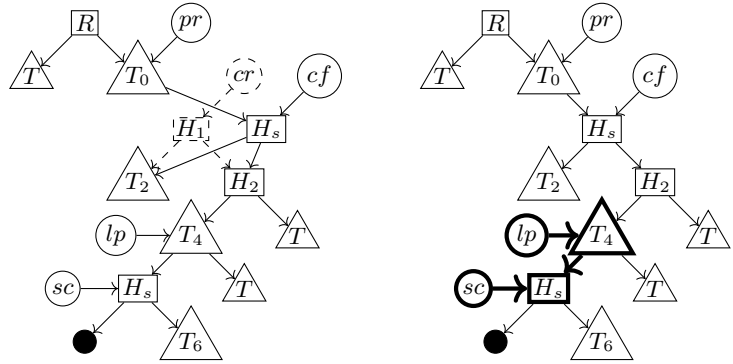
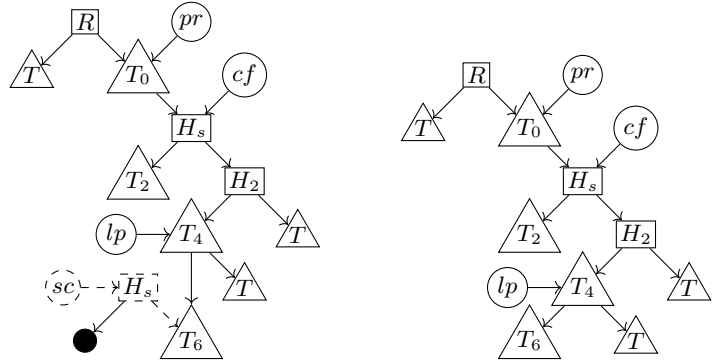
\begin{figure}[!t]
\centering 
\noindent
\begin{subfigure}[b]{.4\linewidth}
\centering
\begin{tikzpicture}[scale=0.5]
\tikzstyle{ref node}=[circle,draw,inner sep=1.5]
\tikzstyle{rref node}=[circle,draw=red,inner sep=1.5]
\tikzstyle{rhollow node}=[rectangle,draw=red,inner sep=1.5]
\tikzstyle{hollow node}=[rectangle,draw,inner sep=1.5]

\tikzstyle{sub node}=[triangle,draw,inner sep=1.5]
\tikzstyle{solid node}=[rectangle,draw,inner sep=1.5]
\tikzset{
  red node/.style = {rectangle,draw=red,inner sep=0.75},
  treenode/.style = {circle, draw=black, align=center, minimum size=0.1cm},
  subtree/.style  = {draw,minimum size=0cm,inner sep=0,regular polygon,regular polygon sides=3},
  rsubtree/.style ={draw=red,minimum size=0cm,inner sep=0,regular polygon,regular polygon sides=3},
  succn/.style = {circle,draw=black,fill=black},
  blue node/.style = {rectangle,draw=green,inner sep=1.5,dashed}
  
}

    \node[hollow node]       (r)     []   {$R$};
    \node[subtree] (t0) [below right of = r] [] {$T_0$};
    \node[subtree]           (t1)    [below left of=r]       {$T$};
    \node[hollow node]           (k)    [below right of=t0]      {$H_1$};
    \node[subtree]             (kl)    [below left of=k]      {$T_{2}$};
    \node[hollow node]          (kr)    [below right of = k]  {$H_2$};
     \node[subtree]          (rlmst)    [below left of = kr]  {$T_4$};
     \node[subtree]  (t5) [below right of =rlmst] {$T$};
    \node[hollow node]             (krl)         [below left of = rlmst]        {$H_s$};
        \node[succn]          (nl)    [below left of = krl]  {};
    \node[subtree]          (krr)       [below right of = kr]           {$T$};
    \node[subtree]          (t6)       [below right of = krl]           {$T_{6}$};
    %%%%%%%%%%%%%%%%%%%%%%%%%%%%%%%%%%%
    \node[ref node] (pr) [above right of= t0] {$pr$};
    \node[ref node] (cr) [above right of= k] {$cr$};
%    \node[fref node] (cf) [above right of = krl] {$cf$};
    \node[ref node] (lp) [above right of = kr] {$lp$};
%    \node[ref node] (sc) [above left of = krl] {$sc$};
    %%%%%%%%%%%%%%%%%%%%%%%%%%%%%%%%%%%%
    \path[->]  
	     (r)     edge node {} (t1)
             (r)     edge node {} (t0)
                (t0)     edge node {} (k)
                (k)    edge node {} (kl)
	     (k)    edge node {} (kr)

         (kr) edge node {} (rlmst)
          (rlmst) edge node {} (krl)
          (rlmst) edge node {} (t5)
             (krl) edge node {} (nl)

                (kr)     edge node {} (krr)
                %%%%%%%%%%%%%%%%%
                (pr) edge node {} (t0)
                (cr) edge node {} (k)
%                (cf) edge node {} (krl)
                (lp) edge node {} (kr)
 %               (sc) edge node {} (krl)
                (krl) edge node {} (t6)
;
\end{tikzpicture}
\caption{The writer traverses subtree $T_0$ to find the heap node $H_1$ with local references $pr$ and $cr$. Black-filled node representing the null node.}
\label{fig:del2.1}
\end{subfigure}\quad
\begin{subfigure}[b]{.4\linewidth}
\centering
\begin{tikzpicture}[scale=0.5]
\tikzstyle{ref node}=[circle,draw,inner sep=1.5]
\tikzstyle{hollow node}=[rectangle,draw,inner sep=1.5]
\tikzstyle{rhollow node}=[rectangle,draw,ultra thick,inner sep=1.5]

\tikzstyle{sub node}=[triangle,draw,inner sep=1.5]
\tikzstyle{solid node}=[rectangle,draw,inner sep=1.5]
\tikzstyle{rref node}=[circle,draw,ultra thick,inner sep=1.5]
\tikzstyle{fref node}=[circle,draw,ultra thick, dotted, inner sep=1.5,dashed]
\tikzset{
  red node/.style = {rectangle,draw=red,inner sep=0.75},
  treenode/.style = {circle, draw=black, align=center, minimum size=0.1cm},
  subtree/.style  = {draw,minimum size=0cm,inner sep=0,regular polygon,regular polygon sides=3},
  rsubtree/.style ={ultra thick, draw,minimum size=0cm,inner sep=0,regular polygon,regular polygon sides=3},
  succn/.style = {circle,draw=black,fill=black},
  blue node/.style = {rectangle,draw,ultra thick, dotted,inner sep=1.5,dashed}
}

    \node[hollow node]       (r)     []   {$R$};
        \node[rsubtree] (t0) [below right of = r] [] {$T_0$};
    \node[subtree]           (t1)    [below left of=r]       {$T$};
    \node[rhollow node]           (k)    [below right of=t0]      {$H_1$};
    \node[blue node]            (kp1) [right of = k]   {$H_s$};    
    \node[subtree]             (kl)    [below left of=k]      {$T_{2}$};
    \node[hollow node]          (kr)    [below right of = k]  {$H_2$};
    \node[subtree]          (rlmst)    [below left of = kr]  {$T_4$};
      \node[subtree]  (t5) [below right of = rlmst] {$T$};
    \node[hollow node]             (krl)         [below left of = rlmst]        {$H_s$};
    \node[succn]          (nl)    [below left of = krl]  {};
    \node[subtree]          (krr)       [below right of = kr]           {$T$};
    
    %%%%%%%%%%%%%%%%%%%%%%%%%%
       \node[rref node] (pr) [above right of= t0] {$pr$};
    \node[rref node] (cr) [above right of= k] {$cr$};
    \node[fref node] (cf) [above right of = kp1] {$cf$};
    \node[ref node] (lp) [ left of = rlmst] {$lp$};
    \node[ref node] (sc) [ left of = krl] {$sc$};
     \node[subtree]          (t6)       [below right of = krl]           {$T_{6}$};
    \path[->]  
	     (r)     edge node {} (t1)
              (r)     edge node {} (t0)
                (t0)     edge[draw, ultra thick] node {} (k)
                
                %(r)     edge node {} (k) 
                (k)    edge node {} (kl)
	     (k)    edge node {} (kr)
	    (kr)    edge node {}  (rlmst)
                (kr)     edge node {} (krr)

         (kr) edge node {} (rlmst)
          (rlmst) edge node {} (krl)
                    (rlmst) edge node {} (t5)
             (krl) edge node {} (nl)

                (kp1)   edge[ultra thick,dotted] node {} (kr)
	     (kp1)    edge[ultra thick,dotted] node {} (kl)
             %%%%%%%%%%%%%%%%%%%%%%%%%%%
                            (pr) edge[draw, ultra thick] node {} (t0)
                (cr) edge[draw, ultra thick] node {} (k)
                (cf) edge[ultra thick,dotted] node {} (kp1)
                %(lp) edge node {} (kr)
                %(sc) edge node {} (krl)
                                (lp) edge node {} (rlmst)
                (sc) edge node {} (krl)

                                (krl) edge node {} (t6)
;
;
\end{tikzpicture}
\caption{Traverse subtree $T_4$ starting from $H_2$ with references $lp$ and $sc$ to find successor $H_s$ of $H_1$. Duplicating $H_s$ as a fresh heap node before replacing $H_1$ with the fresh one.}
\label{fig:del2.2}
\end{subfigure}\quad
\begin{subfigure}[b]{.4\linewidth}
\centering
\begin{tikzpicture}[scale=0.5]
\tikzstyle{ref node}=[circle,draw,inner sep=1.5]
\tikzstyle{hollow node}=[rectangle,draw,inner sep=1.5]
\tikzstyle{rhollow node}=[rectangle,draw=red,inner sep=1.5]

\tikzstyle{sub node}=[triangle,draw,inner sep=1.5]
\tikzstyle{solid node}=[rectangle,draw,inner sep=1.5]
\tikzstyle{rdref node}=[circle,draw,inner sep=1.5,dashed]
\tikzstyle{rref node}=[circle,draw,inner sep=1.5]
\tikzstyle{fref node}=[circle,draw=green,inner sep=1.5,dashed]
\tikzset{
  red node/.style = {rectangle,dashed,draw,inner sep=0.75},
  treenode/.style = {circle, draqw=black, align=center, minimum size=0.1cm},
  subtree/.style  = {draw,minimum size=0cm,inner sep=0,regular polygon,regular polygon sides=3},
  rsubtree/.style ={draw=red,minimum size=0cm,inner sep=0,regular polygon,regular polygon sides=3},
  succn/.style = {circle,draw=black,fill=black},
  blue node/.style = {rectangle,draw=green,inner sep=1.5,dashed}
}

    \node[hollow node]       (r)     []   {$R$};
        \node[subtree] (t0) [below  right of = r] [] {$T_0$};
    \node[subtree]           (t1)    [below left of=r]       {$T$};
    \node[red node]           (k)    [below right of=t0]      {$H_1$};
    \node[hollow node]            (kp1) [right of = k]   {$H_s$};    
    \node[subtree]             (kl)    [below left of=k]      {$T_{2}$};
\node[hollow node]          (kr)    [below right of = k]  {$H_2$};
 
  \node[subtree]          (rlmst)    [below left of = kr]  {$T_4$};
   \node[subtree]  (t5) [below right of = rlmst] {$T$};
    \node[hollow node]             (krl)         [below left of = rlmst]        {$H_s$};
    \node[succn]          (nl)    [below left of = krl]  {};
    \node[subtree]          (krr)       [below right of = kr]           {$T$};
    \node[subtree]          (t6)       [below right of = krl]           {$T_{6}$};
    %%%%%%%%%%%%%%%%%%%%%%%%%%%%%%%%%%
    \node[rref node] (pr) [above right of= t0] {$pr$};
    \node[rdref node] (cr) [above right of= k] {$cr$};
    \node[ref node] (cf) [above right of = kp1] {$cf$};
    \node[ref node] (lp) [ left of = rlmst] {$lp$};
    \node[ref node] (sc) [left of = krl] {$sc$};

    \path[->]  
	     (r)     edge node {} (t1)
              (r)     edge node {} (t0)
                %(t0)     edge node {} (k)
                (t0)     edge node {} (kp1) 
                (k)    edge[dashed] node {} (kl)
	     (k)    edge[dashed] node {} (kr)
	  %  (kr)    edge node {}  (krl)
         (kr) edge node {} (rlmst)
          (rlmst) edge node {} (krl)
                              (rlmst) edge node {} (t5)
             (krl) edge node {} (nl)
                (kr)     edge node {} (krr)
                (kp1)   edge node {} (kr)
	     (kp1)    edge node {} (kl)
             %%%%%%%%%%%%%%%%%%%
                (pr) edge node {} (t0)
                (cr) edge[dashed] node {} (k)
                (cf) edge node {} (kp1)
                (lp) edge node {} (rlmst)
                (sc) edge node {} (krl)
                 (krl) edge node {} (t6)

;
\end{tikzpicture}
\caption{Replace $H_1$ with fresh successor and synchronize with the readers.}
\label{fig:del2.3}
\end{subfigure} \quad
\begin{subfigure}[b]{.4\linewidth}
\centering
\begin{tikzpicture}[scale=0.5]
\tikzstyle{ref node}=[circle,draw,inner sep=1.5]
\tikzstyle{hollow node}=[rectangle,draw,inner sep=1.5]
\tikzstyle{rhollow node}=[rectangle,draw,ultra thick,inner sep=1.5]

\tikzstyle{sub node}=[triangle,draw,inner sep=1.5]
\tikzstyle{solid node}=[rectangle,draw,inner sep=1.5]
\tikzstyle{rdref node}=[circle,draw=red,inner sep=1.5,dashed]
\tikzstyle{rref node}=[circle,draw,ultra thick,inner sep=1.5]
\tikzstyle{fref node}=[circle,draw=green,inner sep=1.5,dashed]
\tikzset{
  red node/.style = {rectangle,draw=red,inner sep=0.75},
  treenode/.style = {circle, draqw=black, align=center, minimum size=0.1cm},
  subtree/.style  = {draw,minimum size=0cm,inner sep=0,regular polygon,regular polygon sides=3},
  rsubtree/.style ={draw,ultra thick,minimum size=0cm,inner sep=0,regular polygon,regular polygon sides=3},
  succn/.style = {circle,draw=black,fill=black},
  blue node/.style = {rectangle,draw=green,inner sep=1.5,dashed}
}

    \node[hollow node]       (r)     []   {$R$};
        \node[subtree] (t0) [below  right of = r] [] {$T_0$};
    \node[subtree]           (t1)    [below left of=r]       {$T$};
   
    \node[hollow node]            (kp1) [below right of = t0]   {$H_s$};    
    \node[subtree]             (kl)    [below left of=k]      {$T_{2}$};
\node[hollow node]          (kr)    [below right of = k]  {$H_2$};
 
  \node[rsubtree]          (rlmst)    [below left of = kr]  {$T_4$};
   \node[subtree]  (t5) [below right of = rlmst] {$T$};
    \node[rhollow node]             (krl)         [below left of = rlmst]        {$H_s$};
    \node[succn]          (nl)    [below left of = krl]  {};
    \node[subtree]          (krr)       [below right of = kr]           {$T$};
    %%%%%%%%%%%%%%%%%%%%%%%%%%%%%%%%%%
    \node[ref node] (pr) [above right of= t0] {$pr$};
    
    \node[ref node] (cf) [above right of = kp1] {$cf$};
    \node[rref node] (lp) [ left of = rlmst] {$lp$};
    \node[rref node] (sc) [left of = krl] {$sc$};
\node[subtree]          (t6)       [below right of = krl]           {$T_{6}$};
    \path[->]  
	     (r)     edge node {} (t1)
              (r)     edge node {} (t0)
                %(t0)     edge node {} (k)
                (t0)     edge node {} (kp1) 
   
	  %  (kr)    edge node {}  (krl)
         (kr) edge node {} (rlmst)
          (rlmst) edge[ultra thick] node {} (krl)
                              (rlmst) edge node {} (t5)
             (krl) edge node {} (nl)
                (kr)     edge node {} (krr)
                (kp1)   edge node {} (kr)
	     (kp1)    edge node {} (kl)
             %%%%%%%%%%%%%%%%%%%
                (pr) edge node {} (t0)
                
                (cf) edge node {} (kp1)
                (lp) edge[ultra thick] node {} (rlmst)
                (sc) edge[ultra thick] node {} (krl)
                                 (krl) edge node {} (t6)

;
\end{tikzpicture}
\caption{Unlinks old successor referenced by $sc$.}
\label{fig:del2.4}
\end{subfigure} \quad
\begin{subfigure}[b]{.4\linewidth}
\centering
\begin{tikzpicture}[scale=0.5]
\tikzstyle{ref node}=[circle,draw,inner sep=1.5]
\tikzstyle{hollow node}=[rectangle,draw,inner sep=1.5]
\tikzstyle{rhollow node}=[rectangle,draw,inner sep=1.5]
\tikzstyle{rdhollow node}=[rectangle,draw,inner sep=1.5,dashed]
\tikzstyle{sub node}=[triangle,draw,inner sep=1.5]
\tikzstyle{solid node}=[rectangle,draw,inner sep=1.5]
\tikzstyle{rdref node}=[circle,draw,inner sep=1.5,dashed]
\tikzstyle{rref node}=[circle,draw,inner sep=1.5]
\tikzstyle{fref node}=[circle,draw=green,inner sep=1.5,dashed]
\tikzset{
  red node/.style = {rectangle,draw,inner sep=0.75},
  treenode/.style = {circle, draqw=black, align=center, minimum size=0.1cm},
  subtree/.style  = {draw,minimum size=0cm,inner sep=0,regular polygon,regular polygon sides=3},
  rsubtree/.style ={draw,minimum size=0cm,inner sep=0,regular polygon,regular polygon sides=3},
  succn/.style = {circle,draw=black,fill=black},
  blue node/.style = {rectangle,draw=green,inner sep=1.5,dashed}
}

    \node[hollow node]       (r)     []   {$R$};
        \node[subtree] (t0) [below  right of = r] [] {$T_0$};
    \node[subtree]           (t1)    [below left of=r]       {$T$};
   
    \node[hollow node]            (kp1) [below right of = t0]   {$H_s$};    
    \node[subtree]             (kl)    [below left of=k]      {$T_{2}$};
\node[hollow node]          (kr)    [below right of = k]  {$H_2$};
 
  \node[rsubtree]          (rlmst)    [below left of = kr]  {$T_4$};
   \node[subtree]  (t5) [below right of = rlmst] {$T$};
    \node[rdhollow node]             (krl)         [below left of = rlmst]        {$H_s$};
    \node[succn]          (nl)    [below left of = krl]  {};
    \node[subtree]          (krr)       [below right of = kr]           {$T$};
    %%%%%%%%%%%%%%%%%%%%%%%%%%%%%%%%%%
    \node[ref node] (pr) [above right of= t0] {$pr$};
    
    \node[ref node] (cf) [above right of = kp1] {$cf$};
    \node[rref node] (lp) [ left of = rlmst] {$lp$};
    \node[rdref node] (sc) [left of = krl] {$sc$};
\node[subtree]          (t6)       [below right of = krl]           {$T_{6}$};
    \path[->]  
	     (r)     edge node {} (t1)
              (r)     edge node {} (t0)
                %(t0)     edge node {} (k)
                (t0)     edge node {} (kp1) 
   
	  %  (kr)    edge node {}  (krl)
         (kr) edge node {} (rlmst)
          (rlmst) edge node {} (t6)
                              (rlmst) edge node {} (t5)
             (krl) edge node {} (nl)
                (kr)     edge node {} (krr)
                (kp1)   edge node {} (kr)
	     (kp1)    edge node {} (kl)
             %%%%%%%%%%%%%%%%%%%
                (pr) edge node {} (t0)
                
                (cf) edge node {} (kp1)
                (lp) edge node {} (rlmst)
                (sc) edge[dashed] node {} (krl)
                                 (krl) edge[dashed] node {} (t6)

;
\end{tikzpicture}
\caption{Safe unlinking of the old successor whose left subtree is null.}
\label{fig:del2.5}
\end{subfigure}\quad
\begin{subfigure}[b]{.4\linewidth}
\centering
\begin{tikzpicture}[scale=1]
\tikzstyle{ref node}=[circle,draw,inner sep=1.5]
\tikzstyle{hollow node}=[rectangle,draw,inner sep=1.5]
\tikzstyle{rhollow node}=[rectangle,draw,inner sep=1.5]
\tikzstyle{rdhollow node}=[rectangle,draw,inner sep=1.5,dashed]
\tikzstyle{sub node}=[triangle,draw,inner sep=1.5]
\tikzstyle{solid node}=[rectangle,draw,inner sep=1.5]
\tikzstyle{rdref node}=[circle,draw,inner sep=1.5,dashed]
\tikzstyle{rref node}=[circle,draw,inner sep=1.5]
\tikzstyle{fref node}=[circle,draw=green,inner sep=1.5,dashed]
\tikzset{
  red node/.style = {rectangle,draw=red,inner sep=0.75},
  treenode/.style = {circle, draqw=black, align=center, minimum size=0.1cm},
  subtree/.style  = {draw,minimum size=0cm,inner sep=0,regular polygon,regular polygon sides=3},
  rsubtree/.style ={draw,minimum size=0cm,inner sep=0,regular polygon,regular polygon sides=3},
  succn/.style = {circle,draw=black,fill=black},
  blue node/.style = {rectangle,draw=green,inner sep=1.5,dashed}
}

    \node[hollow node]       (r)     []   {$R$};
        \node[subtree] (t0) [below right of = r] [] {$T_0$};
    \node[subtree]           (t1)    [below left of=r]       {$T$};
    \node[hollow node]            (kp1) [below right of = t0]   {$H_s$};    
    \node[subtree]             (kl)    [below left of=kp1]      {$T_{2}$};
    \node[hollow node]          (kr)    [below right of = kp1]  {$H_2$};
    \node[rsubtree]          (rlmst)    [below left of = kr]  {$T_4$};
    \node[subtree]  (t5) [below right of = rlmst] {$T$};
    %\node[succn]          (nl)    [below left of = rlmst]  {};
    \node[subtree]          (krr)       [below right of = kr]           {$T$};
    \node[subtree]          (t6)       [below left of = rlmst]           {$T_{6}$};
    %%%%%%%%%%%%%%%%%%%%%%%%%%%%%%%%%%%%%%%%%%
    \node[ref node] (pr) [above right of= t0] {$pr$};
    \node[rref node] (lp) [ left of = rlmst] {$lp$};    
    \node[ref node] (cf) [above right of = kp1] {$cf$};    

    \path[->]  
	     (r)     edge node {} (t1)
              (r)     edge node {} (t0)
         %       (t0)     edge node {} (k)
         (t0)     edge node {} (kp1) 
         (kr)     edge node {} (krr)
         (kr) edge node {} (rlmst)
          (rlmst) edge node {} (t5)
  %       (rlmst) edge node {} (nl)
         (kp1)   edge node {} (kr)
	     (kp1)    edge node {} (kl)

(pr) edge node {} (t0)
                
                (cf) edge node {} (kp1)
                (lp) edge node {} (rlmst)
                (rlmst) edge node {} (t6)
;
\end{tikzpicture}
\captionof{figure}{Reclamation of the old successor.}
\label{fig:del2.6}
\end{subfigure}
\caption{\textsf{Delete} of a heap node with two children in BST~\cite{Arbel:2014:CUR:2611462.2611471}.}
\label{fig:del2}
\end{figure}
We have used our type system to check correct use of RCU primitives in two RCU data structures representative of the broader space.

Figure \ref{fig:rculist} gives the type-annotated code for \lstinline|add| and \lstinline|remove| operations on a linked list implementation of a bag data structure, following McKenney's example~\cite{McKenney2015SomeEO}.
Our technical report~\cite{isotek} contains code for membership checking.

We have also type checked the most challenging part of an RCU binary search tree, the deletion (which also contains the code for a lookup).
Our implementation is a slightly simplified version of the Citrus BST~\cite{Arbel:2014:CUR:2611462.2611471}: their code supports fine-grained locking for multiple writers, while ours supports only one writer by virtue of using our single-writer primitives. For lack of space the annotated code is only in Appendix B of the technical report~\cite{isotek}, but here we emphasise the important aspects our type system via showing its capabilities of typing BST delete method, which also includes looking up for the node to be deleted.% which includes lookup as well, in Section \ref{subsection:type-action}.
%The use of disjunction ($Left|Right$) in field maps and paths is required to capture traversals which follow different fields at different times, such as the lookup in a binary search tree.

In Figure \ref{fig:del2}, we show the steps for deleting the heap node $H_1$. To locate the node $H_1$, as shown in Figure \ref{fig:del2.1}, we first traverse the subtree $T_0$ with references $pr$ and $cr$, where $pr$ is the parent of $cr$ during traversal:
\[ pr:rcuItr(l|r)^{k} \{l|r \rightarrow cr\},\, cr:rcuItr(l|r)^{k}.(l|r) \{\}\]
Traversal of $T_0$ is summarized as $(l|k)^{k}$. The most subtle aspect of the deletion is the final step in the case the node $H_1$ to remove has both children; as shown in Figure \ref{fig:del2.2}, the code must traverse the subtree $T_4$ to locate the next element in collection order: the node $H_s$, the left-most node of $H_1$'s right child ($sc$) and its parent ($lp$):
\[lp:(l|r)^{k}.(l|r).r.(l|r)^{m} \{l|r \rightarrow sc\},\, sc:(l|r)^{k}.(l|r).r.l.(l)^{m}.l\{\}\]
where the traversal of $T_4$ is summarized as $(l|m)^{m}$.

Then $H_s$ is copied into a new \textit{freshly-allocated} node as shown in Figure \ref{fig:del2.2}, which is then used to \emph{replace} node $H_1$ as shown in Figure \ref{fig:del2.3}: the replacement's fields exactly match $H_1$'s except for the data (\textsc{T-Replace} via $\N_1 = \N_2$) as shown in Figure \ref{fig:del2.2}, and the parent is updated to reference the replacement, unlinking $H_1$.

At this point, as shown in Figures \ref{fig:del2.3} and \ref{fig:del2.4}, there are two nodes with the same value in the tree (the \textit{weak} BST property of the Citrus BST~\cite{Arbel:2014:CUR:2611462.2611471}): the replacement node, and what was the left-most node under $H_1$'s right child. This latter (original) node $H_s$ must be unlinked as shown in Figure \ref{fig:del2.5}, which is simpler because by being left-most the left child is null, avoiding another round of replacement (\textsc{T-UnlinkH} via $\forall_{f\in dom(\N_1)} \ldotp f\neq f_2 \implies (\N_1(f) = \textsf{null}$).

Traversing $T_4$ to find successor complicates the reasoning in an interesting way. After the successor node $H_s$ is found in \ref{fig:del2.2}, there are \emph{two} local unlinking operations as shown in Figures \ref{fig:del2.3} and \ref{fig:del2.5}, at different depths of the tree.  This is why the type system must keep separate abstract iteration counts, e.g., $k$ of $(l|r)^{k}$ or $m$ of $(l|r)^{m}$, for traversals in loops --- these indices act like multiple cursors into the data structure, and allow the types to carry enough information to keep those changes separate and ensure neither introduces a cycle.

To the best of our knowledge, we are the first to check such code for memory-safe use of RCU primitives modularly, without appeal to the specific implementation of RCU primitives.
\raggedbottom
\section{Soundness}
\label{sec:soundness}
This section outlines the proof of type soundness -- our full proof appears the accompanying technical report~\cite{isotek}.
We prove type soundness by embedding the type system into an abstract concurrent separation logic called the Views Framework~\cite{views}, which when given certain information about proofs for a specific language (primitives and primitive typing) gives back a full program logic including choice and iteration. As with other work taking this approach~\cite{oopsla12,toplas17}, this consists of several key steps explained in the following subsections, but a high-level informal soundness argument is twofold. First, because the parameters given to the Views framework ensure the Views logic's Hoare triples $\{-\}C\{-\}$ are sound, this proves soundness of the type rules with respect to type denotations. Second, as our denotation of types encodes the property that the post-environment of any type rule accurately characterizes which memory is linked vs. unlinked, etc., and the global invariants ensure all allocated heap memory is reachable from the root or from some thread's stack, this entails that our type system prevents memory leaks.
\subsection{Proof}
\label{sec:proof}
This section provides more details on how the Views Framework~\cite{views} is used to prove soundness, giving the major parameters to the framework and outlining global invariants and key lemmas.
%The Views Framework takes a set of parameters satisfying some properties, and produces a soundness proof for a static reasoning system for a larger programming language.  Among other parameters, the most notable are the choice of machine state, semantics for \emph{atomic} actions (e.g., field writes, or \lstinline|WriteBegin|), and proofs that the reasoning (in our case, type rules) for the atomic actions are sound (in a way chosen by the framework).
%The other critical pieces are a choice for a partial \emph{view} of machine states --- usually an extended machine state with meta-information --- and a relation constraining how other parts of the program can interfere with a view (e.g., modifying a value in the heap, but not changing its type).
%Our type system will be related to the views by giving a denotation of type environments in terms of views, and then proving that for each atomic action shown in \ref{fig:operationalsemrcu} in Section \ref{sec:semantics} and type rule in Figure \ref{fig:tss} Section \ref{subsection:type-action}(also for type rules for \textit{structural} and \textit{reader critical section} in Figure \ref{fig:tssr} Appendix \ref{appendix:readtypes}), given a view in the denotation of the initial type environment of the rule, running the semantics for that action yields a local view in the denotation of the output type environment of the rule. The following works through this in more detail. We define logical states, $\textsf{LState}$ to be

\mypar{Logical State} Section \ref{sec:semantics} defined what Views calls \emph{atomic actions} (the primitive operations) and their semantics on runtime \emph{machine states}.  The Views Framework uses a separate notion of instrumented (logical) state over which the logic is built, related by a concretization function $\lfloor-\rfloor$ taking an instrumented state to the machine states of Section \ref{sec:semantics}.  Most often --- including in our proof ---  the logical state adds useful auxiliary state to the machine state, and the concretization is simply projection.
Thus we define our logical states \textsf{LState} as:
\begin{itemize}
\item A machine state, $\sigma=(s,h,l,rt,R,B)$
\item An observation map, O, of type $ \textsf{Loc} \to \mathcal{P}(\textsf{obs})$
\item Undefined variable map, $U$, of type $\mathcal{P}(\textsf{Var}\times \textsf{TID})$
\item Set of threads, $T$, of type $\mathcal{P}(\textsf{TIDS})$
\item A to-free map (or free list), $F$, of type $\textsf{Loc} \rightharpoonup \mathcal{P}(\textsf{TID})$
\end{itemize}
The thread ID set $T$ includes the thread ID of all running threads.
The free map $F$ tracks which reader threads may hold references to each location. It is not required for execution of code, and for validating an implementation could be ignored, but we use it later with our type system to help prove that memory deallocation is safe.
The (per-thread) variables in the undefined variable map $U$ are those that should not be accessed (e.g., dangling pointers).

The remaining component, the observation map $O$, requires some further explanation.
Each memory allocation / object can be \emph{observed} in one of the following states by a variety of threads, depending on how it was used.
\[\textsf{obs} := \texttt{iterator} \; \mathrm{tid} \mid \texttt{unlinked} \mid \texttt{fresh} \mid \texttt{freeable} \mid \texttt{root}\]
An object can be observed as part of the structure (\texttt{iterator}), removed but possibly accessible to other threads, freshly allocated, safe to deallocate, or the root of the structure.

\mypar{Invariants of RCU Views and Denotations of Types}\label{sec:lemmas} Next, we aim to convey the intuition behind the predicate \textsf{WellFormed} which enforces global invariants on logical states, and how it interacts with the denotations of types (Figure \ref{fig:denotingtypeenviroment}) in key ways.

\textsf{WellFormed} is the conjunction of a number of more specific invariants, which we outline here.
For full details, see Appendix A.2 of the technical report~\cite{isotek}.

\paragraph{The Invariant for Read Traversal} Reader threads access valid heap locations even during the grace period. The validity of their heap accesses ensured by the observations they make over the heap locations --- which can only be \textsf{iterator} as they can only use local \textsf{rcuItr} references. To this end, a \textsf{Readers-Iterators-Only} invariant asserts that reader threads can only observe a heap location as \textsf{iterator}.
\paragraph{Invariants on Grace-Period}
Our logical state includes a ``free list'' auxiliary state tracking which readers are still accessing \emph{each} unlinked node during grace periods. This must be consistent with the bounding thread set $B$ in the machine state, and this consistency is asserted by the \textsf{Readers-In-Free-List} invariant. This is essentially tracking which readers are being ``shown grace'' for each location. The \textsf{Iterators-Free-List} invariant complements this by asserting all readers with such observations on unlinked nodes are in the bounding thread set.

The writer thread can refer to a heap location in the free list with a local reference either in type \textsf{freeable} or \textsf{unlinked}. Once the writer unlinks a heap node, it first observes the heap node as \textsf{unlinked} then \textsf{freeable}. The denotation of \textsf{freeable} is only valid following a grace period: it asserts no readers hold aliases of the \textsf{freeable} reference. The denotation of \textsf{unlinked} permits the either the same (perhaps no readers overlapped) 
or that it is in the to-free list.
\paragraph{Invariants on Safe Traversal against Unlinking}
The write-side critical section must guarantee that no updates to the heap cause invalid memory accesses. 
%Our invariants provide reasoning over the mutated heap through possible observations made by the writer thread. 
The \textsf{Writer-Unlink} invariant asserts that a heap location observed as \textsf{iterator} by the writer thread cannot be observed differently by other threads. The denotation of the writer thread's \textsf{rcuItr} reference, $\llbracket \mathsf{rcuItr}\,\rho \,\N \rrbracket_{tid}$, asserts that 
following a path from the root compatible with $\rho$ reaches the referent, and all are observed as \textsf{iterator}.
%all the heap locations on the path starting from the root to the referent heap location are valid only observed as \textsf{iterator}.

The denotation of a reader thread's \textsf{rcuItr} reference, $\llbracket \mathsf{rcuItr} \rrbracket_{tid}$ and the invariants \textsf{Readers-Iterator-Only}, \textsf{Iterators-Free-List} and \textsf{Readers-In-Free-List} all together assert that a reader thread(which can also be a bounding thread) can view an unlinked heap location(which can be in the free list) only as \textsf{iterator}.
%and it is important in terms of  the preservation of \textsf{Readers-Iterator-Only} invariant for the set of bounding threads against unlinking as they are a subset of the reader threads. 
At the same time, it is essential that reader threads arriving after a node is unlinked cannot access it.
%However, the unreachability of a mutated heap location to newly-arrived reader threads is essential for the \textit{soundness} of the system. 
%A heap location which is observed as \textsf{unlinked}/\textsf{freeable} can only be accessed from a heap location that has already been mutated (can only be observed as \textsf{unlinked}/\textsf{freeable} by the writer as well). By asserting this, \textsf{Unlinked-Reachability} and \textsf{Free-List-Reachability} invariants prevent access by new reader threads.
The invariants \textsf{Unlinked-Reachability} and \textsf{Free-List-Reachability} ensure that any unlinked nodes are reachable only from other unlinked nodes, and never from the root.
\looseness=-1
% but also support reasoning for \textit{batched-deallocation} together with the existence of the logical free list.
%%% CG NOTE: I commented this last part out because this isn't the place to discuss batch deallocation
\paragraph{Invariants on Safe Traversal against Inserting/Replacing} A writer {replacing an existing node with a fresh one} or {inserting a single fresh node} assumes the fresh (before insertion) node is unreachable to readers before it is published/linked.
%The crucial essence in providing memory safety against these actions is enforcing unreachability of the fresh nodes until they are \textit{published/linked}. Unlike in the case of \textit{unlinking} where the reader threads can observe an unlinked heap node as \textsf{iterator} during the grace period, the reader threads cannot observe the fresh heap nodes until they are published. 
The \textsf{Fresh-Writes} invariant asserts that a fresh heap location can only be allocated and referenced by the writer thread. The relation between a freshly allocated heap and the rest of the heap is established by the \textsf{Fresh-Reachable} invariant, which requires that there exists no heap node pointing to the freshly allocated one. This invariant supports the preservation of the tree structure. The \textsf{Fresh-Not-Reader} invariant supports the safe traversal of the reader threads via asserting that they cannot observe a heap location as \textsf{fresh}. 
Moreover, the denotation of the \textsf{rcuFresh} type, $\llbracket \mathsf{rcuFresh}\,\N \rrbracket_{tid}$, enforces that fields in $\N$ point to valid heap locations (observed as \textsf{iterator} by the writer thread).
% to support the soundness of the \textit{framing} relations in \textsc{T-Replace}(discussed in Figure \ref{fig:freshframeout}) and \textsc{T-Insert}.
%Similar to the case for unlinking, the \textsf{Readers-Iterators-Only} must be preserved against inserting/replacing a fresh node and
\paragraph{Invariants on Tree Structure} Our invariants enforce the \textit{tree} structure heap layouts for data structures. The \textsf{Unique-Reachable} invariant asserts that every heap location reachable from root can only be reached with following an unique path. To preserve the tree structure, \textsf{Unique-Root} enforces unreachability of the root from any heap location that is reachable from root itself. 

\begin{figure}[!t]\scriptsize
\[
\begin{array}{l@{\;\;=\;\;}l}
 \llbracket \, x : \textsf{rcuItr}\,\rho\,\N \,  \rrbracket_{tid}
&
\left\{
\begin{array}{l|l}
m \in \mathcal{M}
&(\textsf{iterator} \, tid\in  O(s(x,tid)))  \land (x \notin U)  \\
& \land (\forall_{f_i\in dom(\N)  x_i\in codom(\N) } \ldotp
\left\{\begin{array}{l}  s(x_i,tid) = h(s(x,tid), f_i)  \\
 \land \textsf{iterator}\in O(s(x_i,tid)))\end{array} \right.\\
& \land  (\forall_{\rho', \rho''}\ldotp \rho'.\rho'' = \rho \implies  \textsf{iterator}\,tid \in O(h^{*}(rt,\rho'))) \\
& \land  h^{*}(rt,\rho)= s(x,tid)  \land (l = tid \land s(x,\_) \notin dom(F))) 
\end{array}
\right\}
\\
 \llbracket \, x : \textsf{rcuItr}\,  \rrbracket_{tid}
&
\left\{
\begin{array}{l|l}
m \in \mathcal{M}
&(\textsf{iterator} \, tid\in  O(s(x,tid)))  \land (x \notin U) \land \\
& (tid \in B) \implies \left\{\begin{array}{l}( \exists_{T'\subseteq B}\ldotp \{s(x,tid) \mapsto T'\} \cap F \neq \emptyset) \land \\ \land (tid \in T') \end{array}\right.
\end{array}
\right\}
\\
\llbracket \, x : \textsf{unlinked} \, \rrbracket_{tid}
&
\left\{
\begin{array}{l|l}
m \in \mathcal{M}
&(\textsf{unlinked}\in  O(.s(x,tid)) \land l = tid \land x \notin U) \land \\
& (\exists_{T'\subseteq T}\ldotp s(x,tid) \mapsto T' \in F \implies T' \subseteq B \land tid \notin T' )
\end{array}
\right\}
\\
\llbracket \, x : \textsf{freeable} \, \rrbracket_{tid}
&
\left\{
\begin{array}{l|l}
m \in \mathcal{M}
&\begin{array}{l}\textsf{freeable}\in  O(s(x,tid)) \land l = tid \land x \notin U \land \\
 s(x,tid) \mapsto \{\emptyset\} \in F \end{array}
\end{array}
\right\}
\\
\llbracket \, x : \textsf{rcuFresh} \, \N \, \,  \rrbracket_{tid}
&
\left\{
\begin{array}{l|l}
m \in \mathcal{M}
&(\textsf{fresh}\in  O(s(x,tid)) \land x \notin U  \land s(x,tid) \notin dom(F))\\
&(\forall_{f_i\in dom(\N), x_i\in codom(\N) } \ldotp s(x_i,tid) = h(s(x,tid), f_i) \\
&\land \textsf{iterator}\,tid \in O(s(x_i,tid)) \land s(x_i,tid) \notin dom(F)) 
\end{array}
\right\}
\\
\llbracket  x : \textsf{undef}\rrbracket_{tid} 
&
\left\{
\begin{array}{l|l}
m \in \mathcal{M}
&
(x,tid) \in U \land s(x,tid) \notin dom(F)
\end{array}
\right\}
\\
\llbracket \, x : \textsf{rcuRoot}\rrbracket_{tid}
&
\left\{
\begin{array}{l|l}
m \in \mathcal{M}
& \begin{array}{l}((rt \notin U \land s(x,tid) = rt \land rt \in dom(h) \land \\ 
O(rt) \in \textsf{root} \land s(x,tid) \notin dom(F) ) \end{array}
\end{array}
\right\}
\end{array}
\]
$
\textrm{provided}~h^{*}: (\textsf{Loc} \times \textsf{Path}) \rightharpoonup \textsf{Val}
$
\caption{Type Environments}
\label{fig:denotingtypeenviroment}
\vspace{-2mm}
\end{figure}
\mypar{Type Environments} Assertions in the Views logic are (almost) sets of the logical states that satisfy a validity predicate \textsf{WellFormed}, outlined above:
%A thread's (or scope's) \emph{view} of memory is a subset of the instrumented(logical states), which satisfy certain \textsf{WellFormed}ness criteria relating the physical state and the additional meta-data ($O$, $U$, $T$ and $F$)
\[\mathcal{M} \stackrel{def}{=} \{ m \in (\textsf{MState} \times O \times U \times T \times F) \mid  \textsf{WellFormed}(m) \} \]
%We do our reasoning for soundness over instrumented states and define an erasure relation
%\[\lfloor - \rfloor :\mathsf{MState} \implies \textsf{LState}\]
%that projects instrumented states to the common components with \textsf{MState}.
%In this section we aim to convey an intuition for the {soundness} of the type system through an informal summarization of the global \textsf{WellFormed}ness invariants together with the denotations of the types imposed on the instrumented states in the proof.

Every type environment represents a set of possible views (\textsf{WellFormed} logical states) consistent with the types in the environment. We make this precise with a denotation function
\[\llbracket-\rrbracket\_ : \mathsf{TypeEnv}\rightarrow\mathsf{TID}\rightarrow\mathcal{P}(\mathcal{M})\]
that yields the set of states corresponding to a given type environment. This is defined as the intersection of individual variables' types as in Figure \ref{fig:denotingtypeenviroment}.
%\[\llbracket-:-\rrbracket_- : \mathsf{Var}\rightarrow\mathsf{Type}\rightarrow\mathsf{TID}\rightarrow\mathcal{P}(\mathcal{M})\]
%
%The latter is given in Figure \ref{fig:denotingtypeenviroment}.  To define the former, we first need to state what it means to combine logical machine states.

Individual variables' denotations are extended to context denotations slightly differently depending on whether the environment is a reader or writer thread context: writer threads own the global lock, while readers do not:
\begin{itemize}
\item For read-side as $\llbracket x_1 : T_1, \ldots x_n : T_n \rrbracket_{tid,\textsf{R}} = \llbracket x_1 : T_1 \rrbracket_{tid} \cap \ldots \cap \llbracket x_n : T_n \rrbracket_{tid} \cap \llbracket \textsf{R} \rrbracket_{tid}$ where
$\llbracket \textsf{R} \rrbracket_{tid} = \{ (s,h,l,rt,R,B),O,U,T,F  \mid tid \in R \}$

\item For write-side as $\llbracket x_1 : T_1, \ldots x_n : T_n \rrbracket_{tid,\textsf{M}} = \llbracket x_1 : T_1 \rrbracket_{tid} \cap \ldots \cap \llbracket x_n : T_n \rrbracket_{tid} \cap \llbracket \textsf{M} \rrbracket_{tid}$ where
$\llbracket \textsf{M} \rrbracket_{tid} = \{ (s,h,l,rt,R,B),O,U,T,F  \mid tid = l \}$
%
%\item $\llbracket x_1 : T_1, \ldots x_n : T_n \rrbracket_{tid,\textsf{O}} = \llbracket x_1 : T_1 \rrbracket_{tid} \cap \ldots \cap \llbracket x_n : T_n \rrbracket_{tid} \cap \llbracket \textsf{O} \rrbracket_{tid}$ where
%$\llbracket \textsf{O} \rrbracket_{tid} = \{ (s,h,l,rt,R,B),O,U,T,F  \mid tid \neq l \land tid \notin R \}$.
\end{itemize}

\mypar{Composition and Interference} To support framing (weakening), the Views Framework requires that views form a partial commutative monoid under an operation $\bullet : \mathcal{M} \longrightarrow \mathcal{M} \longrightarrow \mathcal{M}$, provided as a parameter to the framework.
The framework also requires an interference relation $\mathcal{R}\subseteq\mathcal{M}\times\mathcal{M}$ between views to reason about local updates to one view preserving validity of adjacent views (akin to the small-footprint property of separation logic).
Figure \ref{fig:comp} defines our composition operator and the core interference relation $\mathcal{R}_0$ --- the actual inferference between views (between threads, or between a local action and framed-away state) is the reflexive transitive closure of $\mathcal{R}_0$.
Composition is mostly straightforward point-wise union (threads' views may overlap) of each component.
Interference bounds the interference writers and readers may inflict on each other.  Notably, if a view contains the writer thread, other threads may not modify the shared portion of the heap, or release the writer lock.  Other aspects of interference are natural restrictions like that threads may not modify each others' local variables.
\textsf{WellFormed} states are closed under both composition (with another \textsf{WellFormed} state) and interference ($\mathcal{R}$ relates \textsf{WellFormed} states only to other \textsf{WellFormed} states).

\begin{figure}[!t]\scriptsize
    \begin{flalign*}
      \bullet \overset{\mathrm{def}}{=} (\bullet_{\sigma},\bullet_O,\cup,\cup) \;\; (F_1 \bullet_F F_2) \overset{\mathrm{def}}{=}  F_1  \cup F_2 \texttt{   when   } dom(F_1) \cap dom(F_2) = \emptyset \\
      O_{1} \bullet_O O_{2}(loc) \overset{\mathrm{def}}{=}  O_{1}(loc) \cup O_{2}(loc) \;\;   (s_1 \bullet_s s_2) \overset{\mathrm{def}}{=}  s_1 \cup s_2 \texttt{   when   } dom(s_1) \cap dom(s_2) = \emptyset    
      \end{flalign*}
\[(h_1\bullet_h h_2)(o,f)\overset{\mathrm{def}}{=}\left\{
\begin{array}{ll}
\mathrm{undef} & \textrm{if}~h_1(o,f)=v \land h_2(o,f)=v' \land v' \neq v\\
v & \textrm{if}~h_1(o,f)=v \land h_2(o,f)=v\\
v & \textrm{if}~h_1(o,f)=\mathrm{undef}\land h_2(o,f)=v\\
v & \textrm{if}~h_1(o,f)=v\land h_2(o,f)=\mathrm{undef}\\
\mathrm{undef} & \textrm{if}~h_1(o,f)=\mathrm{undef}\land h_2(o,f)=\mathrm{undef}
\end{array}
\right.
\]
\[
\begin{array}{l}
((s,h,l,rt,R,B), O, U, T,F) \mathcal{R}_{0}((s',h',l',rt',R',B'), O', U', T',F') \overset{\mathrm{def}}{=}
\\ \bigwedge\left\{
	\begin{array}{l}
	  l  \in  T \rightarrow (h = h' \land l=l')\\
	  l\in T\rightarrow F=F'\\
	  \forall tid,o\ldotp\textsf{iterator} \, tid \in O(o) \rightarrow o \in dom(h) \\
	  \forall tid,o\ldotp\textsf{iterator} \, tid \in O(o) \rightarrow o \in dom(h') \\
          \forall tid,o\ldotp\textsf{root} \, tid \in O(o) \rightarrow o \in dom(h) \\
	  \forall tid,o\ldotp\textsf{root} \, tid \in O(o) \rightarrow o \in dom(h') \\
	  O = O' \land U = U' \land T = T'\land R =R'\land rt = rt' \\
	  \forall x, t \in T \ldotp s(x,t) = s'(x,t)
	\end{array}
\right\}
\end{array}
\]
\caption{Composition($\bullet$) and Thread Interference Relation($\mathcal{R}_{0}$)}
\label{fig:comp}
\vspace{-2mm}
\end{figure}

%\emph{Separation algebra} is a model to define and axiomatize the \emph{join/composition} operation over a domain which is \emph{set of instrumented states} in our case.

%An important property of composition is that it preserves validity of logical states:
%\begin{lemma}[\textsf{WellFormed} Composition]
%\label{lem:wf-composition}
%Any successful composition of two \textsf{WellFormed} logical states is \textsf{WellFormed}:\[\forall_{x,y,z}\ldotp \mathsf{WellFormed}(x) \implies \mathsf{WellFormed}(y) \implies x\bullet y = z \implies \mathsf{WellFormed(z)}\]
%\end{lemma}
%\begin{proof}
%In Appendix \ref{sec:prooflemmas} Lemma \ref{lem:wf-compositionap}.
%  \end{proof}

\mypar{Stable Environment and Views Shift} The framing/weakening type rule will be translated to a use of the frame rule in the Views Framework's logic.  There separating conjunction is simply the existence of two composable instrumented states:
%Partial separating conjunction then simply requires the existence of two states that compose:
\[ m \in P \ast Q   \stackrel{def}{=} \exists m' \ldotp \exists m'' \ldotp m' \in P \land m'' \in Q \land m \in m' \bullet m''\]
%Different threads' views of the state may overlap (e.g., on shared heap locations, or the reader thread set), but one thread may modify that shared state.  
In order to validate the frame rule in the Views Framework's logic, the assertions in its logic --- sets of well-formed instrumented states --- must be restricted to sets of logical states that are \emph{stable} with respect to expected interference from other threads or contexts, and interference must be compatible in some way with separating conjunction.
Thus a \textsf{View} --- the actual base assertions in the Views logic --- are then:
\[\textsf{View}_{\mathcal{M}} \stackrel{def}{=} \{ M \in \mathcal{P}(\mathcal{M}) | \mathcal{R}(M) \subseteq M\}\]
%Thread interference relation
%\[\mathcal{R} \subseteq \mathcal{M} \times \mathcal{M}\]  defines permissible interference on an instrumented state. The relation must distribute over composition:
Additionally, interference must distribute over composition:
\[ \forall m_{1}, m_{2}, m\ldotp (m_{1} \bullet  m_{2})\mathcal{R}m \Longrightarrow \begin{array}{ll}  \exists  m'_{1} m'_{2} \ldotp m_{1} \mathcal{R} m'_{1} \land m_{2} \mathcal{R} m'_{2} \land  m \in m'_{1} \bullet m'_{2} \end{array}\]
%where $\mathcal{R}$ is transitive-reflexive closure of $\mathcal{R}_{0}$ shown at Figure \ref{fig:comp}. $\mathcal{R}_0$ (and therefore $\mathcal{R}$) also ``preserves'' validity:

%\begin{lemma}[Valid $\mathcal{R}_0$ Interference]\label{lem:inter}
%For any $m$ and $m'$, if $\mathsf{WellFormed}(m)$ and $m\mathcal{R}_0m'$, then $\mathsf{WellFormed}(m')$.
%\end{lemma}
%\begin{proof}
%In Appendix \ref{sec:prooflemmas} Lemma \ref{lem:interap}.
%\end{proof}

Because we use this induced Views logic to prove soundness of our type system by translation, we must ensure any type environment denotes a valid view:
\begin{lemma}[Stable Environment Denotation-M]\label{lemma:stblw}
For any \emph{closed} environment $\Gamma$ (i.e., $\forall x\in\mathsf{dom}(\Gamma)\ldotp, \mathsf{FV}(\Gamma(x))\subseteq\mathsf{dom}(\Gamma)$):
$
\mathcal{R}(\llbracket\Gamma\rrbracket_{\mathsf{M},tid})\subseteq\llbracket\Gamma\rrbracket_{\mathsf{M},tid}
$.
Alternatively, we say that environment denotation is \emph{stable} (closed under $\mathcal{R}$).
\end{lemma}
\begin{proof}
In Appendix A.1 Lemma 7 of the technical report~\cite{isotek}.
\end{proof}
We elide the statement of the analogous result for the read-side critical section, available in Appendix A.1 of the technical report.
%\begin{lemma}[Stable Environment Denotation-R]
%For any \emph{closed} environment $\Gamma$ (i.e., $\forall x\in\mathsf{dom}(\Gamma)\ldotp, \mathsf{FV}(\Gamma(x))\subseteq\mathsf{dom}(\Gamma)$):
%\[
%\mathcal{R}(\llbracket\Gamma\rrbracket_{\mathsf{R},tid})\subseteq\llbracket\Gamma\rrbracket_{\mathsf{R},tid}
%\]
%Alternatively, we say that environment denotation is \emph{stable} (closed under $\mathcal{R}$).
%\end{lemma}
%\begin{proof}
%In Appendix \ref{sec:prooflemmas} Lemma \ref{lem:stblRap}.
%\end{proof}

With this setup done, we we can state the connection between the Views Framework logic induced by earlier parameters, and the type system from Section \ref{sec:tslbl}.
The induced Views logic has a familiar notion of Hoare triple --- $\{ p \} C \{ q \}$ where $p$ and $q$ are elements of $\mathsf{View}_\mathcal{M}$ --- with the usual rules for non-deterministic choice, non-deterministic iteration, sequential composition, and parallel composition, sound given the proof obligations just described above.  It is parameterized by a rule for atomic commands that requires a specification of the triples for primitive operations, and their soundness (an obligation we must prove).  This can then be used to prove that every typing derivation embeds to a valid derivation in the Views Logic, roughly
$\forall\Gamma,C,\Gamma',\mathit{tid}\ldotp\Gamma\vdash C\dashv \Gamma' \Rightarrow \{\llbracket\Gamma\rrbracket_\mathit{tid}\} \llbracket C\rrbracket_\mathit{tid}\{\llbracket\Gamma'\rrbracket_\mathit{tid}\}$
once for the writer type system, once for the readers.

\begin{figure}[t]\scriptsize
$
\begin{array}{l}
\downarrow{\mathsf{if}\;(x.f==y)\;C_1\;C_2}\downarrow\mathit{tid} \overset{\mathrm{def}}{=} z=x.f;((\mathsf{assume}(z=y);C_1)+(\mathsf{assume}(z\neq y);C_2))
\end{array}
$
$
\llbracket\texttt{assume}(\mathcal{S})\rrbracket (s)\overset{\mathrm{def}}{=}\left\{
\begin{array}{ll}
\{ s\} & \textrm{if}~s \in \mathcal{S}\\
\emptyset & \textrm{Otherwise}
\end{array}
\right.
$
$
\downarrow{\mathsf{while}\;(e)\;C}\downarrow \overset{\mathrm{def}}{=} \left(\mathsf{assume}(e);C\right)^{*};(\mathsf{assume}(\lnot e ));
$
$
\inferrule
{
\{P\} \cap \{\lceil\mathcal{S} \rceil \}  \sqsubseteq \{Q\}
}
{
 \{P\} \texttt{assume}\left(\mathcal{S}\right)\{Q\}
}
$
\textsf{ where } $\lceil \mathcal{S} \rceil = \{m | \lfloor m \rfloor \cap \mathcal{S} \neq \emptyset \}
$
\caption{Encoding branch conditions with \textsf{assume}(b)}
\label{fig:asm}
\end{figure}

There are two remaining subtleties to address.  First,
commands $C$ also require translation: the Views Framework has only non-deterministic branches and loops, so the standard versions from our core language must be encoded.  The approach to this is based on a standard idea in verification, which we show here for conditionals as shown in Figure \ref{fig:asm}. $\textsf{assume}(b)$ is a standard idea in verification semantics~\cite{Barnett:2005:BMR:2090458.2090481,Muller:2016:VVI:2963187.2963190}, which ``does nothing'' (freezes) if the condition $b$ is false, so its postcondition in the Views logic can reflect the truth of $b$.  \textsf{assume} in Figure \ref{fig:asm} adapts this for the Views Framework as in other Views-based proofs~\cite{oopsla12,toplas17}, specifying sets of machine states as a predicate. We write boolean expressions as shorthand for the set of machine states making that expression true. With this setup done, the top-level soundness claim then requires proving -- once for the reader type system, once for the writer type system -- that every valid source typing derivation corresponds to a valid derivation in the Views logic:
$
    \forall \Gamma,C,\Gamma', \Gamma\vdash_M C \dashv \Gamma' \Rightarrow \{\db{\Gamma}\}\downarrow C\downarrow\{\db{\Gamma'}\}
$.

Second, we have not addressed a way to encode subtyping.  One might hope this corresponds to a kind of implication, and therefore subtyping corresponds to consequence.  Indeed, this is how we (and prior work~\cite{oopsla12,toplas17}) address subtyping in a Views-based proof.
Views defines the notion of \emph{view shift}\footnote{This is the same notion present in later program logics like Iris~\cite{krebbers2017essence}, though more recent variants are more powerful.} ($\sqsubseteq$) as a way to reinterpret a set of instrumented states as a new (compatible) set of instrumented states, offering a kind of logical consequence, used in a rule of consequence in the Views logic:
\[ p \sqsubseteq q \stackrel{def}{=} \forall m\in\mathcal{M} \ldotp \lfloor p * \{ m \} \rfloor \subseteq \lfloor q* \mathcal{R}(\{m\})\rfloor\]

We are now finally ready to prove the key lemmas of the soundness proof, relating subtying to view shifts, proving soundness of the primitive actions, and finally for the full type system.  These proofs occur once for the writer type system, and once for the reader; we show here only the (more complex) writer obligations:

%We express the reinterpretation of $p$ as $q$ with action $\alpha$ ensuring that the operation interpretation of the action satisfies the specification:$ p \sqsubseteq q \stackrel{def}{=} \forall m\in\mathcal{M} \ldotp \lfloor p * \{ m \} \rfloor \subseteq \lfloor q* \mathcal{R}(\{m\})\rfloor$. Because the Views framework handles soundness for the structural rules (sequencing, parallel composition, etc.), there are really only three types of proof obligations for us to prove.  First, we must prove that the non-trivial command translations (i.e., for conditionals and while loops) embed correctly in the Views logic, which is straightforward.  Second, we must show that for our environment subtyping, if $\Gamma<:\Gamma'$, then $\llbracket{\Gamma}\rrbracket\sqsubseteq\llbracket\Gamma'\rrbracket$.  And finally, we must prove that each atomic action's type rule corresponds to a valid semantic judgment in the Views Framework:
%\[
%\forall m\ldotp   \llbracket \alpha \rrbracket ( \lfloor \llbracket \Gamma_{1} \rrbracket_{tid}  * \{m\} \rfloor) \subseteq
%  \lfloor \llbracket \Gamma_{2} \rrbracket_{tid} * \mathcal{R}(\{m\}) \rfloor
%  \]
%The use of $*$ validates the frame rule and makes this obligation akin to an interference-tolerant version of the small footprint property from traditional separation logics~\cite{Reynolds:2002:SLL:645683.664578,Calcagno:2007:LAA:1270399.1271718}.
\begin{lemma}[Axiom of Soundness for Atomic Commands]
  \label{lem:axmsoundness}
For each axiom, $\Gamma_{1} \vdash_{\textsf{M}} \alpha \dashv \Gamma_{2}$, we show
$
\forall m\ldotp   \llbracket \alpha \rrbracket  (\lfloor \llbracket \Gamma_{1} \rrbracket_{tid}  * \{m\} \rfloor )\subseteq  \lfloor \llbracket \Gamma_{2} \rrbracket_{tid} * \mathcal{R}(\{m\}) \rfloor
$
\end{lemma}
\begin{proof}
By case analysis on $\alpha$. Details in Appendix A.1 of the techical report~\cite{isotek}. % Lemma \ref{lem:axmsoundnessap}
\end{proof}
\begin{lemma}[Context-SubTyping-M]\label{lem:cntxsubt-mcln}
$ \Gamma \subt \Gamma'  \implies \llbracket \Gamma \rrbracket_{M,tid} \sqsubseteq \llbracket  \Gamma' \rrbracket_{M,tid} $
\end{lemma}
\begin{proof}
Induction on the subtyping derivation, then inducting on the single-type subtype relation for the first variable in the non-empty context case.
\end{proof}

%Type soundness proceeds according to the requirements of the Views Framework, primarily embedding each type judgment into the Views logic:
%\begin{lemma}[Views Embedding for Read-Side]
%  \label{lemma:embedr}
%\[  \forall\Gamma,C,\Gamma',\mathit{t}\ldotp\Gamma\vdash_R C\dashv \Gamma' \Rightarrow
%\llbracket\Gamma\rrbracket_\mathit{t}\cap\llbracket{R}\rrbracket_t\vdash \llbracket C\rrbracket_\mathit{t}\dashv\llbracket\Gamma'\rrbracket_\mathit{t}\cap\llbracket{R}\rrbracket_t
%\]
%\end{lemma}
%\begin{proof}
%In Appendix \ref{sec:prooflemmas} Lemma \ref{lemma:embedrap}.
%  \end{proof}
\begin{lemma}[Views Embedding for Write-Side]
  \label{lemma:embedw}
\\\mbox{$\qquad
\forall\Gamma,C,\Gamma',\mathit{t}\ldotp\Gamma\vdash_M C\dashv \Gamma' \Rightarrow
\llbracket\Gamma\rrbracket_\mathit{t}\cap\llbracket{M}\rrbracket_t\vdash \llbracket C\rrbracket_\mathit{t}\dashv\llbracket\Gamma'\rrbracket_\mathit{t}\cap\llbracket{M}\rrbracket_t
$}
\end{lemma}
\begin{proof}
By induction on the typing derivation, appealing to Lemma \ref{lem:axmsoundness} for primitives, Lemma \ref{lem:cntxsubt-mcln} and consequence for subtyping, and otherwise appealing to structural rules of the Views logic and inductive hypotheses.
Full details in Appendix A.1 of the technical report~\cite{isotek}. %Lemma \ref{lemma:embedwap}.
  \end{proof}

The corresponding obligations and proofs for the read-side critical section type system are similar in statement and proof approach, just for the read-side type judgments and environment denotations.
\raggedbottom
\section{Related Work}
\label{sec:relatedwork}
Our type system builds on a great deal of related work on RCU implementations and models; and general concurrent program verification (via program logics, model checking, and type systems).

\mypar{Modeling RCU and Memory Models}
Alglave et al.~\cite{Alglave:2018:FSC:3173162.3177156} propose a memory model to be assumed by the platform-independent parts of the Linux kernel, regardless of the underlying hardware's memory model.
As part of this, they give the first formalization of what it means for an RCU implementation to be correct (previously this was difficult to state, as the guarantees in principle could vary by underlying CPU architecture). Essentially, that reader critical sections do not span grace periods. They prove by hand that the Linux kernel RCU implementation~\cite{DBLP:conf/cav/AlglaveKT13} satisfies this property. According to the fundamental requirements of \textsf{RCU} ~\cite{abssem}, our model in Section \ref{sec:semantics} can be considered as a valid RCU implementation satisfying all requirements for an RCU implementation(assuming sequential consistency) aside from one performance optimization, \textit{Read-to-Write Upgrade}, which is important in practice but not memory-safety centric -- see the technical report~\cite{isotek} for detailed discussion on satisfying RCU requirements.
\begin{itemize}
  \item \textit{Grace-Period and Memory-Barrier Guarantee}: To reclaim a heap location, a mutator thread must synchronize with all of the reader threads with overlapping read-side critical sections to guarantee that none of the updates to the memory cause invalid memory accesses.
The operational semantics enforce a \textit{protocol} on the mutator thread's actions. First it unlinks a node from the data structure; the local type for that reference becomes \textsf{unlinked}. Then it waits for current reader threads to exit, after which the local type is \textsf{freeable}. Finally, it may safely reclaim the memory, after which the local type is \textsf{undef}.
The semantics prevent the writer from reclaiming too soon by adding the heap location to the free list of the state, which is checked dynamically by the actual free operation. We discuss the grace period and unlinking invariants in our system in Section \ref{sec:lemmas}.
\item \textit{Publish-Subscribe Guarantee}: Fresh heap nodes cannot be observed by the reader threads until they are published. As we see in the operational semantics, once a new heap location is allocated it can only be referenced by a local variable of type \textsf{fresh}. Once published, the local type for that reference becomes \textsf{rcuItr}, indicating it is now safe for the reader thread to access it with local references in \textsf{rcuItr} type. We discuss the related type assertions for inserting/replacing(Figures \ref{fig:frframeout}-\ref{fig:freshframeout}) a fresh node in Section \ref{subsection:type-rules} and the related invariants in Section \ref{sec:lemmas}.
\item \textit{RCU Primitives Guaranteed to Execute Unconditionally}: Unconditional execution of RCU Primitives are provided by the definitions in our operational semantics for our RCU primitives(e.g. \lstinline|ReadBegin|, \lstinline|ReadEnd|, \lstinline|WriteBegin| and \lstinline|WriteEnd|) as their executions do not consider failure/retry.
\item \textit{Guaranteed Read-to-Write Upgrade}: This is a performance optimization which allows the reader threads to upgrade the read-side critical section to the write-critical section by acquiring the lock after a traversal for a data element and ensures that the upgrading-reader thread exit the read-critical section before calling RCU synchronization. This optimization also allows sharing the traversal code between the critical sections. \textit{Read-to-Write} is an important optimization in practice but largely orthogonal to memory-safety. Current version of our system provides a strict separation of \textit{traverse-and-update} and \textit{traverse-only} intentions through the type system(e.g. different iterator types and rules for the RCU Write/Read critical sections) and the programming primitives. As a future work, we want to extend our system to support this performance optimization.
\end{itemize}

To the best of our knowledge, ours is the first abstract \emph{operational} model for a Linux kernel-style RCU implementation -- others are implementation-specific~\cite{Mandrykin:2016:TDV:3001219.3001297} or axiomatic like Alglave et al.'s.

Tassarotti et al. model a well-known way of implementing RCU synchronization without hurting readers' performance, \textsf{Quiescent State Based Reclamation}(QSBR)~\cite{urcu_ieee} where synchronization between the writer thread and reader threads provided via per-thread counters. Tassarotti et al.~\cite{verrcu} uses a protocol based program logic based on separation and ghost variables called \textsf{GPS}~\cite{Turon:2014:GNW:2660193.2660243} to verify a user-level implementation of \textsf{RCU} with a singly linked list client under \emph{release-acquire} semantics, which is a weaker memory model than sequential-consistency. They require \textit{release-writes} and \textit{acquire-reads} to the QSRB counters for proper synchronization in between the mutator and the reader threads. This protocol is exactly what we enforce over the logical observations of the mutator thread: from \textsf{unlinked} to \textsf{freeable}. 
Tassarotti et al.'s synchronization for linking/publishing new nodes occurs in a similar way to ours, so we anticipate it would be possible to extend our type system in the future for similar weak memory models.

\mypar{Program Logics} Fu et al.~\cite{shao_temp} extend Rely-Guarantee and Separation-Logic~\cite{vafeiadis07,Feng:2007:RCS:1762174.1762193,Feng:2009:LRR:1480881.1480922} with the \textit{past-tense} temporal operator to eliminate the need for using a history variable and lift the standard separation conjunction to assert over on execution histories. Gotsman et al.~\cite{Gotsman:2013:VCM:2450268.2450289} take assertions from temporal logic to separation logic~\cite{vafeiadis07} to capture the essence of epoch-based memory reclamation algorithms and have a simpler proof than what Fu et al. have~\cite{shao_temp} for Michael's non-blocking stack~\cite{Michael:2004:HPS:987524.987595} implementation under a sequentially consistent memory model.

Tassarotti et al.~\cite{verrcu} use \textit{abstract-predicates} -- e.g. WriterSafe -- that are specialized to the singly-linked structure in their evaluation.  This means reusing their ideas for another structure, such as a binary search tree, would require revising many of their invariants.  By contrast, our types carry similar information (our denotations are similar to their definitions), but are reusable across at least singly-linked and tree data structures (Section \ref{sec:eval}). 
Their proofs of a linked list also require managing assertions about RCU implementation resources, while these are effectively hidden in the type denotations in our system.
On the other hand, their proofs ensure full functional correctness.  Meyer and Wolff~\cite{myr} make a compelling argument that separating memory safety from correctness if profitable, and we provide such a decoupled memory safety argument. 

\mypar{Realizing our RCU Model} A direct implementation of our semantics would yield unacceptable performance, since both entering (\lstinline|ReadBegin|) and exiting (\lstinline|ReadEnd|) modify shared data structures for the \textit{bounding-threads} and \textit{readers} sets. A slight variation on our semantics would use a bounding set  that tracked such a snapshot of counts, and a vector of per-thread counts in place of the reader set. Blocking grace period completion until the snapshot was strictly older than all current reader counts would be more clearly equivalent to these implementations. Our current semantics are simpler than this alternative, while also equivalent. 
%We also know that if we had this slight variation in our model then all actions (traversal and reads) from a reader thread would \textit{reduce} to a single \textit{atomic} action: read of a data value. However, we do not prove the \textit{atomicity} of the readers actions within the scope of this paper.
%%% NOTE: this is actually false in general: since the readers can still see changes to the structure by the writer, it's possible for their executions to be non-linearizable; if they're doing something like least element greater than foo in a binary tree, concurrent writer modifications can yield lookup results you wouldn't see in a serialized execution.

\mypar{Model Checking} Kokologiannakis et al.~\cite{Kokologiannakis:2017:SMC:3092282.3092287} use model-checking to test the core of \textsf{Tree RCU} in Linux kernel. Liang et al.~\cite{LiangMKM16} use model-checking to verify the \emph{grace period} guarantee of \textsf{Tree RCU}. Both focus on validating a particular RCU implementation, whereas we focus on verifying memory safety of clients independent of implementation. Desnoyers et al.~\cite{urcu_ieee} use the \textsf{SPIN} model checker to verify a user-mode implementation of RCU and this requires manual translation from C to SPIN modeling language.
In addition to being implementation-specific, they require test harness code, validating its behavior rather than real client code.

\mypar{Type Systems}
Howard et al.~\cite{Howard:2011:RES:2001252.2001267,Cooper2015RelativisticPI} present a \textsf{Haskell} library called \emph{Monadic RP} which provides types and relativistic programming constructs for write/read critical sections which enforce correct usage of relativistic programming pattern. They also have only checked a linked list. They claim  handling trees (look-up followed by update) as a future work~\cite{Howard:2011:RES:2001252.2001267}. Thus our work is the first type system for ensuring correct use of RCU primitives that is known to handle more complex structures than linked lists.
\section{Conclusions}
\label{sec:concls}
We presented the first type system that ensures code uses RCU memory management safely, and which is significantly simpler than full-blown verification logics. To this end, we gave the first general operational model for RCU-based memory management. Based on our suitable abstractions for RCU in the operational semantics we are the first showing that decoupling the \textit{memory-safety} proofs of RCU clients from the underlying reclamation model is possible. Meyer et al.~\cite{myr} took similar approach for decoupling the \textit{correctness} verification of the data structures from the underlying reclamation model under the assumption of the \textit{memory-safety} for the data structures. We demonstrated the applicability/reusability of our types on two examples: a linked-list based bag~\cite{McKenney2015SomeEO} and a binary search tree~\cite{Arbel:2014:CUR:2611462.2611471}. To our best knowledge, we are the first presenting the \textit{memory-safety} proof for a tree client of RCU. We managed to prove type soundness by embedding the type system into an abstract concurrent separation logic called the Views Framework~\cite{views} and encode many RCU properties as either type-denotations or global invariants over abstract RCU state. By doing this, we managed to discharge these invariants once as a part of soundness proof and did not need to prove them for each different client.
\paragraph{}
\mypar{Acknowledgements.} We are grateful to Matthew Parkinson for guidance and productive discussions on the early phase of this project. We also thank to Nik Sultana and Klaus V. Gleissenthall for their helpful comments and suggestions for improving the paper. 
\raggedbottom
\bibliographystyle{splncs04}
\bibliography{references/paper}
\newpage
\appendix
%!TEX root = ../paper.tex
\section{Complete Soundness Proof of Atoms and Structural Program Statements}
\subsection{Complete Constructions for \textsf{Views}}
\label{sec:prooflemmas}
To prove soundness we use the Views Framework~\cite{views}.
The Views Framework takes a set of parameters satisfying some properties, and produces a soundness proof for a static reasoning system for a larger programming language.  Among other parameters, the most notable are the choice of machine state, semantics for \emph{atomic} actions (e.g., field writes, or \lstinline|WriteBegin|), and proofs that the reasoning (in our case, type rules) for the atomic actions are sound (in a way chosen by the framework).
The other critical pieces are a choice for a partial \emph{view} of machine states --- usually an extended machine state with meta-information --- and a relation constraining how other parts of the program can interfere with a view (e.g., modifying a value in the heap, but not changing its type).
Our type system will be related to the views by giving a denotation of type environments in terms of views, and then proving that for each atomic action shown in \ref{fig:operationalsemrcu} in Section \ref{sec:semantics} and type rule in Figures \ref{fig:tss} Section \ref{subsection:type-action} and \ref{fig:tssr} Appendix \ref{appendix:readtypes}, given a view in the denotation of the initial type environment of the rule, running the semantics for that action yields a local view in the denotation of the output type environment of the rule. The following works through this in more detail. We define logical states, $\textsf{LState}$ to be
\begin{itemize}
\item A machine state, $\sigma=(s,h,l,rt,R,B)$;
\item An observation map, O, of type $ \textsf{Loc} \to \mathcal{P}(\textsf{obs})$
\item Undefined variable map, $U$, of type $\mathcal{P}(\textsf{Var}\times \textsf{TID})$
\item Set of threads, $T$, of type $\mathcal{P}(\textsf{TIDS})$
\item A to-free map(or free list), $F$, of type $\textsf{Loc} \rightharpoonup \mathcal{P}(\textsf{TID})$
\end{itemize}
The free map $F$ tracks which reader threads may hold references to each location. It is not required for execution of code, and for validating an implementation could be ignored, but we use it later with our type system to help prove that memory deallocation is safe.

Each memory region can be observed in one of the following type states within a snapshot taken at any time
\[\textsf{obs} := \texttt{iterator} \; \mathrm{tid} \mid \texttt{unlinked} \mid \texttt{fresh} \mid \texttt{freeable} \mid \texttt{root}\]

We are interested in \textsf{RCU} typed of heap domain which we define as:
\[\textsf{RCU} = \{ o \mid \textsf{ftype}(f) = \textsf{RCU} \land \exists o' \ldotp h(o',f) = o \}\]

A thread's (or scope's) \emph{view} of memory is a subset of the instrumented(logical states), which satisfy certain well-formedness criteria relating the physical state and the additional meta-data ($O$, $U$, $T$ and $F$)
\[\mathcal{M} \stackrel{def}{=} \{ m \in (\textsf{MState} \times O \times U \times T \times F) \mid  \textsf{WellFormed}(m) \} \]

We do our reasoning for soundness over instrumented states and define an erasure relation
\[\lfloor - \rfloor :\mathsf{MState} \implies \textsf{LState}\]

that projects instrumented states to the common components with \textsf{MState}.
\begin{figure}\scriptsize
\[
\begin{array}{l@{\;\;=\;\;}l}
 \llbracket \, x : \textsf{rcuItr}\,\rho\,\N \,  \rrbracket_{tid}
&
\left\{
\begin{array}{l|l}
m\in \mathcal{M}
&(\textsf{iterator} \, tid\in  O(s(x,tid)))  \land (x \notin U)  \\
& \land (\forall_{f_i\in dom(\N)  x_i\in codom(\N) } \ldotp
\left\{\begin{array}{l}  s(x_i,tid) = h(s(x,tid), f_i)  \\
 \land \textsf{iterator}\in O(s(x_i,tid)))\end{array} \right.\\
& \land  (\forall_{\rho', \rho''}\ldotp \rho'.\rho'' = \rho \implies  \textsf{iterator}\,tid \in O(h^{*}(rt,\rho'))) \\
& \land  h^{*}(rt,\rho)= s(x,tid)  \land (l = tid \land s(x,\_) \notin dom(F))) 
\end{array}
\right\}
\\
 \llbracket \, x : \textsf{rcuItr}\,  \rrbracket_{tid}
&
\left\{
\begin{array}{l|l}
m\in \mathcal{M}
&(\textsf{iterator} \, tid\in  O(s(x,tid)))  \land (x \notin U) \land \\
& (tid \in B) \implies \left\{\begin{array}{l}( \exists_{T'\subseteq B}\ldotp \{s(x,tid) \mapsto T'\} \cap F \neq \emptyset) \land \\ \land (tid \in T') \end{array}\right.
\end{array}
\right\}
\\
\llbracket \, x : \textsf{unlinked} \, \rrbracket_{tid}
&
\left\{
\begin{array}{l|l}
m\in \mathcal{M}
&(\textsf{unlinked}\in  O(.s(x,tid)) \land l = tid \land x \notin U) \land \\
& (\exists_{T'\subseteq T}\ldotp s(x,tid) \mapsto T'\in F \implies T' \subseteq B \land tid \notin T' )
\end{array}
\right\}
\\
\llbracket \, x : \textsf{freeable} \, \rrbracket_{tid}
&
\left\{
\begin{array}{l|l}
m\in \mathcal{M}
&\begin{array}{l}\textsf{freeable}\in  O(s(x,tid)) \land l = tid \land x \notin U \land \\
 s(x,tid) \mapsto \{\emptyset\} \in F \end{array}
\end{array}
\right\}
\\
\llbracket \, x : \textsf{rcuFresh} \, \N \, \,  \rrbracket_{tid}
&
\left\{
\begin{array}{l|l}
m\in \mathcal{M}
&(\textsf{fresh}\in  O(s(x,tid)) \land x \notin U  \land s(x,tid) \notin dom(F))\\
&(\forall_{f_i\in dom(\N), x_i\in codom(\N) } \ldotp s(x_i,tid) = h(s(x,tid), f_i) \\
&\land \textsf{iterator}\,tid \in O(s(x_i,tid)) \land s(x_i,tid) \notin dom(F)) 
\end{array}
\right\}
\\
\llbracket  x : \textsf{undef}\rrbracket_{tid} 
&
\left\{
\begin{array}{l|l}
m\in \mathcal{M}
&
(x,tid) \in U \land s(x,tid) \notin dom(F)
\end{array}
\right\}
\\
\llbracket \, x : \textsf{rcuRoot}\rrbracket_{tid}
&
\left\{
\begin{array}{l|l}
m\in \mathcal{M}
& \begin{array}{l}((rt \notin U \land s(x,tid) = rt \land rt \in dom(h) \land \\ 
O(rt) \in \textsf{root} \land s(x,tid) \notin dom(F) ) \end{array}
\end{array}
\right\}
\end{array}
\]
$
\textrm{provided}~h^{*}: (\textsf{Loc} \times \textsf{Path}) \rightharpoonup \textsf{Val}
$
\caption{Type Environments}
\label{fig:denotingtypeenviromentap}
\vspace{-2mm}
\end{figure}

Every type environment represents a set of possible views (well-formed logical states) consistent with the types in the environment.  We make this precise with a denotation function
\[\llbracket-\rrbracket\_ : \mathsf{TypeEnv}\rightarrow\mathsf{TID}\rightarrow\mathcal{P}(\mathcal{M})\]
that yields the set of states corresponding to a given type environment. This is defined in terms of denotation of individual variable assertions
\[\llbracket-:-\rrbracket_- : \mathsf{Var}\rightarrow\mathsf{Type}\rightarrow\mathsf{TID}\rightarrow\mathcal{P}(\mathcal{M})\]

The latter is given in Figure \ref{fig:denotingtypeenviromentap}.  To define the former, we first need to state what it means to combine logical machine states.

Composition of instrumented states is an operation
\[\bullet : \mathcal{M} \longrightarrow \mathcal{M} \longrightarrow \mathcal{M}\]
that is commutative and associative, and defined component-wise in terms of composing physical states, observation maps, undefined sets, and thread sets as shown in Figure \ref{fig:compap}
\begin{figure}\scriptsize
$  \bullet  = (\bullet_{\sigma},\bullet_O,\cup,\cup)$ \quad
  $O_{1} \bullet_O O_{2}(loc) \overset{\mathrm{def}}{=}  O_{1}(loc) \cup O_{2}(loc)$\quad
  $ (s_1 \bullet_s s_2) \overset{\mathrm{def}}{=}  s_1 \cup s_2 \texttt{   when   } dom(s_1) \cap dom(s_2) = \emptyset$ \\
  $(F_1 \bullet_F F_2) \overset{\mathrm{def}}{=}  F_1  \cup F_2 \texttt{   when   } dom(F_1) \cap dom(F_2) = \emptyset$ \\
$(h_1\bullet_h h_2)(o,f)\overset{\mathrm{def}}{=}\left\{
\begin{array}{ll}
\mathrm{undef} & \textrm{if}~h_1(o,f)=v \land h_2(o,f)=v' \land v' \neq v\\
v & \textrm{if}~h_1(o,f)=v \land h_2(o,f)=v\\
v & \textrm{if}~h_1(o,f)=\mathrm{undef}\land h_2(o,f)=v\\
v & \textrm{if}~h_1(o,f)=v\land h_2(o,f)=\mathrm{undef}\\
\mathrm{undef} & \textrm{if}~h_1(o,f)=\mathrm{undef}\land h_2(o,f)=\mathrm{undef}
\end{array}
\right.
$
$
\begin{array}{l}
((s,h,l,rt,R,B), O, U, T,F) \mathcal{R}_{0}((s',h',l',rt',R',B'), O', U', T',F') \overset{\mathrm{def}}{=}
\\ \bigwedge\left\{
	\begin{array}{l}
	  l  \in  T \rightarrow (h = h' \land l=l')\\
	  l\in T\rightarrow F=F'\\
	  \forall tid,o\ldotp\textsf{iterator} \, tid \in O(o) \rightarrow o \in dom(h) \\
	  \forall tid,o\ldotp\textsf{iterator} \, tid \in O(o) \rightarrow o \in dom(h') \\
          \forall tid,o\ldotp\textsf{root} \, tid \in O(o) \rightarrow o \in dom(h) \\
	  \forall tid,o\ldotp\textsf{root} \, tid \in O(o) \rightarrow o \in dom(h') \\
	  O = O' \land U = U' \land T = T'\land R =R'\land rt = rt' \\
	  \forall x, t \in T \ldotp s(x,t) = s'(x,t)
	\end{array}
\right\}
\end{array}
$
\caption{Composition($\bullet$) and Thread Interference Relation($\mathcal{R}_{0}$)}
\label{fig:compap}
\vspace{-2mm}
\end{figure}
%\emph{Separation algebra} is a model to define and axiomatize the \emph{join/composition} operation over a domain which is \emph{set of instrumented states} in our case.
An important property of composition is that it preserves validity of logical states:
\begin{lemma}[Well Formed Composition]
\label{lem:wf-compositionap}
Any successful composition of two well-formed logical states is well-formed:\[\forall_{x,y,z}\ldotp \mathsf{WellFormed}(x) \implies \mathsf{WellFormed}(y) \implies x\bullet y = z \implies \mathsf{WellFormed(z)}\]
\end{lemma}
\begin{proof}
By assumption, we know that \textsf{Wellformed}(x) and \textsf{Wellformed}(y) hold. We need to show that composition of two well-formed states preserves well-formedness which is to show that for all $z$ such that $x\bullet y = z$, \textsf{Wellformed}(z) holds.
  Both $x$ and $y$ have components $((s_x,h_x,l_x,rt_x,R_x,B_x),O_x,U_x,T_x,F_x)$ and $((s_y,h_y,l_y,rt_y,R_y,B_y),O_y,U_y,T_y,F_y)$, respectively. $\bullet_s$ operator over stacks $s_x$ and $s_y$ enforces $dom(s_x) \cap dom(s_y) = \emptyset$ which enables to make sure that wellformed mappings in $s_x$ does not violate wellformed mappings in $s_y$ when we union these mappings for $s_z$. Same argument applies for $\bullet_F$ operator over $F_x$ and $F_y$. Disjoint unions of wellformed $R_x$ with wellformed $R_y$ and wellformed $B_x$ with wellformed $B_y$ preserves wellformedness in composition as it is disjoint union of different wellformed elements of sets. Wellformed unions of $O_x$ with $O_y$,  $U_x$ with $U_y$  and $T_x$ with $T_y$ preserve wellformedness. When we compose $h_x(s(x,tid),f)$ and $h_y(s(x,l),f)$, it is easy to show that we preserve wellformedness if both threads agree on the heap location. Otherwise, if the heap location is undefined for one thread but a value for the other thread then composition considers the value. If a heap location is undefined for both threads then this heap location is also undefined for the location. All the cases for heap composition still preserves the wellformedness from the assumption that $x$ and $y$ are wellformed. 
  \end{proof}
We define separation on elements of type contexts
\begin{itemize}
\item For read-side as $\llbracket x_1 : T_1, \ldots x_n : T_n \rrbracket_{tid,\textsf{R}} = \llbracket x_1 : T_1 \rrbracket_{tid} \cap \ldots \cap \llbracket x_n : T_n \rrbracket_{tid} \cap \llbracket \textsf{R} \rrbracket_{tid}$ where
$\llbracket \textsf{R} \rrbracket_{tid} = \{ (s,h,l,rt,R,B),O,U,T,F  \mid tid \in R \}$

\item For write-side as $\llbracket x_1 : T_1, \ldots x_n : T_n \rrbracket_{tid,\textsf{M}} = \llbracket x_1 : T_1 \rrbracket_{tid} \cap \ldots \cap \llbracket x_n : T_n \rrbracket_{tid} \cap \llbracket \textsf{M} \rrbracket_{tid}$ where
$\llbracket \textsf{M} \rrbracket_{tid} = \{ (s,h,l,rt,R,B),O,U,T,F  \mid tid = l \}$

\item $\llbracket x_1 : T_1, \ldots x_n : T_n \rrbracket_{tid,\textsf{O}} = \llbracket x_1 : T_1 \rrbracket_{tid} \cap \ldots \cap \llbracket x_n : T_n \rrbracket_{tid} \cap \llbracket \textsf{O} \rrbracket_{tid}$ where
$\llbracket \textsf{O} \rrbracket_{tid} = \{ (s,h,l,rt,R,B),O,U,T,F  \mid tid \neq l \land tid \notin R \}$.
\end{itemize}

Partial separating conjunction then simply requires the existence of two states that compose:
\[ m \in P \ast Q   \stackrel{def}{=} \exists m' \ldotp \exists m'' \ldotp m' \in P \land m'' \in Q \land m \in m' \bullet m''\]
Different threads' views of the state may overlap (e.g., on shared heap locations, or the reader thread set), but one thread may modify that shared state.  The Views Framework restricts its reasoning to subsets of the logical views that are \emph{stable} with respect to expected interference from other threads or contexts.  We define the interference as (the transitive reflexive closure of) a binary relation $\mathcal{R}$ on $\mathcal{M}$, and a \textsf{View} in the formal framework is then:
\[\textsf{View}_{\mathcal{M}} \stackrel{def}{=} \{ M \in \mathcal{P}(\mathcal{M}) | \mathcal{R}(M) \subseteq M\}\]
Thread interference relation
\[\mathcal{R} \subseteq \mathcal{M} \times \mathcal{M}\]  defines permissible interference on an instrumented state. The relation must distribute over composition:
\[ \forall m_{1}, m_{2}, m\ldotp (m_{1} \bullet  m_{2})\mathcal{R}m \Longrightarrow \begin{array}{ll}  \exists  m'_{1} m'_{2} \ldotp m_{1} \mathcal{R} m'_{1} \land m_{2} \mathcal{R} m'_{2} \land  m \in m'_{1} \bullet m'_{2} \end{array}\]
where $\mathcal{R}$ is transitive-reflexive closure of $\mathcal{R}_{0}$ shown at Figure \ref{fig:compap}. $\mathcal{R}_0$ (and therefore $\mathcal{R}$) also ``preserves'' validity:
\begin{lemma}[Valid $\mathcal{R}_0$ Interference]\label{lem:interap}
For any $m$ and $m'$, if $\mathsf{WellFormed}(m)$ and $m\mathcal{R}_0m'$, then $\mathsf{WellFormed}(m')$.
\end{lemma}
\begin{proof}
  By assumption, we know that $m = (s,h,l,rt,R,B),O,U,T,F)$ is wellformed. We also know that $m'= (s',h',l',rt',R',B'),O',U',T',F')$ is related to $m$ via $R_{0}$. By assumptions in $R_{0}$ and semantics, we know that $O$,$R$,$T$ and $U$ which means that these components do not have any effect on wellformedness of the $m$. In addition, change on stack, $s$, does not affect the wellformedness as
\[\forall x,t \in T \ldotp s(x,t) = s'(x,t) \]
Moreover, from semantics we know that $l$ and $h$ can only be changed by writer thread and from $R_0$
\[l  \in  T \rightarrow (h = h' \land l=l')\]
\[  l\in T\rightarrow F=F'\]
and by assumptions from the lemma($\mathsf{WellFormed}(m)$.\textbf{RINFL}) we can conclude that $F$,$l$ and $h$ do not have effect on wellformedness of the $m$.
\end{proof}
\begin{lemma}[Stable Environment Denotation-M]\label{lemma:stblwap}
For any \emph{closed} environment $\Gamma$ (i.e., $\forall x\in\mathsf{dom}(\Gamma)\ldotp, \mathsf{FV}(\Gamma(x))\subseteq\mathsf{dom}(\Gamma)$):
\[
\mathcal{R}(\llbracket\Gamma\rrbracket_{\mathsf{M},tid})\subseteq\llbracket\Gamma\rrbracket_{\mathsf{M},tid}
\]
Alternatively, we say that environment denotation is \emph{stable} (closed under $\mathcal{R}$).
\end{lemma}
\begin{proof}
  By induction on the structure of $\Gamma$.  The empty case holds trivially.  In the other case where $\Gamma=\Gamma',x:T$, we have by the inductive hypothesis that
  \[\llbracket\Gamma'\rrbracket_{\mathsf{M},tid}\] is stable, and must show that
  \[\llbracket\Gamma'\rrbracket_{\mathsf{M},tid}\cap\llbracket{x:\tau}\rrbracket_{tid}\] is as well.  This latter case proceeds by case analysis on $T$.

We know that $O$, $U$, $T$, $R$, $s$ and $rt$ are preserved by $R_0$. By unfolding the type environment in the assumption we know that $tid = l$. So we can derive conclusion for preservation of $F$ and $h$ and $l$ by
\[l  \in  T \rightarrow (h = h' \land l=l')\]
\[  l\in T\rightarrow F=F'\]
Cases in which denotations, $\llbracket x:T \rrbracket$, touching these \emph{R$_0$ preserved} maps are trivial to show.
  \begin{case} - \textsf{unlinked}, \textsf{undef}, $\textsf{rcuFresh}\, \N$ and \textsf{freeable} trivial.
\begin{case} - $\textsf{rcuItr}\,\rho\,\N$: All the facts we know so far from $R_0$, $tid=l$ and additional fact we know from $R_0$:
  \[\forall tid,o\ldotp\textsf{iterator} \, tid \in O(o) \rightarrow o \in dom(h) \]
  \[\forall tid,o\ldotp\textsf{iterator} \, tid \in O(o) \rightarrow o \in dom(h')\]
  prove this case.
  \end{case}
  \end{case}
  \begin{case} - \textsf{root}: All the facts we know so far from $R_0$, $tid=l$ and additional fact we know from $R_0$:
    \[ \forall tid,o\ldotp\textsf{root} \, tid \in O(o) \rightarrow o \in dom(h) \]
    \[ \forall tid,o\ldotp\textsf{root} \, tid \in O(o) \rightarrow o \in dom(h') \]
    prove this case.
    \end{case}
\end{proof}
\begin{lemma}[Stable Environment Denotation-R]
  \label{lem:stblRap}
For any \emph{closed} environment $\Gamma$ (i.e., $\forall x\in\mathsf{dom}(\Gamma)\ldotp, \mathsf{FV}(\Gamma(x))\subseteq\mathsf{dom}(\Gamma)$):
\[
\mathcal{R}(\llbracket\Gamma\rrbracket_{\mathsf{R},tid})\subseteq\llbracket\Gamma\rrbracket_{\mathsf{R},tid}
\]
Alternatively, we say that environment denotation is \emph{stable} (closed under $\mathcal{R}$).
\end{lemma}
\begin{proof}
Proof is similar to one for Lemma \ref{lemma:stblwap} where there is only one simple case, $\llbracket x:\textsf{rcuItr} \rrbracket$. 
\end{proof}

The Views Framework defines a program logic (Hoare logic) with judgments of the form $\{ p \} C \{ q \}$ for views p and q and commands $C$. Commands include atomic actions, and soundness of such judgments for atomic actions is a parameter to the framework. The framework itself provides for soundness of rules for sequencing, fork-join parallelism, and other general rules.
To prove type soundness for our system, we define a denotation of \emph{type judgments} in terms of the Views logic, and show that every valid typing derivation translates to a valid derivation in the Views logic:
\[\forall\Gamma,C,\Gamma',\mathit{tid}\ldotp\Gamma\vdash_{M,R} C\dashv \Gamma' \Rightarrow \{\llbracket\Gamma\rrbracket_\mathit{tid}\} \llbracket C\rrbracket_\mathit{tid}\{\llbracket\Gamma'\rrbracket_\mathit{tid}\}\]
The antecedent of the implication is a type judgment(shown in Figure \ref{fig:tss} Section \ref{subsection:type-rules}, Figure \ref{fig:type-judgements-for-cf} Section \ref{subsection:typwrt} and Figure \ref{fig:tssr} Appendix \ref{appendix:readtypes}) and the conclusion is a judgment in the Views logic. The environments are translated to views ($\mathsf{View}_\mathcal{M}$) as previously described. Commands $C$ also require translation, because the Views logic is defined for a language with non-deterministic branches and loops, so the standard versions from our core language must be encoded.  The approach to this is based on a standard idea in verification, which we show here for conditionals as shown in Figure \ref{fig:asmap}. $\textsf{assume}(b)$ is a standard construct in verification semantics~\cite{Barnett:2005:BMR:2090458.2090481} ~\cite{Muller:2016:VVI:2963187.2963190}, which ``does nothing'' (freezes) if the condition $b$ is false, so its postcondition in the Views logic can reflect the truth of $b$.  This is also the approach used in previous applications of the Views Framework~\cite{oopsla12,toplas17}.
\begin{figure}\scriptsize
$
\begin{array}{l}
\llbracket{\mathsf{if}\;(x.f==y)\;C_1\;C_2}\rrbracket_\mathit{tid} \overset{\mathrm{def}}{=} z=x.f;((\mathsf{assume}(z=y);C_1)+(\mathsf{assume}(z\neq y);C_2))
\end{array}
$
$
\llbracket\texttt{assume}(\mathcal{S})\rrbracket (s)\overset{\mathrm{def}}{=}\left\{
\begin{array}{ll}
\{ s\} & \textrm{if}~s \in \mathcal{S}\\
\emptyset & \textrm{Otherwise}
\end{array}
\right.
$
$
\llbracket{\mathsf{while}\;(e)\;C}\rrbracket \overset{\mathrm{def}}{=} \left(\mathsf{assume}(e);C\right)^{*};(\mathsf{assume}(\lnot e ));
$
$
\inferrule
{
\{P\} \cap \{\lceil\mathcal{S} \rceil \}  \sqsubseteq \{Q\}
}
{
 \{P\} \texttt{assume}\left(b\right)\{Q\}
}
$
\textsf{ where } $\lceil \mathcal{S} \rceil = \{m | \lfloor m \rfloor \cap \mathcal{S} \neq \emptyset \}$
\caption{Encoding of \textsf{assume}(b)}
\label{fig:asmap}
\end{figure}

The framework also describes a useful concept called the view shift operator $\subseteq$, that describes a way to reinterpret a set of instrumented states as a new set of instrumented states. This operator enables us to define an abstract notion of executing a small step of the program. We express the step from $p$ to $q$ with action $\alpha$ ensuring that the operation interpretation of the action satisfies the specification:$ p \sqsubseteq q \stackrel{def}{=} \forall m\in\mathcal{M} \ldotp \lfloor p * \{ m \} \rfloor \subseteq \lfloor q* \mathcal{R}(\{m\})\rfloor$. Because the Views framework handles soundness for the structural rules (sequencing, parallel composition, etc.), there are really only three types of proof obligations for us to prove.  First, we must prove that the non-trivial command translations (i.e., for conditionals and while loops) embed correctly in the Views logic, which is straightforward.  Second, we must show that for our environment subtyping, if $\Gamma<:\Gamma'$, then $\llbracket{\Gamma}\rrbracket\sqsubseteq\llbracket\Gamma'\rrbracket$.  And finally, we must prove that each atomic action's type rule corresponds to a valid semantic judgment in the Views Framework:
\[
\forall m\ldotp   \llbracket \alpha \rrbracket ( \lfloor \llbracket \Gamma_{1} \rrbracket_{tid}  * \{m\} \rfloor) \subseteq
  \lfloor \llbracket \Gamma_{2} \rrbracket_{tid} * \mathcal{R}(\{m\}) \rfloor
  \]
The use of $*$ validates the frame rule and makes this obligation akin to an interference-tolerant version of the small footprint property from traditional separation logics~\cite{Reynolds:2002:SLL:645683.664578,Calcagno:2007:LAA:1270399.1271718}.
\begin{lemma}[Axiom of Soundness for Atoms]
  \label{lem:axmsoundnessap}
For each axiom, $\Gamma_{1} \vdash_{\textsf{RMO}} \alpha \dashv \Gamma_{2}$, we must show
\[
\forall m\ldotp   \llbracket \alpha \rrbracket  (\lfloor \llbracket \Gamma_{1} \rrbracket_{tid}  * \{m\} \rfloor )\subseteq  \lfloor \llbracket \Gamma_{2} \rrbracket_{tid} * \mathcal{R}(\{m\}) \rfloor
\]
\end{lemma}
\begin{proof}
By case analysis on the atomic action $\alpha$ followed by inversion on typing derivation. All the cases proved as different lemmas in Section \ref{lem:lematom}.
\end{proof}

Type soundness proceeds according to the requirements of the Views Framework, primarily embedding each type judgment into the Views logic:
\begin{lemma}[Views Embedding for Read-Side]
  \label{lemma:embedrap}
\[  \forall\Gamma,C,\Gamma',\mathit{t}\ldotp\Gamma\vdash_R C\dashv \Gamma' \Rightarrow
\llbracket\Gamma\rrbracket_\mathit{t}\cap\llbracket{R}\rrbracket_t\vdash \llbracket C\rrbracket_\mathit{t}\dashv\llbracket\Gamma'\rrbracket_\mathit{t}\cap\llbracket{R}\rrbracket_t
\]
\end{lemma}
\begin{proof}
  Proof is similar to the one for Lemma \ref{lemma:embedwap} except the denotation for type system definition is $\llbracket R \rrbracket_t = \{\{((s,h,l,rt,R,B),O,U,T,F)| t \in R \}$ which shrinks down the set of all logical states to the one that can only be defined by types($\textsf{rcuItr}$) in read type system. 
  \end{proof}
\begin{lemma}[Views Embedding for Write-Side]
  \label{lemma:embedwap}
  \[
\forall\Gamma,C,\Gamma',\mathit{t}\ldotp\Gamma\vdash_M C\dashv \Gamma' \Rightarrow
\llbracket\Gamma\rrbracket_\mathit{t}\cap\llbracket{M}\rrbracket_t\vdash \llbracket C\rrbracket_\mathit{t}\dashv\llbracket\Gamma'\rrbracket_\mathit{t}\cap\llbracket{M}\rrbracket_t
\]
\end{lemma}
\begin{proof}
  Induction on derivation of $\Gamma \vdash_M C \dashv \Gamma'$ and then inducting on the type of first element of the environment. For the nonempty case, $\Gamma'',x:T$ we do case analysis on $T$. Type environment for write-side actions includes only: $\textsf{rcuItr} \, \rho \, \N$, $\textsf{undef}$, $\textsf{rcuFresh}$, $\textsf{unlinked}$ and $\textsf{freeable}$. Denotations of these types include the constraint $t=l$ and other constraints specific to the type's denotation. The set of logical state defined by the denotation of the type is \emph{subset} of intersection of the set of logical states defined by $\llbracket M \rrbracket_{t} \cap \llbracket x:T \rrbracket_{t}$ which shrinks down the logical states defined by  $\llbracket M \rrbracket_{t}= \{((s,h,l,rt,R,B),O,U,T,F)| t = l\}$ to the set of logical states defined by denotation $\llbracket x:T \rrbracket_t$.
  \end{proof}
Because the intersection of the environment denotation with the denotations for the different critical sections remains a valid view, the Views Framework provides most of this proof for free, given corresponding lemmas for the \emph{atomic actions} $\alpha$:
\[\begin{array}{l}
\forall \alpha,\Gamma_1,\Gamma_2\ldotp \Gamma_1\vdash_R\alpha\dashv\Gamma_2 \Rightarrow
\\
\quad\forall m\ldotp   \llbracket \alpha \rrbracket  (\lfloor \llbracket \Gamma_{1} \rrbracket_{\textsf{R},tid}  * \{m\} \rfloor )\subseteq  \lfloor \llbracket \Gamma_{2} \rrbracket_{\textsf{R},tid} * \mathcal{R}(\{m\}) \rfloor
\end{array}\]
\[
\begin{array}{l}
\forall \alpha,\Gamma_1,\Gamma_2\ldotp \Gamma_1\vdash_M\alpha\dashv\Gamma_2 \Rightarrow
\\
\quad\forall m\ldotp   \llbracket \alpha \rrbracket  (\lfloor \llbracket \Gamma_{1} \rrbracket_{\textsf{M},tid}  * \{m\} \rfloor )\subseteq  \lfloor \llbracket \Gamma_{2} \rrbracket_{\textsf{M},tid} * \mathcal{R}(\{m\}) \rfloor
\end{array}
\]
% \[
% \forall\Gamma,\alpha,\Gamma',\mathit{t}\ldotp\Gamma\vdash_R \alpha\dashv \Gamma' \Rightarrow
%     \llbracket\Gamma\rrbracket_\mathit{t}\cap\llbracket{R}\rrbracket_t\vdash \llbracket \alpha\rrbracket_\mathit{t}\dashv\llbracket\Gamma'\rrbracket_\mathit{t}\cap\llbracket{R}\rrbracket_t
% \]
% \[
% \forall\Gamma,\alpha,\Gamma',\mathit{t}\ldotp\Gamma\vdash_M \alpha\dashv \Gamma' \Rightarrow
%     \llbracket\Gamma\rrbracket_\mathit{t}\cap\llbracket{M}\rrbracket_t\vdash \llbracket \alpha\rrbracket_\mathit{t}\dashv\llbracket\Gamma'\rrbracket_\mathit{t}\cap\llbracket{M}\rrbracket_t
% \]
$\alpha$ ranges over any atomic command, such as a field access or variable assignment.

%%%A number of rules in the formal system (Section \todo{ref}) are present in \emph{both} systems, and we would like to reuse their proofs.  To support this, we prove the following lemma, which permits us to ignore which critical section such rules are considered in, and lift that ``ignorant'' proof to either critical section:
Denoting a type environment $\llbracket\Gamma\rrbracket_{\mathsf{M},tid}$, unfolding the definition one step, is merely $\llbracket\Gamma\rrbracket_{tid}\cap\llbracket\mathsf{M}\rrbracket_{tid}$.  In the type system for write-side critical sections, this introduces extra boilerplate reasoning to prove that each action preserves lock ownership.  To simplify later cases of the proof, we first prove this convenient lemma.
\begin{lemma}[Write-Side Critical Section Lifting]
\label{lem:crit-liftingap}
For each $\alpha$ whose semantics does not affect the write lock, if
\[\forall m\ldotp   \llbracket \alpha \rrbracket  (\lfloor \llbracket \Gamma_{1} \rrbracket_{tid}  * \{m\} \rfloor )\subseteq  \lfloor \llbracket \Gamma_{2} \rrbracket_{tid} * \mathcal{R}(\{m\}) \rfloor\]
then
%\[\forall m\ldotp   \llbracket \alpha \rrbracket  (\lfloor \llbracket \Gamma_{1} \rrbracket_{\textsf{R},tid}  * \{m\} \rfloor )\subseteq  \lfloor \llbracket \Gamma_{2} \rrbracket_{\textsf{R},tid} * \mathcal{R}(\{m\}) \rfloor\]
\[\forall m\ldotp   \llbracket \alpha \rrbracket  (\lfloor \llbracket \Gamma_{1} \rrbracket_{\textsf{M},tid}  * \{m\} \rfloor )\subseteq  \lfloor \llbracket \Gamma_{2} \rrbracket_{\textsf{M},tid} * \mathcal{R}(\{m\}) \rfloor\]
\end{lemma}
\begin{proof}
Each of these shared actions $\alpha$ preserves the lock  component of the physical state, the only component constrained by $\llbracket-\rrbracket_{M,tid}$ beyond $\llbracket-\rrbracket_{tid}$.
For the read case, we must prove from the assumed subset relationship that for an aritrary $m$:
\[\llbracket \alpha \rrbracket  (\lfloor \llbracket \Gamma_{1} \rrbracket_{tid}\cap\llbracket\textsf{M}\rrbracket_{tid}  * \{m\} \rfloor )\subseteq  \lfloor \llbracket \Gamma_{2} \rrbracket_{tid}\cap\llbracket\textsf{M}\rrbracket_{tid} * \mathcal{R}(\{m\}) \rfloor\]
By assumption, transitivity of $\subseteq$, and the semantics for the possible $\alpha$s,
the left side of this containment is already a subset of
\[\lfloor \llbracket \Gamma_{2} \rrbracket_{tid} * \mathcal{R}(\{m\}) \rfloor\]
What remains is to show that the intersection with $\llbracket\mathsf{M}\rrbracket_{tid}$ is preserved by the atomic action.
This follows from the fact that none of the possible $\alpha$s modifies the global lock.
\end{proof}
\subsection{Complete Memory Axioms}
\label{sec:memaxioms}
\begin{enumerate}
\item{Ownership} invariant in Figure \ref{fig:ownership} invariant asserts that none of the heap nodes can be observed as undefined by any of those threads.
\begin{figure}[!htb]
\[
\textbf{OW}(\sigma,O,U,T,F) =
\left\{
\begin{array}{ll}
 	 & \forall_{o,o'f,f'} \ldotp  \sigma.h(o,f) = v \land \sigma.h(o',f') = v \\
	& \land v \in \textsf{OID}  \land  \textsf{FType}(f) = \textsf{RCU} \implies \\
 	&\left\{
		\begin{array}{cl}
			&  o=o' \land f=f' \\
			& \lor\textsf{unlinked} \in O(o) \\
		  & \lor \textsf{unlinked} \in O(o') \\
                  & \lor \textsf{freeable} \in O(o)\\
                  & \lor \textsf{freeable} \in O(o')\\
		        & \lor \textsf{fresh} \in O(o)) \\
                        & \lor \textsf{fresh} \in O(o')
		\end{array}
		\right.%\} 
\end{array}
\right.%\}
\]
\caption{Ownership}
\label{fig:ownership}
\end{figure}
\item{Reader-Writer-Iterators-CoExistence} invariant in Figure \ref{fig:rwitrexistence} asserts that  if a heap location is not undefined then all reader threads and the writer thread can observe the heap location as \textsf{iterator} or the writer thread can observe heap as \textsf{fresh}, \textsf{unlinked} or \textsf{freeable}.
\begin{figure}[!htb]
\[
\textbf{RWOW}(\sigma,O,U,T,F)=
\left\{
\begin{array}{ll}
	 \forall x,tid,o \ldotp \sigma.s(x,tid) = o  
         \land (x,tid) \notin U \implies \\ \left\{
		\begin{array}{ll}
 			& \textsf{iterator } \, tid \in O(o) \\
 		  & \lor (\sigma.l=tid \land (\textsf{unlinked} \in O(o) ) \\
		  & \lor (\sigma.l=tid  \land \textsf{freeable} \in O(o)))  \\
       & \lor (\sigma.l=tid \land \textsf{fresh} \in O(o))     
		\end{array}
	\right.%\}
\end{array}
\right.%\}
\]
\caption{Reader-Writer-Iterator-Coexistence-Ownership}
\label{fig:rwitrexistence}
\end{figure}
\item{Alias-With-Root} invariant in Figure \ref{fig:aliaswithroot} asserts that the unique root location can only be aliased with thread local references through which the unique root location is observed as \textsf{iterator}.
\begin{figure}[!htb]
\[
\textbf{AWRT}(\sigma,O,U,T,F)=
\left\{
(\forall_{y,tid} \ldotp h^{*}(\sigma.rt,\epsilon) = s(y,tid)  \implies \textsf{iterator}\, tid \in O(s(y,tid)))
	\right.%\}
\]
\caption{Alias with Unique Root}
\label{fig:aliaswithroot}
\end{figure}
\item{Iterators-Free-List} invariant in Figure \ref{fig:itrfreelist} asserts that if a heap location is observed as \textsf{iterator} and it is the free list then the observer thread is in the set of bounding threads.
\begin{figure}[!htb]
\[
\textbf{IFL}(\sigma,O,U,T,F)=
\left\{
	\forall tid, o \ldotp  \textsf{iterator} \, tid \in O(o)  \land \forall_{T'\subseteq T} \ldotp \sigma.F([o \mapsto T'])  \implies tid \in T' 
\right.%\}
\]
\caption{Iterators-Free-List}
\label{fig:itrfreelist}
\end{figure}
\item{Unlinked-Reachability} invariant in Figure \ref{fig:unlinkreach} asserts that if a heap node is observed as \textsf{unlinked} then all heap locations from which you can reach to the \textsf{unlinked} one are also unlinked or in the free list.
\begin{figure}[!htb]
\[
\textbf{ULKR}(\sigma,O,U,T,F) =
\left\{
\begin{array}{ll}
  \forall o \ldotp \textsf{unlinked} \in O(o) \implies \\
  \left\{
	\begin{array}{cl}
	  \forall o',f '\ldotp \sigma.h(o',f') = o \implies \\
          \left\{
			\begin{array}{cl}
				&\textsf{unlinked} \in O(o') \lor\\
				& \textsf{freeable} \in O(o')
			\end{array}
		\right.%\}
	\end{array}
	\right.%\}
\end{array}
\right.%\}
\]
\caption{Unlinked-Reachability}
\label{fig:unlinkreach}
\end{figure}
\item{Free-List-Reachability} invariant in Figure \ref{fig:freelistchain} asserts that if a heap location is in the free list then all heap locations from which you can reach to the one in the free list are also in the free list.
\begin{figure}[!htb]
\[
\textbf{FLR}(\sigma,O,U,T,F) =
\left\{
\begin{array}{ll}
  \forall o \ldotp F([o \mapsto T]) \implies \\
  \left\{
   		\begin{array}{cl}
		  \forall o',f' \ldotp \sigma.h(o',f') = o \implies \\
                \left\{
   \begin{array}{cl}
   	 		&\exists_{T'\subseteq T} \ldotp  F([o'\mapsto T']) 
   \end{array}
   \right.
                  \end{array}
 	\right.%\}
\end{array}
\right.%\}
\]
\caption{Free-List-Reachability}
\label{fig:freelistchain}
\end{figure}
\item{Writer-Unlink} invariant in Figure \ref{fig:mutunliked} asserts that the writer thread cannot observe a heap location as \textsf{unlinked}.
\begin{figure}[!htb]
\[
\textbf{WULK}(\sigma,O,U,T,F) =
\left\{
\begin{array}{ll}
\forall o \ldotp \textsf{iterator} \, \sigma.l \in O(o)  \implies \textsf{unlinked} \notin O(o)  \land \textsf{freeable} \notin O(o)  \land \textsf{undef} \notin O(o)
\end{array}
\right.%\}
\]
\caption{Writer-Unlink}
\label{fig:mutunliked}
\end{figure}
\item{Fresh-Reachable} invariant in Figure \ref{fig:freach} asserts that there exists no heap location that can reach to a freshly allocated heap location together with fact on nonexistence of aliases to it.
\begin{figure}[!htb]
  \[
  \textbf{FR}(\sigma,O,U,T,F) =
 \begin{array}{ll}
   \forall_{tid,x,o} \ldotp (\sigma.s(x,tid) = o \land \textsf{fresh} \in O(o) ) \implies \\
   \left\{
	\begin{array}{cl}
          (\forall_{y,o',f',tid'}. (h(o',f') \neq o) \lor  (s(y,tid) \neq  o \\
          \lor (tid'\neq tid \implies s(y,tid') \neq o ))
        \end{array}
        \right\}
        \end{array}
\]
\caption{Fresh-Reachable}
\label{fig:freach}
\end{figure}

\item{Fresh-Writer} invariant in Figure \ref{fig:fmut} asserts that heap allocation can be done only by writer thread.
  \begin{figure}[!htb]
  \[
\textbf{WF}(\sigma,O,U,T,F) =
 \forall_{tid,x,o} \ldotp (\sigma.s(x,tid) = o \land \textsf{fresh} \in O(o) ) \implies  tid = \sigma.l 
\]
\caption{Fresh-Writer}
\label{fig:fmut}
  \end{figure}
\item{Fresh-Not-Reader} invariant in Figure \ref{fig:fnotreader} asserts that a heap location allocated freshly cannot be observed as \textsf{unlinked} or \textsf{iterator}.
  \begin{figure}[!htb]
  \[
\textbf{FNR}(\sigma,O,U,T,F) =
 \forall_{o} \ldotp (\textsf{fresh} \in O(o) ) \implies  (\forall_{x,tid} \ldotp \textsf{iterator}\,tid  \notin O(o)  ) \land \textsf{unlinked} \notin O(o) 
\]
\caption{Fresh-Not-Reader}
\label{fig:fnotreader}
  \end{figure}
\item{Fresh-Points-Iterator} invariant in Figure \ref{fig:fsinglefield} states that any field of fresh allocated object can only be set to point heap node which can be observed as \textsf{iterator} (not \textsf{unlinked or freeable}).  This invariant captures the fact $\N = \N'$ asserted in the type rule for fresh node linking(\textsc{T-Replace}). 
  \begin{figure}[!htb]
  \[
\textbf{FPI}(\sigma,O,U,T,F) =
 \forall_{o} \ldotp (\textsf{fresh} \in O(o)  \land \exists_{f,o'}\ldotp h(o,f)=o') \implies (\forall_{tid} \ldotp  \textsf{iterator}\,tid \in O(o')) 
\]
\caption{Fresh-Points-Iterator}
\label{fig:fsinglefield}
  \end{figure}
   
\item{Writer-Not-Reader} invariant in Figure \ref{fig:wnr} asserts that a writer thread identifier can not be a reader thread identifier.
  \begin{figure}[!htb]
    \[
\textbf{WNR}(\sigma,O,U,T,F) =
\left\{
\begin{array}{ll}
 \sigma.l \notin \sigma.R
\end{array}
\right.%\}
\]
    \caption{Writer-Not-Reader}
\label{fig:wnr}
  \end{figure}
  \item{Readers-Iterator-Only} invariant in the Figure \ref{fig:riter} asserts that a reader threads can only make \textsf{iterator} observation on a heap location.
  \begin{figure}[!htb]
    \[
\textbf{RITR}(\sigma,O,U,T,F) =
\left\{
\begin{array}{ll}
 \forall_{tid \in \sigma.R,o}\ldotp \textsf{iterator} \, tid \in O(o) 
\end{array}
\right.%\}
\]
    \caption{Readers-Iterator-Only}
\label{fig:riter}
  \end{figure}
\item{Readers-In-Free-List} invariant in Figure \ref{fig:readerinflist} asserts that for any mapping from a location to a set of threads in the free list we know the fact that  this set of threads is a subset of bounding threads( which itself is subset of reader threads).
\begin{figure}[!htb]
  \[
\textbf{RINFL}(\sigma,O,U,T,F) =
\left\{
\begin{array}{ll}
  \forall_{o} \ldotp F([o \mapsto T]) \implies T \subseteq \sigma.B
\end{array}
\right.%\}
\]
\caption{Readers-In-Free-List}
\label{fig:readerinflist}
\end{figure}
\item{Heap-Domain} invariant in the Figure \ref{fig:dreachable} defines the domain of the heap.
\begin{figure}[!htb]
\[
\textbf{HD}(\sigma,O,U,T,F) =
\forall_{o,f',o'} \ldotp  \sigma.h(o,f) = o' \implies  o' \in dom(\sigma.h)
\]
    \caption{Heap-Domain}
\label{fig:dreachable}
\end{figure}
\item{Unique-Root} invariant in Figure \ref{fig:uroot} asserts that a heap location which is observed as \textsf{root} has no incoming edges from any nodes in the domain of the heap and all nodes accessible from root is is observed as \textsf{iterator}. This invariant is part of enforcement for \emph{acyclicity}.
  \begin{figure}[!htb]
\[
\textbf{UNQRT}(\sigma,O,U,T,F) =
  \left\{ \begin{array}{ll}
    \forall_{\rho \neq \epsilon} \ldotp \textsf{iterator} \, tid \in O( h^{*}(\sigma.rt,\rho)\\
    \land \lnot(\exists_{f'} \ldotp \sigma.rt = h( h^{*}(\sigma.rt,\rho),f'))
\end{array} \right\}
\]
    \caption{Unique-Root}
\label{fig:uroot}
\end{figure}
\item{Unique-Reachable} invariant in Figure \ref{fig:upath} asserts that every node is reachable from root node with an unique path. This invariant is a part of acyclicity(tree structure) enforcement on the heap layout of the data structure.
  \begin{figure}[!htb]
\[
\textbf{UNQR}(\sigma,O,U,T,F) =
\left\{
\begin{array}{ll}
    \forall_{\rho, \rho'}  \ldotp  h^{*}(\sigma.rt,\rho) \neq h^{*}(\sigma.rt, \rho') \implies  \rho \neq \rho' 
\end{array}
\right.%\}
\]
\caption{Unique-Reachable}
\label{fig:upath}
\end{figure}
\end{enumerate}
 Each of these memory invariants captures different aspects of validity of the memory under \textsf{RCU} setting, $\textsf{WellFormed}(\sigma,O,U,T,F) $, is defined as conjunction of all memory axioms.

 \subsection{Soundness Proof of Atoms}
 \label{lem:lematom}
 In this section, we do proofs to show the soundness of each type rule for each atomic actions.
 \begin{lemma}[\textsc{Unlink}]\small
   \label{lemma:unlink}
\begin{align*}
  \llbracket x.f_1:=r \rrbracket (\lfloor \llbracket \Gamma, \, x:\mathsf{rcuItr}\,\rho\,\N ( [f_1 \rightharpoonup  z]), z:\mathsf{rcuItr} \, \rho' \,\N'([ f_2 \rightharpoonup r]), \;
  r:\mathsf{rcuItr} \, \rho'' \,\N'' \rrbracket_{M,tid} * \{m\}\rfloor)     \subseteq \\
   \lfloor \llbracket \Gamma, \,  x:\mathsf{rcuItr} \, \rho \, \N( f_1 \rightharpoonup z \setminus r),\;
                                                              z:\mathsf{unlinked}, \;
                                                              r:\mathsf{rcuItr} \, \rho' \, \N'' \rrbracket  * \mathcal{R}(\{m\})\rfloor
\end{align*}
 \end{lemma}
 \begin{proof}
We assume
\begin{gather}\label{ahu1}
  \begin{aligned}
    (\sigma, O, U, T, F) \, \in &  \llbracket \Gamma, \, x:\mathsf{rcuItr}\,\rho\,\N ,\, z:\mathsf{rcuItr} \, \rho' \,\N', \\
    &r:\mathsf{rcuItr} \, \rho'' \,\N'' \rrbracket_{M,tid} * \{m\}
    \end{aligned} \\
\textsf{WellFormed}(\sigma,O,U,T,F)
\label{ahu2}
\end{gather}

From assumptions in the type rule of \textsc{T-UnlinkH} we assume that
\begin{gather}
  \rho.f_1=\rho' \text{ and } \rho'.f_2=\rho'' \text{ and } \N(f_1) = z \text{ and } \N'(f_2)=r
    \label{ahu4} \\
\forall_{f\in dom(\N')} \ldotp f\neq f_2 \implies  \N'(f) = \textsf{null}
      \label{ahu5} \\
\begin{array}{l}\footnotesize
\forall_{n\in \Gamma,m,\N''', p''',f}\ldotp n:\textsf{rcuItr}\,\rho'''\,\N'''([f\rightharpoonup m]) \implies %\arcr
\left\{\begin{array}{l}
((\neg\mathsf{MayAlias}(\rho''',\{\rho,\rho',\rho''\})  ) \arcr
\land (m\not\in\{z,r\} ) ) \arcr
\land (\forall_{\rho''''\neq \epsilon} \ldotp \neg\mathsf{MayAlias}(\rho''', \rho''.\rho'''') )
\end{array}\right.
\end{array}
        \label{ahu6}
\end{gather}
We split the composition in  \ref{ahu1} as 
\begin{gather} \label{ahu11}
  \begin{aligned}
    (\sigma_1, O_{1}, U_{1}, T_{1},F_1 ) \in  &\llbracket \Gamma, \, x:\mathsf{rcuItr}\,\rho\,\N ,\, z:\mathsf{rcuItr} \, \rho' \,\N',\\
    &r:\mathsf{rcuItr} \, \rho'' \,\N'' \rrbracket_{M,tid} \end{aligned}\\
\label{ahu12}
(\sigma_2, O_{2}, U_{2}, T_{2},F_2) = m
\\
\label{ahusig}
\sigma_1 \bullet_s \sigma_2 = \sigma \\
\label{ahu13}
O_{1} \bullet_{O} O_{2} = O
\\
\label{ahu14}
U_{1} \cup U_{2} = U
\\
\label{ahu15}
T_{1} \cup T_{2} = T
\\
\label{ahff}
F_{1} \uplus F_2 = F
\\
\label{ahu16}
\textsf{WellFormed}(\sigma_1,O_{1},U_{1},T_{1},F_1)
\\
\label{ahu17}
\textsf{WellFormed}(\sigma_2,O_{2},U_{2},T_{2},F_2)
\end{gather}

We must show  $\exists_{\sigma'_1,\sigma'_2, O'_{1}, O'_{2}, U'_{1}, U'_{2}, T'_{1}, T'_{2}, F'_1,F'_2}$ such that
\begin{gather}\label{phu5}
\begin{aligned}
(\sigma_1',O'_{1},U'_{1}, T'_{1},F'_1)  \in \llbracket \Gamma, \,  x:\mathsf{rcuItr} \, \rho \,\N([f_1 \rightharpoonup r]) ,\, z:\mathsf{unlinked}, r:\mathsf{rcuItr} \, \rho' \, \N'' \rrbracket_{M,tid}
\end{aligned}\\
\\
\label{phuN}
\N(f_1)= r
\\
\label{phu6}
(\sigma_2',O'_{2},U'_{2}, T'_{2}, F'_2) \in \mathcal{R}(\{m\})
\\
\label{ahusig'}
\sigma'_1 \bullet_s \sigma'_2 = \sigma' \\
\label{phu7}
O'_{1} \bullet_{O} O'_{2} = O'
\\
\label{phu8}
U'_{1} \cup U'_{2} = U'
\\
\label{phu9}
T'_{1} \cup T'_{2} = T'
\\
\label{phff}
F'_{1} \uplus F'_2 = F'
\\
\label{phu10}
\textsf{WellFormed}(\sigma_1',O'_{1},U'_{1},T'_{1},F'_1) \\
\label{phu11}
\textsf{WellFormed}(\sigma_2',O'_{2},U'_{2},T'_{2},F'_2)
\end{gather}
We also know from operational semantics that the machine state has changed as
\begin{gather}\label{ahus}
\sigma_1' =  \sigma_1[h(s(x,tid),f_1 ) \mapsto s(r,tid) ]
\end{gather}
and \ref{ahus} is determined by operational semantics. 

The only change in the observation map is on $s(y,tid)$ from $\textsf{iterator}\;tid$ to $\textsf{unliked}$
\begin{gather}\label{ahus1}
  O'_1 =  O_1(s(y,tid))[\textsf{iterator}\;tid \mapsto \textsf{unlinked}]
\end{gather}
\ref{ahus2} follows from \ref{ahu1}
\begin{gather}\label{ahus2}
  T_1 = \{tid\} \text{ and } tid = \sigma.l
\end{gather}

$\sigma'_1$ is determined by operational semantics. The undefined map, free list and $T_1$ need not change so we can pick $U'_1$ as $U_1$, $T'_1$ as $T_1$ and $F'_1$ as $F_1$. Assuming \ref{ahu11} and choices on maps makes $(\sigma_1',O'_{1},U'_{1}, T'_{1},F'_1)$ in denotation
\[\llbracket \Gamma, \,  x:\mathsf{rcuItr} \, \rho \,\N([ f \rightharpoonup r]), z:\mathsf{unlinked}, r:\mathsf{rcuItr} \, \rho' \, \N'' \rrbracket_{M,tid}\]

In the rest of the proof, we prove \ref{phu10}, \ref{phu11} and show the composition of $(\sigma'_1, O'_1, U'_1,T'_1,F'_1)$ and  $(\sigma'_2, O'_2, U'_2,T'_2,F'_2)$. To prove \ref{phu10}, we need to show that each of the memory axioms in Section \ref{sec:memaxioms} holds for the state $(\sigma',O'_1,U_1',T_1',F'_1)$.

Let $o_x$ be $\sigma.s(x,tid)$, $o_y$ be $\sigma.s(y,tid)$ and $o_z$ be $\sigma.s(z,tid)$.
\begin{case}\label{unqr} - \textbf{UNQR} \ref{ahus3} and \ref{ahus4} follow from framing assumption(\ref{ahu4}-\ref{ahu6}), denotations of the precondition(\ref{ahu11}) and \ref{ahu16}.\textbf{UNQR}
  \begin{gather}\label{ahus3}
    \rho \neq \rho' \neq \rho'' 
  \end{gather}
and   
  \begin{gather}\label{ahus4}
    o_x \neq o_y \neq o_z 
  \end{gather}
  where $o_x$, $o_y$ and $o_z$ are equal to $\sigma.h^{*}(\sigma.rt,\rho)$, $\sigma.h^{*}(\sigma.rt,\rho,f_1)$ and $\sigma.h^{*}(\sigma.rt,\rho.f_1.f_2)$ respectively and they($o_x,o_y,o_z$ and $\rho,\rho'$) are unique.

We must prove 
\begin{gather}\label{phuc1}
     h'^{*}(\sigma.rt,\rho) \neq h'^{*}(\sigma.rt, \rho.f_1)  \implies \rho \neq \rho.f_1 
\end{gather}
to show that uniqueness is preserved.

We know from operational semantics that root has not changed so
\[\sigma.rt = \sigma'.rt\]

From denotations (\ref{phu5}) we know that all heap locations reached by following $\rho$ and $\rho.f_1$ are observed as $\textsf{iterator}\, tid$ including the final reached heap locations($\textsf{iterator}\,tid \in O'_1(\sigma'.h^{*}(\sigma.rt,\rho))$ and $\textsf{iterator}\,tid \in O_1'(\sigma'.h^{*}(\sigma.rt,\rho.f_1))$). \ref{phuN} is determined directly by operational semantics.

$\textsf{unlinked}\in O'_1(o_y)$ follows from \ref{ahus1} and \ref{ahus} which makes path $\rho.f_1.f_2$ invalid(from denotation(\ref{phu5}), all heap locations reaching to  $O_1'(o_r)$ from root($\sigma.rt$) are observed as  $\textsf{iterator}\, tid$ so this proves that  $\textsf{unlinked}\in O'_1(o_y)$)  cannot be observed on the path to the $o_r$ which implies that $f_2$ cannot be part of the path and uniqueness of the paths to $o_x$ and $o_r$ is preserved. So we conclude \ref{ahus5} and \ref{ahus6}
  \begin{gather}\label{ahus5}
    \rho \neq \rho'
  \end{gather}
    \begin{gather}\label{ahus6}
    o_x \neq o_y \neq o_z 
  \end{gather}
from which \ref{phuc1} follows.
\end{case}
\begin{case} - \textbf{OW} By \ref{ahu16}.\textbf{OW}, \ref{ahus}, \ref{ahus1}.
\end{case}
\begin{case} - \textbf{RWOW} By \ref{ahu16}.\textbf{RWOW}, \ref{ahus} and \ref{ahus1}.
\end{case}
\begin{case} - \textbf{IFL} By \ref{ahu16}.\textbf{WULK}, \ref{ahu16}.\textbf{RINFL}, \ref{ahu16}.\textbf{IFL}, \ref{ahus1} and choice of $F'_1$.
\end{case}
\begin{case} - \textbf{FLR} By choice of $F'_1$ and \ref{ahu16}. 
\end{case}
\begin{case} - \textbf{WULK} By \ref{phu5},  \ref{ahus1} and \ref{ahus2}.
\end{case}
\begin{case} - \textbf{WF}, \textbf{FPI} and \textbf{FR} Trivial.
\end{case}
\begin{case} - \textbf{AWRT} By \ref{phu5}.
\end{case}
\begin{case} - \textbf{HD} By \ref{phu10}.\textbf{OW}(proved), \ref{ahu16}.\textbf{HD}  and \ref{ahus}. 
\end{case}
\begin{case} - \textbf{WNR} By \ref{ahu16}.\textbf{WNR}, \ref{ahus}, \ref{ahus1} and \ref{ahus2}.
\end{case}
\begin{case} - \textbf{RINFL} By \ref{phu5}, \ref{ahu16}.\textbf{RINFL}, choice of $F'_1$ and \ref{ahus}.
\end{case}
\begin{case} - \textbf{ULKR} We must prove \ref{phuc2}
  \begin{gather}\label{phuc2}
    \begin{aligned}
      \forall_{o',f'} \ldotp \sigma'.h(o',f') = o_y \implies & \textsf{unlinked} \in O'_1(o')  \\
      &\lor (\textsf{freeable} \in O'_1(o'))
      \end{aligned}
\end{gather}
which follows from \ref{phu5}, \ref{ahu16}.\textbf{OW}, operational semantics(\ref{ahus}) and \ref{ahus1}. If $o'$ were observed as \textsf{iterator} then that would conflict with \ref{phu10}.\textbf{UNQR}. 
\end{case}
\begin{case} - \textbf{UNQRT}: By \ref{ahu16}.\textbf{UNQRT}, \ref{ahus1} and \ref{ahus}.
\end{case}
To prove \ref{phu6} we need to show interference relation
\[(\sigma, O_2, U_2, T_2,F_2) \mathcal{R} (\sigma', O'_2, U'_2, T'_2,F'_2)  \]
which by definition means that we must show 
\begin{gather}\label{phu17}
  \sigma_2.l  \in  T_2 \rightarrow (\sigma_2.h =\sigma'_2.h \land \sigma_2.l=\sigma'_2.l)\\
  \label{phu18}
  l\in T_2\rightarrow F_2=F_2'\\
  \label{phu20}
  \forall tid,o\ldotp\textsf{iterator} \, tid \in O_2(o) \rightarrow o \in dom(\sigma_2.h) \\
  \label{phu21}
  \forall tid,o\ldotp\textsf{iterator} \, tid \in O_2(o) \rightarrow o \in dom(\sigma'_2.h) \\
  \label{phu22}
  O_2 = O_2' \land U_2 = U_2' \land T_2 = T_2' \land \sigma_2.R = \sigma'_2.R \land \sigma_2.rt = \sigma'_2.rt \\
  \label{phu23}
  \forall x, t \in T_2 \ldotp \sigma_2.s(x,t) = \sigma'_2.s(x,t) \\
  \label{phurt1}
  \forall tid,o\ldotp\textsf{root} \, tid \in O(o) \rightarrow o \in dom(h) \\
  \label{phurt2}
  \forall tid,o\ldotp\textsf{root} \, tid \in O(o) \rightarrow o \in dom(h') 
\end{gather}
To prove all relations (\ref{phu17}-\ref{phu23}) we assume \ref{ahus2} which is to assume $T_2$ as subset of reader threads. Let $\sigma'_2$ be $\sigma_2$. $O_2$ need not change so we pick $O'_2$  as $O_2$. Since $T_2$ is subset of reader threads, we pick $T_2$ as  $T'_2$. We pick $F'_2$ as $F_2$.

\ref{phu17} and \ref{phu18} follow from \ref{ahus2} and choice of $F_2'$. \ref{phurt1}, \ref{phurt2} and \ref{phu22} are determined by choice of $\sigma'_2$, operational semantic and choices made on maps related to the assertions.

By assuming \ref{ahu17} we show \ref{phu11}. \ref{phu20} and \ref{phu21} follow trivially. \ref{phu23} follows from choice of $\sigma'_2$, \ref{ahus} and \ref{ahus2}. 

To prove \ref{phu7} consider two cases: $O'_1 \cap O'_2 = \emptyset$ and $O'_1 \cap O'_2 \neq \emptyset$. The first case is trivial. The second case is where we consider 
\[
\textsf{iterator}\,tid \in O'_2(o_y)
\]
We also know from \ref{ahus1} that
\[\textsf{unliked} \in O'_1(o_y)\]
Both together with \ref{ahu13} and \ref{phu5} proves \ref{phu7}.

To show \ref{ahusig'} we consider two cases: $\sigma'_1.h \cap \sigma'_2.h = \emptyset$ and $\sigma'_1.h \cap \sigma'_2.h \neq \emptyset$. First is trivial. Second follows from \ref{phu10}.\textbf{OW}-\textbf{HD} and \ref{phu11}.\textbf{OW}-\textbf{HD}. \ref{phu8}, \ref{phu9} and \ref{phff} are trivial by choices on related maps and semantics of composition operations on them. All compositions shown let us to derive conclusion for $(\sigma'_1, O'_1, U'_1, T'_1,F'_1) \bullet (\sigma'_2, O'_2, U'_2, T'_2,F'_2) $.
   \end{proof}
 \begin{lemma}[\textsc{Replace}]
   \label{lemma:linkf}
   \begin{align*}
  \llbracket p.f := n \rrbracket (\lfloor \llbracket \Gamma,\,
 p:\textsf{rcuItr}\, \rho \, \N \, ,
  r:\textsf{rcuItr}\, \rho' \, \N' \, , n:\textsf{rcuFresh} \, \N''\rrbracket_{M,tid} * \{m\}\rfloor)  \subseteq \\
  \lfloor \llbracket \Gamma \,,p:\textsf{rcuItr}\, \rho \, \N([f \rightharpoonup r \setminus  n]) \, , n:\textsf{rcuItr}\, \rho' \, \N'' \, ,r:\unlinked \rrbracket  * \mathcal{R}(\{m\})\rfloor
\end{align*}
 \end{lemma}
 \begin{proof}
 We assume
\begin{gather}\label{ahu1f}
  \begin{aligned}
    (\sigma, O, U, T,F) \, \in &  \llbracket \Gamma,\,
 p:\textsf{rcuItr}\, \rho \, \N \, ,
  r:\textsf{rcuItr}\, \rho' \, \N' \, , n:\textsf{rcuFresh} \, \N'' \rrbracket_{M,tid} * \{m\}
    \end{aligned} \\
\textsf{WellFormed}(\sigma,O,U,T,F)
\label{ahu2f}
\end{gather}
From assumptions in the type rule of \textsc{T-Replace} we assume that
\begin{gather}
\textsf{FV}(\Gamma) \cap \{p,r,n\}  =\emptyset 
  \label{ahu3f} \\
\rho.f  = \rho' \text{ and } \N(f) = r
    \label{ahu4f} \\
\N' = \N''
\label{ahu5f} \\
\begin{aligned}
\forall_{x \in \Gamma, \N''', \rho'',f',y} \ldotp (x:\textsf{rcuItr}\,\rho''\,\N'''([f'\rightharpoonup y])) \implies (\neg\mathsf{MayAlias}(\rho'',\{\rho,\rho'\}) \land (y\neq o  ))  
\label{ahu6f}
  \end{aligned}
\end{gather}
We split the composition in  \ref{ahu1f} as 
\begin{gather} \label{ahu11f}
  \begin{aligned}
    (\sigma_1, O_{1}, U_{1}, T_{1} ,F_1) \in & \llbracket \Gamma,\,
 p:\textsf{rcuItr}\, \rho \, \N \, ,
  r:\textsf{rcuItr}\, \rho' \, \N' \, , n:\textsf{rcuFresh} \, \N'' \rrbracket_{M,tid} \end{aligned}\\
\label{ahu12f}
(\sigma_2, O_{2}, U_{2}, T_{2},F_2) = m
\\
\label{ahu13f}
O_{1} \bullet_{O} O_{2} = O
\\
\label{ahusigf}
\sigma_1 \bullet_s \sigma_2 = \sigma \\
\label{ahu14f}
U_{1} \cup U_{2} = U
\\
\label{ahu15f}
T_{1} \cup T_{2} = T
\\
\label{ahufff}
F_1 \uplus F_2 = F
\\
\label{ahu16f}
\textsf{WellFormed}(\sigma_1,O_{1},U_{1},T_{1},F_1)
\\
\label{ahu17f}
\textsf{WellFormed}(\sigma_2,O_{2},U_{2},T_{2},F_2)
\end{gather}
We must show $\exists_{\sigma'_1, \sigma'_2, O'_{1}, O'_{2}, U'_{1}, U'_{2}, T'_{1}, T'_{2}, F'_1, F'_2}$ such that
\begin{gather}\label{phu5f}
\begin{aligned}
(\sigma_1',O'_{1},U'_{1}, T'_{1},F'_1)  \in \llbracket p:\textsf{rcuItr}\, \rho \, \N \, , n:\textsf{rcuItr}\, \rho' \, \N'' \, , r:\unlinked\, ,  \Gamma \rrbracket_{M,tid}
\end{aligned}
\\
\label{phuNf}
\N(f)=n 
\\
\label{phu6f}
(\sigma_2',O'_{2},U'_{2}, T'_{2}, F'_2) \in \mathcal{R}(\{m\})
\\
\label{phu7f}
O'_{1} \bullet_{O} O'_{2} = O'
\\
\label{ahusigf'}
\sigma'_1 \bullet_s \sigma'_2 = \sigma' \\
\label{phu8f}
U'_{1} \cup U'_{2} = U'
\\
\label{phu9f}
T'_{1} \cup T'_{2} = T'
\\
\label{phufff}
F'_1 \uplus F'_2 = F'
\\
\label{phu10f}
\textsf{WellFormed}(\sigma_1',O'_{1},U'_{1},T'_{1},F'_1) \\
\label{phu11f}
\textsf{WellFormed}(\sigma_2',O'_{2},U'_{2},T'_{2},F'_2)
\end{gather}
We also know from operational semantics that the machine state has changed as
\begin{gather}\label{ahusf}
\sigma_1' =  \sigma_1[h(s(p,tid),f ) \mapsto s(n,tid) ]
\end{gather}
\ref{phuNf} is determined directly from operational semantics.

We know that changes in observation map are
\begin{gather}\label{ahus1f}
O'_1 =  O_1(s(r,tid))[\textsf{iterator}\;tid \mapsto \textsf{unlinked}]
\end{gather}
and
\begin{gather}\label{ahus2f}
O'_1 =  O_1(s(n,tid))[\textsf{fresh} \mapsto \textsf{iterator}\,tid]
\end{gather}

\ref{ahus3f} follows from \ref{ahu1f}
\begin{gather}\label{ahus3f}
  T_1 = \{tid\} \text{ and } tid = \sigma.l
\end{gather}
Let $T'_1$ be $T_1$, $F'_1$ be $F_1$ and $\sigma'_1$ be determined by operational semantics. The undefined map need not change so we can pick $U'_1$ as $U_1$. Assuming \ref{ahu11f} and choices on maps makes $(\sigma_1',O'_{1},U'_{1}, T'_{1})$ in denotation
\[\llbracket p:\textsf{rcuItr}\, \rho \, \N(f \rightharpoonup r \setminus n) \, , n:\textsf{rcuItr}\, \rho' \, \N'' \, , r:\unlinked\, ,  \Gamma \rrbracket_{M,tid}\]

In the rest of the proof, we prove \ref{phu10f}, \ref{phu11f} and show the composition of $(\sigma'_1, O'_1, U'_1,T'_1,F'_1)$ and  $(\sigma'_2, O'_2, U'_2,T'_2,F'_2)$. To prove \ref{phu10f}, we need to show that each of the memory axioms in Section \ref{sec:memaxioms} holds for the state $(\sigma',O'_1,U_1',T_1',F'_1)$.

\begin{case}\label{unqrf} - \textbf{UNQR} 
Let $o_p$ be $\sigma.s(p,tid)$, $o_r$ be $\sigma.s(r,tid)$ and $o_n$ be $\sigma.s(n,tid)$. 
  \ref{ahus3f} and \ref{ahus4f} follow from framing assumption(\ref{ahu3f}-\ref{ahu6f}), denotations of the precondition(\ref{ahu11f}), \ref{ahu16}.\textbf{FR} and \ref{ahu16f}.\textbf{UNQR}
  \begin{gather}\label{ahusbf}
    \rho \neq \rho.f \neq \forall_{\N'([f_i \rightharpoonup x_i])}\ldotp \rho.f.f_i
  \end{gather}
and   
  \begin{gather}\label{ahus4f}
    o_p \neq o_r \neq o_n \neq o_i \text{  where  } o_i=h(o_r,f_i)
  \end{gather}
where $o_p$,  $o_r$ are  $\sigma.h^{*}(\sigma.rt,\rho)$, $\sigma.h^{*}(\sigma.rt,\rho.f)$ respectively and they(heap locations in \ref{ahus4f} and paths in \ref{ahusbf}) are unique(From \ref{ahu16f}.\textbf{FR}, we assume that there exists no field alias/path alias to heap location freshly allocated $o_n$). 

We must prove 
\begin{gather}\label{phuc1f}
    \rho \neq \rho.f \neq \rho.f.f_i \iff \sigma'.h^{*}(\sigma.rt,\rho) \neq \sigma'.h^{*}(\sigma.rt, \rho.f) )  \neq \sigma'.h^{*}(\sigma.rt, \rho.f.f_i) )
\end{gather}

We know from operational semantics that root has not changed so
\[\sigma.rt = \sigma'.rt\]

From denotations (\ref{phu5f}) we know that all heap locations reached by following $\rho$ and $\rho.f$ are observed as $\textsf{iteartor}\,tid$ including the final reached heap locations($\textsf{iterator}\,tid \in O'_1(\sigma'.h^{*}(\sigma.rt,\rho))$, $\textsf{iterator}\,tid \in O'_1(\sigma'.h^{*}(\sigma.rt,\rho.f))$ and $\textsf{iterator}\,tid \in O'_1(\sigma'.h^{*}(\sigma.rt,\rho.f.f_i))$). The preservation of uniqueness follows from \ref{ahus1f}, \ref{ahus2f}, \ref{ahusf} and \ref{ahu16f}.\textbf{FR}.

from which we conclude \ref{ahus5f} and \ref{ahus6f}
  \begin{gather}\label{ahus5f}
    \rho \neq \rho.f \neq \rho.f.f_i
  \end{gather}
    \begin{gather}\label{ahus6f}
    o_p \neq o_n \neq o_r 
  \end{gather}
from which \ref{phuc1f} follows.
\end{case}
\begin{case} - \textbf{OW} By \ref{ahu16f}.\textbf{OW}, \ref{ahusf}, \ref{ahus1f} and \ref{ahus2f}.
\end{case}
\begin{case} - \textbf{RWOW} By \ref{ahu16f}.\textbf{RWOW}, \ref{ahusf}, \ref{ahus1f} and \ref{ahus2f}
\end{case}
\begin{case} - \textbf{AWRT} Trivial.
\end{case}
\begin{case} - \textbf{IFL}  By \ref{ahu16f}.\textbf{WULK}, \ref{ahus1f}, \ref{ahus2f} choice of $F'_1$ and operational semantics.
\end{case}
\begin{case} - \textbf{FLR} By choice of $F'_1$ and \ref{ahu16f}. 
\end{case}
\begin{case} - \textbf{FPI} By \ref{phu5f}.
\end{case}
\begin{case} - \textbf{WULK} Determined by operational semantics By \ref{ahu16f}.\textbf{WULK}, \ref{ahus1f}, \ref{ahus2f} and operational semantics.
\end{case}
\begin{case} - \textbf{WF} and \textbf{FR} Trivial.
\end{case}
\begin{case} - \textbf{HD} 
\end{case}
\begin{case} - \textbf{WNR} By \ref{ahus3f} and operational semantics.
\end{case}
\begin{case} - \textbf{RINFL} Determined by operational semantics(\ref{ahusf}) and \ref{ahu16f}.\textbf{RINFL}.
\end{case}
\begin{case} - \textbf{ULKR} We must prove 
  \begin{gather}\label{phuc2f}
\begin{aligned}
\forall_{o',f'} \ldotp \sigma'.h(o',f') = o_r \implies &\textsf{unlinked} \in O'_1(o') \\
 & \textsf{freeable} \in O'_1(o')
                \end{aligned}
\end{gather}
which follows from \ref{phu5f}, \ref{ahu16f}.\textbf{OW} and determined by operational semantics(\ref{ahusf}), \ref{ahus1f}, \ref{ahus2f}.  If $o'$ were observed as \textsf{iterator} then that would conflict with \ref{phu10f}.\textbf{UNQR}.
\end{case}
\begin{case} - \textbf{UNQRT} By \ref{ahu16f}.\textbf{UNQRT}, \ref{ahus1f}, \ref{ahus2f} and \ref{ahusf}.
\end{case}

To prove \ref{phu6f}, we need to show  interference relation
\[(\sigma, O_2, U_2, T_2,F_2) \mathcal{R} (\sigma', O'_2, U'_2, T'_2,F'_2)  \]
which by definition means that we must show 
\begin{gather}\label{phu17f}
  \sigma_2.l  \in  T_2 \rightarrow (\sigma_2.h =\sigma'_2.h \land \sigma_2.l=\sigma'_2.l)\\
  \label{phu18f}
  l\in T_2\rightarrow F_2=F_2'\\
  \label{phu20f}
  \forall tid,o\ldotp\textsf{iterator} \, tid \in O_2(o) \rightarrow o \in dom(\sigma_2.h) \\
  \label{phu21f}
  \forall tid,o\ldotp\textsf{iterator} \, tid \in O_2(o) \rightarrow o \in dom(\sigma'_2.h) \\
  \label{phu22f}
  O_2 = O_2' \land U_2 = U_2' \land T_2 = T_2' \land \sigma_2.R = \sigma'_2.R \land \sigma_2.rt = \sigma'_2.rt \\
  \label{phu23f}
  \forall x, t \in T_2 \ldotp \sigma_2.s(x,t) = \sigma'_2.s(x,t) \\
    \label{phufrt1}
  \forall tid,o\ldotp\textsf{root} \, tid \in O(o) \rightarrow o \in dom(h) \\
  \label{phufrt2}
  \forall tid,o\ldotp\textsf{root} \, tid \in O(o) \rightarrow o \in dom(h') 
\end{gather}
To prove all relations (\ref{phu17f}-\ref{phu23f}) we assume \ref{ahus3f} which is to assume $T_2$ as subset of reader threads. Let $\sigma'_2$ be $\sigma_2$, $F'_2$ be $F_2$. $O_2$ need not change so we pick $O'_2$  as $O_2$. Since $T_2$ is subset of reader threads, we pick $T_2$ as  $T'_2$. By assuming \ref{ahu17f}  we show \ref{phu11f}. \ref{phu20f} and \ref{phu21f} follow trivially. \ref{phu23f} follows from choice of $\sigma'_2$,  \ref{ahusf} and \ref{ahus3f}.

\ref{phu17f} and \ref{phu18f} follow from \ref{ahus3f} and choice of $F_2'$. \ref{phu22f}, \ref{phufrt1} and \ref{phufrt2} are determined by  choice of $\sigma'_2$, operational semantics and choices made on maps related to the assertions.

To prove \ref{phu7f} consider two cases: $O'_1 \cap O'_2 = \emptyset$ and $O'_1 \cap O'_2 \neq \emptyset$. The first case is trivial. The second case is where we consider \ref{intersect1} and \ref{intersect2}
\begin{gather}\label{intersect1}
\textsf{iterator}\,tid \in O'_2(o_r)
\end{gather}
From \ref{ahus1f} we know that
\[\textsf{unliked} \in O'_1(o_r)\]
Both together with \ref{ahu13f} and \ref{phu5f} proves \ref{phu7f}.

For case \ref{intersect2}
\begin{gather}\label{intersect2}
\textsf{fresh} \in O_2(o_n)
\end{gather}
From \ref{ahus2f} we know that
\[\textsf{iterator}\, tid \in O'_1(o_n)\]
Both together with \ref{ahu13f} and \ref{phu5f} proves \ref{phu7f}.

To show \ref{ahusigf'} we consider two cases: $\sigma'_1 \cap \sigma'_2 = \emptyset$ and $\sigma'_1 \cap \sigma'_2 \neq \emptyset$. First is trivial. Second follows from \ref{phu10f}.\textbf{OW}-\textbf{HD} and \ref{phu11f}.\textbf{OW}-\textbf{HD}. \ref{phu8f}, \ref{phufff} and \ref{phu9f} are trivial by choices on related maps and semantics of the composition operators for these maps. All compositions shown  let us to derive conclusion for $(\sigma'_1, O'_1, U'_1, T'_1,F'_1) \bullet (\sigma'_2, O'_2, U'_2, T'_2,F'_2) $.
 \end{proof}
 \begin{lemma}[\textsc{Insert}]
   \label{lemma:linkflk}
   \begin{align*}
  \llbracket p.f := n \rrbracket (\lfloor \llbracket \Gamma,\,
 p:\textsf{rcuItr}\, \rho \, \N \, ,
  r:\textsf{rcuItr}\, \rho_1 \, \N_2 \, , n:\textsf{rcuFresh} \, \N_1 \rrbracket_{M,tid} * \{m\}\rfloor)  \subseteq \\
  \lfloor \llbracket \Gamma \,,p:\textsf{rcuItr}\, \rho \, \N([f \rightharpoonup r \setminus  n]) \, , n:\textsf{rcuItr}\, \rho_1 \, \N_1 \, ,r:\textsf{rcuItr}\,\rho_2\,\N_2 \rrbracket  * \mathcal{R}(\{m\})\rfloor
\end{align*}
 \end{lemma}
 \begin{proof}
 We assume
\begin{gather}\label{ahu1flk}
  \begin{aligned}
    (\sigma, O, U, T,F) \, \in &  \llbracket \Gamma,\,
 p:\textsf{rcuItr}\, \rho \, \N \, ,
  r:\textsf{rcuItr}\, \rho_1 \, \N_2 \, , n:\textsf{rcuFresh} \, \N_1 \rrbracket_{M,tid} * \{m\}
    \end{aligned} \\
\textsf{WellFormed}(\sigma,O,U,T,F)
\label{ahu2flk}
\end{gather}
From assumptions in the type rule of \textsc{T-Insert} we assume that
\begin{gather}
\textsf{FV}(\Gamma) \cap \{p,r,n\}  =\emptyset 
  \label{ahu3flk} \\
\rho.f  = \rho_1 \text{ and } \rho.f_4  = \rho_2 \text{ and } \N(f) = r
    \label{ahu4flk} \\
\N(f) = \N_1(f_4) \text{ and } \forall_{f_2\in dom(\N_1)} \ldotp f_4 \neq f_2 \implies \N_1(f_2)=\textsf{null}
\label{ahu5flk} \\
\begin{aligned}
  \forall_{x \in \Gamma, \N_3, \rho_3,f_1,y} \ldotp (x:\textsf{rcuItr}\,\rho_3\,\N_3([f_1\rightharpoonup y])) \implies(\forall_{\rho_4\neq \epsilon} \ldotp \neg\mathsf{MayAlias}(\rho_3, \rho.\rho_4) )
\label{ahu6flk}
  \end{aligned}
\end{gather}
We split the composition in  \ref{ahu1flk} as 
\begin{gather} \label{ahu11flk}
  \begin{aligned}
    (\sigma_1, O_{1}, U_{1}, T_{1} ,F_1) \in & \llbracket \Gamma,\,
 p:\textsf{rcuItr}\, \rho \, \N \, ,
  r:\textsf{rcuItr}\, \rho_1 \, \N_2 \, , n:\textsf{rcuFresh} \, \N_1 \rrbracket_{M,tid} \end{aligned}\\
\label{ahu12flk}
(\sigma_2, O_{2}, U_{2}, T_{2},F_2) = m
\\
\label{ahu13flk}
O_{1} \bullet_{O} O_{2} = O
\\
\label{ahusigflk}
\sigma_1 \bullet_s \sigma_2 = \sigma \\
\label{ahu14flk}
U_{1} \cup U_{2} = U
\\
\label{ahu15flk}
T_{1} \cup T_{2} = T
\\
\label{ahuffflk}
F_1 \uplus F_2 = F
\\
\label{ahu16flk}
\textsf{WellFormed}(\sigma_1,O_{1},U_{1},T_{1},F_1)
\\
\label{ahu17flk}
\textsf{WellFormed}(\sigma_2,O_{2},U_{2},T_{2},F_2)
\end{gather}
We must show $\exists_{\sigma'_1, \sigma'_2, O'_{1}, O'_{2}, U'_{1}, U'_{2}, T'_{1}, T'_{2}, F'_1, F'_2}$ such that
\begin{gather}\label{phu5flk}
\begin{aligned}
(\sigma_1',O'_{1},U'_{1}, T'_{1},F'_1)  \in \llbracket p:\textsf{rcuItr}\, \rho \, \N([f\rightharpoonup r \setminus n]) \, , n:\textsf{rcuItr}\, \rho_1 \, \N_1 \, , r:\textsf{rcuItr}\, \rho_2\, \N_2\, ,  \Gamma \rrbracket_{M,tid}
\end{aligned}
\\
\label{phu6flk}
(\sigma_2',O'_{2},U'_{2}, T'_{2}, F'_2) \in \mathcal{R}(\{m\})
\\
\label{phu7flk}
O'_{1} \bullet_{O} O'_{2} = O'
\\
\label{ahusigf'lk}
\sigma'_1 \bullet_s \sigma'_2 = \sigma' \\
\label{phu8flk}
U'_{1} \cup U'_{2} = U'
\\
\label{phu9flk}
T'_{1} \cup T'_{2} = T'
\\
\label{phuffflk}
F'_1 \uplus F'_2 = F'
\\
\label{phu10flk}
\textsf{WellFormed}(\sigma_1',O'_{1},U'_{1},T'_{1},F'_1) \\
\label{phu11flk}
\textsf{WellFormed}(\sigma_2',O'_{2},U'_{2},T'_{2},F'_2)
\end{gather}
We also know from operational semantics that the machine state has changed as
\begin{gather}\label{ahusflk}
\sigma_1' =  \sigma_1[h(s(p,tid),f ) \mapsto s(n,tid) ]
\end{gather}

We know that changes in observation map are
\begin{gather}\label{ahus1flk}
O'_1 =  O_1(s(n,tid))[\textsf{fresh} \mapsto \textsf{iterator}\;tid ]
\end{gather}
\ref{ahus3flk} follows from \ref{ahu1flk}
\begin{gather}\label{ahus3flk}
  T_1 = \{tid\} \text{ and } tid = \sigma.l
\end{gather}
Let $T'_1$ be $T_1$, $F'_1$ be $F_1$ and $\sigma'_1$ be determined by operational semantics. The undefined map need not change so we can pick $U'_1$ as $U_1$. Assuming \ref{ahu11flk} and choices on maps makes $(\sigma_1',O'_{1},U'_{1}, T'_{1})$ in denotation
\[\llbracket p:\textsf{rcuItr}\, \rho \, \N(f \rightharpoonup r \setminus n) \, , n:\textsf{rcuItr}\, \rho_1 \, \N_1 \, , r:\textsf{rcuItr}\,\rho_2\,\N_2\, ,  \Gamma \rrbracket_{M,tid}\]

In the rest of the proof, we prove \ref{phu10flk}, \ref{phu11flk} and show the composition of $(\sigma'_1, O'_1, U'_1,T'_1,F'_1)$ and  $(\sigma'_2, O'_2, U'_2,T'_2,F'_2)$. To prove \ref{phu10flk}, we need to show that each of the memory axioms in Section \ref{sec:memaxioms} holds for the state $(\sigma',O'_1,U_1',T_1',F'_1)$.

Proofs for \textbf{OW}, \textbf{RWOW}, \textbf{AWRT}, \textbf{IFL}, \textbf{WULK}, \textbf{FLR}, \textbf{FPI}, \textbf{WF}, \textbf{FR}, \textbf{HD}, \textbf{WNR}, \textbf{RINFL} and \textbf{ULKR}. The proof of \textbf{UNQR} is similar to the ones  we did for Lemma  \ref{lemma:linkf} and Lemma \ref{lemma:unlink} with a simpler fact to prove: we assume framing conditions \ref{ahu3flk}-\ref{ahu6flk} together with the \ref{ahu16flk}.\textbf{UNQR} and \ref{ahu16flk}.\textbf{FR} which makes \ref{phu10flk}\textbf{UNQR} trivial.

To prove \ref{phu6flk}, we need to show  interference relation
\[(\sigma, O_2, U_2, T_2,F_2) \mathcal{R} (\sigma', O'_2, U'_2, T'_2,F'_2)  \]
which by definition means that we must show 
\begin{gather}\label{phu17flk}
  \sigma_2.l  \in  T_2 \rightarrow (\sigma_2.h =\sigma'_2.h \land \sigma_2.l=\sigma'_2.l)\\
  \label{phu18flk}
  l\in T_2\rightarrow F_2=F_2'\\
  \label{phu20flk}
  \forall tid,o\ldotp\textsf{iterator} \, tid \in O_2(o) \rightarrow o \in dom(\sigma_2.h) \\
  \label{phu21flk}
  \forall tid,o\ldotp\textsf{iterator} \, tid \in O_2(o) \rightarrow o \in dom(\sigma'_2.h) \\
  \label{phu22flk}
  O_2 = O_2' \land U_2 = U_2' \land T_2 = T_2' \land \sigma_2.R = \sigma'_2.R \land \sigma_2.rt = \sigma'_2.rt \\
  \label{phu23flk}
  \forall x, t \in T_2 \ldotp \sigma_2.s(x,t) = \sigma'_2.s(x,t) \\
    \label{phufrt1lk}
  \forall tid,o\ldotp\textsf{root} \, tid \in O(o) \rightarrow o \in dom(h) \\
  \label{phufrt2lk}
  \forall tid,o\ldotp\textsf{root} \, tid \in O(o) \rightarrow o \in dom(h') 
\end{gather}

To prove all relations (\ref{phu17f}-\ref{phu23f}) we assume \ref{ahus3flk} which is to assume $T_2$ as subset of reader threads. Let $\sigma'_2$ be $\sigma_2$, $F'_2$ be $F_2$. $O_2$ need not change so we pick $O'_2$  as $O_2$. Since $T_2$ is subset of reader threads, we pick $T_2$ as  $T'_2$. By assuming \ref{ahu17flk}  we show \ref{phu11flk}. \ref{phu20flk} and \ref{phu21flk} follow trivially. \ref{phu23flk} follows from choice of $\sigma'_2$ and \ref{ahus3flk}.

\ref{phu17flk} and \ref{phu18flk} follow from \ref{ahus3flk} and choice of $F_2'$. \ref{phu22flk}, \ref{phufrt1lk} and \ref{phufrt2lk} are determined by  choice of $\sigma'_2$, operational semantics and choices made on maps related to the assertions.

\ref{phu7flk} follows from assumptions \ref{ahus1flk}, \ref{ahu13flk} and choice of $O'_2$ as $O_2$.

To show \ref{ahusigf'lk} we consider two cases: $\sigma'_1 \cap \sigma'_2 = \emptyset$ and $\sigma'_1 \cap \sigma'_2 \neq \emptyset$. First is trivial. Second follows from \ref{phu10flk}.\textbf{OW}-\textbf{HD} and \ref{phu11flk}.\textbf{OW}-\textbf{HD}. \ref{phu8flk}, \ref{phuffflk} and \ref{phu9flk} are trivial by choices on related maps and semantics of the composition operators for these maps. All compositions shown  let us to derive conclusion for $(\sigma'_1, O'_1, U'_1, T'_1,F'_1) \bullet (\sigma'_2, O'_2, U'_2, T'_2,F'_2) $.
 \end{proof}
 \begin{lemma}[\textsc{ReadStack}]
   \label{lemma:readstack}
\begin{align*}
  \llbracket z:=x \rrbracket (\lfloor \llbracket \Gamma\,, z:\_ \, ,\rcuitrT{x}{G}{k}{k+1}{\_} \rrbracket_{M,tid} * \{m\}\rfloor)  \subseteq \\
                                                              \lfloor \llbracket \Gamma\,, \rcuitrT{x}{G}{k}{k+1}{\_}\, , \rcuitrT{z}{G}{k}{k+1}{\_}  \rrbracket  * \mathcal{R}(\{m\})\rfloor
\end{align*}
 \end{lemma}
 \begin{proof}
We assume
\begin{gather}\label{ahu1sr}
  \begin{aligned}
    (\sigma, O, U, T,F) \, \in &  \llbracket \Gamma\,, z:\_ \, ,\rcuitrT{x}{G}{k}{k+1}{\_} \rrbracket_{M,tid} * \{m\}
    \end{aligned} \\
\textsf{WellFormed}(\sigma,O,U,T,F)
\label{ahu2sr}
\end{gather}
From the assumption in the type rule of \textsc{T-ReadS} we assume that
\begin{gather}
\textsf{FV}(\Gamma) \cap \{z\}  =\emptyset 
  \label{ahu3sr}
\end{gather}
We split the composition in  \ref{ahu1sr} as 
\begin{gather} \label{ahu11sr}
  \begin{aligned}
    (\sigma, O_{1}, U_{1}, T_{1} ,F_1) \in & \llbracket  \Gamma\,, z:\_ \, ,\rcuitrT{x}{G}{k}{k+1}{\_} \rrbracket_{M,tid} \end{aligned}\\
\label{ahu12sr}
(\sigma, O_{2}, U_{2}, T_{2}, F_2) = m
\\
\label{ahusigsr}
\sigma_1 \bullet \sigma_2 = \sigma
\\
\label{ahu13sr}
O_{1} \bullet_{O} O_{2} = O
\\
\label{ahu14sr}
U_{1} \cup U_{2} = U
\\
\label{ahu15sr}
T_{1} \cup T_{2} = T
\\
\label{ahusrf}
F_1 \uplus F_2 = F
\\
\label{ahu16sr}
\textsf{WellFormed}(\sigma_1,O_{1},U_{1},T_{1},F_1)
\\
\label{ahu17sr}
\textsf{WellFormed}(\sigma_2,O_{2},U_{2},T_{2},F_2)
\end{gather}
We must show $\exists_{\sigma'_1, \sigma'_2, O'_{1}, O'_{2}, U'_{1}, U'_{2}, T'_{1}, T'_{2}, F'_1 , F'_2}$ such that
\begin{gather}\label{phu5sr}
\begin{aligned}
(\sigma',O'_{1},U'_{1}, T'_{1}, F'_1)  \in \llbracket \Gamma\,, \rcuitrT{x}{G}{k}{k+1}{\_}\, , \rcuitrT{z}{G}{k}{k+1}{\_}  \rrbracket_{M,tid}
\end{aligned}\\
\label{phu6sr}
(\sigma',O'_{2},U'_{2}, T'_{2}, F'_2) \in \mathcal{R}(\{m\})
\\
\label{ahusigsr'}
\sigma'_1 \bullet \sigma'_2 = \sigma'
\\
\label{phu7sr}
O'_{1} \bullet_{O} O'_{2} = O'
\\
\label{phu8sr}
U'_{1} \cup U'_{2} = U'
\\
\label{phu9sr}
T'_{1} \cup T'_{2} = T'
\\
\label{phusrf}
F'_1 \uplus F'_2 = F'
\\
\label{phu10sr}
\textsf{WellFormed}(\sigma_1',O'_{1},U'_{1},T'_{1},F'_1) \\
\label{phu11sr}
\textsf{WellFormed}(\sigma_2',O'_{2},U'_{2},T'_{2}, F'_2)
\end{gather}

Let $s(x,tid)$ be $o_x$. We also know from operational semantics that the machine state has changed as 
\begin{gather}\label{ahussr}
\sigma' =  \sigma[s(z,tid) \mapsto o_x ]
\end{gather}

We know that there exists no change in the observation of heap locations
\begin{gather}\label{ahus1sr}
O'_1 =  O_1
\end{gather}

\ref{ahus2sr} follows from \ref{ahu1sr}
\begin{gather}\label{ahus2sr}
  T_1 = \{tid\} \text{ and } tid = \sigma.l
\end{gather}

$\sigma'_1$ is determined by operational semantics. The undefined map, $T_1$ and free list need not change so we can pick $U'_1$ as $U_1$, $T'_1$  as $T_1$ and $F'_1$ as $F_1$. Assuming \ref{ahu11sr} and choices on maps makes $(\sigma_1',O'_{1},U'_{1}, T'_{1},F'_1)$ in denotation
\[\llbracket \Gamma\,, \rcuitrT{x}{G}{k}{k+1}{\_}\, , \rcuitrT{z}{G}{k}{k+1}{\_}  \rrbracket_{M,tid}\]

In the rest of the proof, we prove \ref{phu10sr}, \ref{phu11sr} and show the composition of $(\sigma'_1, O'_1, U'_1,T'_1,F'_1)$ and  $(\sigma'_2, O'_2, U'_2,T'_2,F'_2)$. \ref{phu10sr} follows from \ref{ahu16sr} trivially.

To prove \ref{phu6sr}, we need to show interference relation
\[(\sigma, O_2, U_2, T_2,F_2) \mathcal{R} (\sigma', O'_2, U'_2, T'_2,F'_2)  \]
which by definition means that we must show 
\begin{gather}\label{phu17sr}
  \sigma_2.l  \in  T_2 \rightarrow (\sigma_2.h =\sigma'_2.h \land \sigma_2.l=\sigma'_2.l)\\
  \label{phu18sr}
  l\in T_2\rightarrow F_2=F_2'\\
  \label{phu20sr}
  \forall tid,o\ldotp\textsf{iterator} \, tid \in O_2(o) \rightarrow o \in dom(\sigma_2.h) \\
  \label{phu21sr}
  \forall tid,o\ldotp\textsf{iterator} \, tid \in O_2(o) \rightarrow o \in dom(\sigma'_2.h) \\
  \label{phu22sr}
  O_2 = O_2' \land U_2 = U_2' \land T_2 = T_2'\land \sigma_2.R_2 = \sigma_2'.R_2 \land \sigma_2.rt = \sigma'_2.rt \\
  \label{phu23sr}
  \forall x, t \in T_2 \ldotp \sigma_2.s(x,t) = \sigma'_2.s(x,t)  \\
    \label{phusrrt1}
  \forall tid,o\ldotp\textsf{root} \, tid \in O(o) \rightarrow o \in dom(h) \\
  \label{phusrrt2}
  \forall tid,o\ldotp\textsf{root} \, tid \in O(o) \rightarrow o \in dom(h') 
\end{gather}
To prove all relations (\ref{phu17sr}-\ref{phu23sr}) we assume \ref{ahus2sr} which is to assume $T_2$ as subset of reader threads. Let $\sigma'_2$ be $\sigma_2$. $O_2$ need not change so we pick $O'_2$  as $O_2$. We pick $F'_2$ as $F_2$. Since $T_2$ is subset of reader threads, we pick $T_2$ as  $T'_2$. By assuming \ref{ahu17sr}  we show \ref{phu11sr}. \ref{phu20sr}, \ref{phu21sr}, \ref{phusrrt1} and \ref{phusrrt2} follow trivially. \ref{phu23sr} follows from choice of $\sigma'_2$ and \ref{ahussr}(determined by operational semantics).

 \ref{phu17sr} and \ref{phu18sr} follow from \ref{ahus2sr} and choice of $F_2'$. \ref{phu22sr}, \ref{phusrrt1} and \ref{phusrrt2} are determined by choice of $\sigma'_2$, operational semantics and choices made on maps related to the assertions.

\ref{phu7sr}-\ref{phusrf} are  trivial by choices on related maps and semantics of the composition operators for these maps. \ref{ahusigsr'} follows trivially from \ref{ahusigsr}. All compositions shown  let us to derive conclusion for $(\sigma'_1, O'_1, U'_1, T'_1,F'_1) \bullet (\sigma'_2, O'_2, U'_2, T'_2,F'_2) $ trivial.
 \end{proof}
  \begin{lemma}[\textsc{ReadHeap}]
   \label{lemma:readheap}
\begin{align*}
  \llbracket z:=x.f \rrbracket (\lfloor \llbracket \Gamma\, , z:\_ \, ,  x:\textsf{rcuItr} \, \rho \, \N  \rrbracket_{M,tid} * \{m\}\rfloor)  \subseteq \\
                                                              \lfloor \llbracket \Gamma\,,  x:\textsf{rcuItr} \, \rho \, \N[f\mapsto z]\, ,z:\textsf{rcuItr} \, \rho' \, \N_{\emptyset}  \rrbracket  * \mathcal{R}(\{m\})\rfloor
\end{align*}
 \end{lemma}
 \begin{proof}
We assume
\begin{gather}\label{ahu1hr}
  \begin{aligned}
    (\sigma, O, U, T,F) \, \in &  \llbracket \Gamma\,, z:\textsf{rcuItr}\,\_ \, ,\rcuitrT{x}{G}{k}{k+1}{\_} \rrbracket_{M,tid} * \{m\}
    \end{aligned} \\
\textsf{WellFormed}(\sigma,O,U,T,F)
\label{ahu2hr}
\end{gather}
From the assumption in the type rule of \textsc{T-ReadH} we assume that
\begin{gather}
\textsf{FV}(\Gamma) \cap \{z\}  =\emptyset 
\label{ahu3hr} \\
\rho.f = \rho' 
\label{ahu4hr} \\
\end{gather}
We split the composition in  \ref{ahu1hr} as 
\begin{gather} \label{ahu11hr}
  \begin{aligned}
    (\sigma_1, O_{1}, U_{1}, T_{1} ,F_1) \in & \llbracket  \Gamma\,, z:\textsf{rcuItr}\,\_ \, ,\rcuitrT{x}{G}{k}{k+1}{\_} \rrbracket_{M,tid} \end{aligned}\\
\label{ahu12hr}
(\sigma_2, O_{2}, U_{2}, T_{2},F_2) = m
\\
\label{ahusighr}
\sigma_1 \bullet \sigma_2 = \sigma
\\
\label{ahu13hr}
O_{1} \bullet_{O} O_{2} = O
\\
\label{ahu14hr}
U_{1} \cup U_{2} = U
\\
\label{ahu15hr}
T_{1} \cup T_{2} = T
\\
\label{ahuhrf}
F_1 \uplus F_2 = F
\\
\label{ahu16hr}
\textsf{WellFormed}(\sigma_1,O_{1},U_{1},T_{1})
\\
\label{ahu17hr}
\textsf{WellFormed}(\sigma_2,O_{2},U_{2},T_{2})
\end{gather}
We must show $\exists_{\sigma'_1, \sigma'_2, O'_{1}, O'_{2}, U'_{1}, U'_{2}, T'_{1}, T'_{2},F'_1 ,F'_2}$ such that
\begin{gather}\label{phu5hr}
\begin{aligned}
(\sigma',O'_{1},U'_{1}, T'_{1},F_1)  \in \llbracket \Gamma\,,  x:\textsf{rcuItr} \, \rho \, \N[f\mapsto z]\, ,z:\textsf{rcuItr} \, \rho' \, \N_{\emptyset} \rrbracket_{M,tid}
\end{aligned}
\\
\label{phuNhr}
\N(f) = z
\\
\label{phu6hr}
(\sigma',O'_{2},U'_{2}, T'_{2}, F_2) \in \mathcal{R}(\{m\})
\\
\label{ahusighr'}
\sigma'_1 \bullet \sigma'_2 = \sigma'
\\
\label{phu7hr}
O'_{1} \bullet_{O} O'_{2} = O'
\\
\label{phu8hr}
U'_{1} \cup U'_{2} = U'
\\
\label{phu9hr}
T'_{1} \cup T'_{2} = T'
\\
\label{phuhrf}
F'_1 \uplus F'_2 = F'
\\
\label{phu10hr}
\textsf{WellFormed}(\sigma_1',O'_{1},U'_{1},T'_{1},F'_1) \\
\label{phu11hr}
\textsf{WellFormed}(\sigma_2',O'_{2},U'_{2},T'_{2},F'_2)
\end{gather}

Let $h(s(z,tid),f)$ be $o_z$. We also know from operational semantics that the machine state has changed as
\begin{gather}\label{ahushr}
\sigma_1' =  \sigma_1[s(x,tid) \mapsto o_z]
\end{gather}
\ref{phuNhr} is determined directly from operational semantics.

We know that there exists no change in the observation of heap locations
\begin{gather}\label{ahus1hr}
O'_1 =  O_1
\end{gather}

\ref{ahus2hr} follows from \ref{ahu1hr}
\begin{gather}\label{ahus2hr}
  T_1 = \{tid\} \text{ and } tid = \sigma.l
\end{gather}

$\sigma'_1$ is determined by operational semantics. The undefined map, free list and $T_1$ need not change so we can pick $U'_1$ as $U_1$, $F'_1$ as $F_1$ and $T'_1$ and $T_1$. Assuming \ref{ahu11hr} and choices on maps makes $(\sigma_1',O'_{1},U'_{1}, T'_{1},F'_1)$ in denotation
\[ \llbracket \Gamma\,,  x:\textsf{rcuItr} \, \rho \, \N[f\mapsto z]\, ,z:\textsf{rcuItr} \, \rho' \, \N_{\emptyset} \rrbracket_{M,tid}\]

In the rest of the proof, we prove \ref{phu10hr}, \ref{phu11hr} and show the composition of $(\sigma'_1, O'_1, U'_1,T'_1,F'_1)$ and  $(\sigma'_2, O'_2, U'_2,T'_2,F'_2)$. 

To prove \ref{phu6hr}, we need to show that each of the memory axioms in Section \ref{sec:memaxioms} holds for the state $(\sigma',O'_1,U_1',T_1',F'_1)$.
\begin{case} - \textbf{UNQR} By \ref{ahushr}, \ref{ahu16hr}.\textbf{UNQR} and $\sigma.rt = \sigma.rt'$.
\end{case}
\begin{case} - \textbf{OW} By \ref{ahushr}, \ref{ahus1hr} and \ref{ahu16hr}.\textbf{OW}
\end{case}
\begin{case} - \textbf{RWOW} By \ref{ahushr}, \ref{ahus1hr} and \ref{ahu16hr}.\textbf{RWOW}
\end{case}
\begin{case} - \textbf{AWRT} Trivial. 
\end{case}
\begin{case} - \textbf{IFL}  By \ref{phu5hr}, \ref{ahu16hr}.\textbf{WULK}, \ref{ahus1hr}, choice of $F'_1$ and operational semantics.
\end{case}
\begin{case} - \textbf{FLR}  By operational semantics(\ref{ahushr}), choice for $F'_1$ and \ref{ahu16hr}. 
\end{case}
\begin{case} - \textbf{WULK}  By \ref{ahu16hr}.\textbf{WULK}, \ref{ahus1hr} and operational semantics($\sigma.l = \sigma.l'$).
\end{case}
\begin{case} - \textbf{WF}, \textbf{FNR}, \textbf{FPI} and \textbf{FR} Trivial.
\end{case}
\begin{case} - \textbf{HD} 
\end{case}
\begin{case} - \textbf{WNR} By \ref{ahus2hr} and operational semantics($\sigma.l = \sigma.l'$).
\end{case}
\begin{case} - \textbf{RINFL} By operational semantics(\ref{ahushr}) bounding threads have not changed. We choose $F'_1$ as $F_1$. These two together with \ref{ahu16hr} shows \textbf{RINFL}. 
\end{case}
\begin{case} - \textbf{ULKR} Trivial. 
\end{case}
\begin{case} - \textbf{UNQRT}  By \ref{ahu16hr}.\textbf{UNQRT}, \ref{ahus1hr} and \ref{ahushr}.
\end{case}

To prove \ref{phu6hr}, we need to show interference relation
\[(\sigma, O_2, U_2, T_2,F_2) \mathcal{R} (\sigma', O'_2, U'_2, T'_2,F'_2)  \]
which by definition means that we must show 
\begin{gather}\label{phu17hr}
  \sigma_2.l  \in  T_2 \rightarrow (\sigma_2.h =\sigma'_2.h \land \sigma_2.l=\sigma'_2.l)\\
  \label{phu18hr}
  l\in T_2\rightarrow F_2=F_2'\\
  \label{phu20hr}
  \forall tid,o\ldotp\textsf{iterator} \, tid \in O_2(o) \rightarrow o \in dom(\sigma_2.h) \\
  \label{phu21hr}
  \forall tid,o\ldotp\textsf{iterator} \, tid \in O_2(o) \rightarrow o \in dom(\sigma'_2.h) \\
  \label{phu22hr}
  O_2 = O_2' \land U_2 = U_2' \land T_2 = T_2'\land \sigma_2.R_2 = \sigma_2'.R_2 \land \sigma_2.rt = \sigma'_2.rt \\
  \label{phu23hr}
  \forall x, t \in T_2 \ldotp \sigma_2.s(x,t) = \sigma'_2.s(x,t) \\
 \label{phuhrrt1}
  \forall tid,o\ldotp\textsf{root} \, tid \in O(o) \rightarrow o \in dom(h) \\
  \label{phuhrrt2}
  \forall tid,o\ldotp\textsf{root} \, tid \in O(o) \rightarrow o \in dom(h') 
\end{gather}
To prove all relations (\ref{phu17hr}-\ref{phu23hr}) we assume \ref{ahus2hr} which is to assume $T_2$ as subset of reader threads.  Let $\sigma'_2$ be $\sigma_2$ and $F'_2$ be $F_2$. $O_2$ need not change so we pick $O'_2$  as $O_2$. Since $T_2$ is subset of reader threads, we pick $T_2$ as  $T'_2$. By assuming \ref{ahu17hr}  we show \ref{phu11hr}. \ref{phu20hr} and \ref{phu21hr} follows trivially. \ref{phu23hr} follows from choice of $\sigma'_2$ and \ref{ahushr}(determined by operational semantics).

\ref{phu17hr} and \ref{phu18hr} follow from \ref{ahus2hr} and choice of $F_2'$.\ref{phu22hr}, \ref{phuhrrt1} and \ref{phuhrrt2} are determined by choice of $\sigma'_2$, operational semantics and choices made on maps related to the assertions. 

\ref{phu7hr}-\ref{phuhrf} are trivial by choices on related maps and semantics of the composition operators for these maps. \ref{ahusighr'} follows trivially from \ref{ahusighr}. All compositions shown let us to derive conclusion for $(\sigma'_1, O'_1, U'_1, T'_1,F'_1) \bullet (\sigma'_2, O'_2, U'_2, T'_2,F'_2) $.
 \end{proof}
 \begin{lemma}[\textsc{WriteFreshField}]
   \label{lemma:writef}
   \begin{align*}
  \llbracket p.f := z \rrbracket (\lfloor \llbracket \Gamma, p:\textsf{rcuFresh}\,\N'_{f,\emptyset},\; x:\textsf{rcuItr}\;\; \rho \;\; \N \rrbracket_{M,tid} * \{m\}\rfloor)  \subseteq \\
  \lfloor \llbracket \Gamma \,, \rcunf{p}{f}{z} \, , x:\textsf{rcuItr}\, \rho \, \N([f\rightharpoonup z]) \rrbracket  * \mathcal{R}(\{m\})\rfloor
\end{align*}
 \end{lemma}
 \begin{proof}
 We assume
\begin{gather}\label{ahu1wf}
  \begin{aligned}
    (\sigma, O, U, T,F) \, \in &  \llbracket \Gamma, p:\textsf{rcuFresh}\,\N'_{f,\emptyset},\; x:\textsf{rcuItr}\;\; \rho \;\; \N \rrbracket_{M,tid} * \{m\}
    \end{aligned} \\
\textsf{WellFormed}(\sigma,O,U,T,F)
\label{ahu2wf}
\end{gather}
From the assumption in the type rule of \textsc{T-WriteFH} we assume that
\begin{gather}
z:\textsf{rcuItr}\,\rho.f\,\_ \text{ and } \N(f) = z \text{ and } f \notin dom(\N')
\label{ahu4wf} 
\end{gather}

We split the composition in  \ref{ahu1wf} as 
\begin{gather} \label{ahu11wf}
  \begin{aligned}
    (\sigma, O_{1}, U_{1}, T_{1},F_1 ) \in & \llbracket  \Gamma, p:\textsf{rcuFresh}\,\N'_{f,\emptyset},\; x:\textsf{rcuItr}\;\; \rho \;\; \N \rrbracket_{M,tid} \end{aligned}\\
\label{ahu12wf}
(\sigma, O_{2}, U_{2}, T_{2}, F_2) = m
\\
\label{ahusigwf}
\sigma_1 \bullet \sigma_2 = \sigma
\\
\label{ahu13wf}
O_{1} \bullet_{O} O_{2} = O
\\
\label{ahu14wf}
U_{1} \cup U_{2} = U
\\
\label{ahu15wf}
T_{1} \cup T_{2} = T
\\
\label{ahuwff}
F_1 \uplus F_2 = F
\\
\label{ahu16wf}
\textsf{WellFormed}(\sigma_1,O_{1},U_{1},T_{1},F_1)
\\
\label{ahu17wf}
\textsf{WellFormed}(\sigma_2,O_{2},U_{2},T_{2}, F_2)
\end{gather}
We must show $\exists_{\sigma'_1, \sigma'_2, O'_{1}, O'_{2}, U'_{1}, U'_{2}, T'_{1}, T'_{2},F'_1 , F'_2}$ such that
\begin{gather}\label{phu5wf}
\begin{aligned}
(\sigma',O'_{1},U'_{1}, T'_{1},F'_1)  \in \llbracket \Gamma \,, \rcunf{p}{f}{z} \, , x:\textsf{rcuItr}\, \rho \, \N([f\rightharpoonup z]) \rrbracket_{M,tid}
\end{aligned}
\\
\label{phuNwf}
\N(f) = z \land \N'(f) = z
\\
\label{phu6wf}
(\sigma',O'_{2},U'_{2}, T'_{2},F'_2) \in \mathcal{R}(\{m\})
\\
\label{ahusigwf'}
\sigma'_1 \bullet \sigma'_2 = \sigma'
\\
\label{phu7wf}
O'_{1} \bullet_{O} O'_{2} = O'
\\
\label{phu8wf}
U'_{1} \cup U'_{2} = U'
\\
\label{phu9wf}
T'_{1} \cup T'_{2} = T'
\\
\label{phuwff}
F'_1 \uplus F'_2 = F'
\\
\label{phu10wf}
\textsf{WellFormed}(\sigma_1',O'_{1},U'_{1},T'_{1}, F'_1) \\
\label{phu11wf}
\textsf{WellFormed}(\sigma_2',O'_{2},U'_{2},T'_{2}, F'_2)
\end{gather}

We also know from operational semantics that the machine state has changed as
\begin{gather}\label{ahuswf}
\sigma' =  \sigma[h(s(p,tid),f) \mapsto s(z,tid)]
\end{gather}

There exists no change in the observation of heap locations
\begin{gather}\label{ahus1wf}
O'_1 =  O_1
\end{gather}

\ref{ahus2wf} follows from \ref{ahu1wf}
\begin{gather}\label{ahus2wf}
  T_1 = \{tid\} \text{ and } tid = \sigma.l
\end{gather}

$\sigma'_1$ is determined by operational semantics. The undefined map, free list, $T_1$ need not change so we can pick $U'_1$ as $U_1$, $T'_1$ as $T_1$ and $F'_1$ as $F_1$. Assuming \ref{ahu11wf} and choices on maps makes $(\sigma_1',O'_{1},U'_{1}, T'_{1})$ in denotation
\[\llbracket \Gamma \,, \rcunf{p}{f}{z} \, , x:\textsf{rcuItr}\, \rho \, \N([f\rightharpoonup z]) \rrbracket_{M,tid}\]

In the rest of the proof, we prove \ref{phu10wf} and \ref{phu11wf} and show the composition of $(\sigma'_1, O'_1, U'_1,T'_1,F'_1)$ and  $(\sigma'_2, O'_2, U'_2,T'_2,F'_2)$. To prove \ref{phu10wf}, we need to show that each of the memory axioms in Section \ref{sec:memaxioms} holds for the state $(\sigma',O'_1,U_1',T_1',F'_1)$.
\begin{case} - \textbf{UNQR} By \ref{ahu16wf}.\textbf{UNQR}, \ref{phu10wf}.\textbf{FR}(proved) and $\sigma.rt = \sigma.rt'$.
\end{case}
\begin{case} - \textbf{OW} By \ref{ahuswf},\ref{ahus1wf} and \ref{ahu16wf}.\textbf{OW}
\end{case}
\begin{case} - \textbf{RWOW} By \ref{ahuswf}, \ref{ahus1wf} and \ref{ahu16wf}.\textbf{RWOW}
\end{case}
\begin{case} - \textbf{AWRT} Trivial. 
\end{case}
\begin{case} - \textbf{IFL}  By \ref{ahu16wf}.\textbf{WULK}, \ref{ahus1wf}, choice of $F'_1$ and operational semantics.
\end{case}
\begin{case} - \textbf{FLR}  By operational semantics(\ref{ahuswf}), choice of $F'_1$ and \ref{ahu16wf}. 
\end{case}
\begin{case} - \textbf{WULK}  By \ref{ahu16wf}.\textbf{WULK}, \ref{ahus1wf} and operational semantics($\sigma.l = \sigma.l'$).
\end{case}
\begin{case} - \textbf{WF} By \ref{ahu16wf}.\textbf{WF}, \ref{ahus2wf}, \ref{ahus1wf} and operational semantics(\ref{ahuswf}).
\end{case}
\begin{case} - \textbf{FR} By \ref{ahu16wf}.\textbf{FR}, \ref{ahus2wf}, \ref{ahus1wf} and operational semantics(\ref{ahuswf}).
\end{case}
\begin{case} - \textbf{FNR} By \ref{ahu16wf}.\textbf{FNR}, \ref{ahus2wf}, \ref{ahus1wf} and operational semantics(\ref{ahuswf}).
\end{case}
\begin{case} - \textbf{FPI} By \ref{ahu16wf}.\textbf{FPI}, \ref{ahu11wf} and \ref{ahu4wf}
\end{case} 
\begin{case} - \textbf{HD} 
\end{case}
\begin{case} - \textbf{WNR} By \ref{ahus2wf} and operational semantics($\sigma.l = \sigma.l'$).
\end{case}
\begin{case} - \textbf{RINFL} By operational semantics(\ref{ahuswf} - bounding threads have not changed), choice of $F'_1$ and \ref{ahu16wf}. 
\end{case}
\begin{case} - \textbf{ULKR} Trivial. 
\end{case}
\begin{case} - \textbf{UNQRT} By \ref{ahu16wf}.\textbf{UNQRT}, \ref{ahus1wf} and \ref{ahuswf}.
\end{case}

To prove \ref{phu6wf}, we need to show interference relation
\[(\sigma, O_2, U_2, T_2,F_2) \mathcal{R} (\sigma', O'_2, U'_2, T'_2,F'_2)  \]
which by definition means that we must show 
\begin{gather}\label{phu17wf}
  \sigma_2.l  \in  T_2 \rightarrow (\sigma_2.h =\sigma'_2.h \land \sigma_2.l=\sigma'_2.l)\\
  \label{phu18wf}
  l\in T_2\rightarrow F_2=F_2'\\
  \label{phu20wf}
  \forall tid,o\ldotp\textsf{iterator} \, tid \in O_2(o) \rightarrow o \in dom(\sigma_2.h) \\
  \label{phu21wf}
  \forall tid,o\ldotp\textsf{iterator} \, tid \in O_2(o) \rightarrow o \in dom(\sigma'_2.h) \\
  \label{phu22wf}
  O_2 = O_2' \land U_2 = U_2' \land T_2 = T_2'\land \sigma_2.R = \sigma'_2.R \land \sigma_2.rt = \sigma'_2.rt \\
  \label{phu23wf}
  \forall x, t \in T_2 \ldotp \sigma_2.s(x,t) = \sigma'_2.s(x,t)\\
  \label{phuwfrt1}
  \forall tid,o\ldotp\textsf{root} \, tid \in O(o) \rightarrow o \in dom(h) \\
  \label{phuwfrt2}
  \forall tid,o\ldotp\textsf{root} \, tid \in O(o) \rightarrow o \in dom(h') 
\end{gather}
To prove all relations (\ref{phu17wf}-\ref{phu23wf}) we assume \ref{ahus2wf} which is to assume $T_2$ as subset of reader threads and \ref{ahu17wf}. Let $\sigma'_2$ be $\sigma_2$ and $F'_2$ be $F_2$. $O_2$ need not change so we pick $O'_2$  as $O_2$. Since $T_2$ is subset of reader threads, we pick $T_2$ as  $T'_2$. By assuming \ref{ahu17wf}  we show \ref{phu11wf}. \ref{phu20wf} and \ref{phu21wf} follows trivially. \ref{phu23wf} follows from choice of $\sigma'_2$ and \ref{ahuswf}(determined by operational semantics).

\ref{phu17wf} and \ref{phu18wf} follow from \ref{ahus2wf} and choice of $F_2'$.  \ref{phu22wf} are determined by operational semantics, choice of $\sigma'_2$ and choices made on maps related to the assertions.

\ref{phu8wf}-\ref{phuwff} are trivial by choices on related maps and semantics of the composition operators for these maps. \ref{phuwfrt1} and \ref{phuwfrt2} follow from choice of $\sigma'_2$.

$O'_1 \bullet O'_2 $ follows from \ref{ahu13wf}, \ref{ahus1wf} and choice of $O_2$. 

We assume $\sigma_1.h \bullet \sigma_2.h$. We know from \ref{ahu4wf} that $f \notin dom(\N')$. From \ref{phu5wf}, \ref{phu10wf}-\ref{phu11wf}.\textbf{FNR}, \ref{phu10wf}-\ref{phu11wf}.\textbf{RITR} and \ref{phu10wf}-\ref{phu11wf}.\textbf{WNR} we show $\sigma'_1.h \bullet \sigma'_2.h$ (with choices for other maps in the machine state we show \ref{ahusighr'}). All compositions shown  let us to derive conclusion for $(\sigma'_1, O'_1, U'_1, T'_1,F'_1) \bullet (\sigma'_2, O'_2, U'_2, T'_2,F'_2) $.
 \end{proof}
\begin{lemma}[\textsc{Sycn}]
   \label{lemma:syncstop}
   \begin{align*}
  \llbracket  \textsf{SyncStart};\textsf{SyncStop} \rrbracket (\lfloor \llbracket \Gamma  \rrbracket_{M,tid} * \{m\}\rfloor)  \subseteq \\
                                                              \lfloor \llbracket \Gamma[\overline{x:\textsf{freeable}/x:\textsf{unlinked}}] \rrbracket  * \mathcal{R}(\{m\})\rfloor
\end{align*}
 \end{lemma}
 \begin{proof}
 We assume
\begin{gather}\label{ahu1st}
  \begin{aligned}
    (\sigma, O, U, T,F) \, \in &  \llbracket \Gamma \rrbracket_{M,tid} * \{m\}
    \end{aligned} \\
\textsf{WellFormed}(\sigma,O,U,T,F)
\label{ahu2st}
\end{gather}

We split the composition in  \ref{ahu1st} as 
\begin{gather} \label{ahu11st}
  \begin{aligned}
    (\sigma, O_{1}, U_{1}, T_{1},F_1 ) \in & \llbracket  \Gamma \rrbracket_{M,tid} \end{aligned}\\
  \label{ahu12st}
(\sigma, O_{2}, U_{2}, T_{2},F_2) = m
\\
\label{ahusigst}
\sigma_1 \bullet \sigma_2 = \sigma
\\
\label{ahu13st}
O_{1} \bullet_{O} O_{2} = O
\\
\label{ahu14st}
U_{1} \cup U_{2} = U
\\
\label{ahu15st}
T_{1} \cup T_{2} = T
\\
\label{ahustF}
F_1 \uplus F_2 = F
\\
\label{ahu16st}
\textsf{WellFormed}(\sigma,O_{1},U_{1},T_{1},F_1)
\\
\label{ahu17st}
\textsf{WellFormed}(\sigma,O_{2},U_{2},T_{2},F_2)
\end{gather}
We must show $\exists_{\sigma'_1, \sigma'_2, O'_{1}, O'_{2}, U'_{1}, U'_{2}, T'_{1}, T'_{2}},F'_1 ,F'_2$ such that
\begin{gather}\label{phu5st}
\begin{aligned}
(\sigma',O'_{1},U'_{1}, T'_{1},F'_1)  \in \llbracket \Gamma[\overline{x:\textsf{freeable}/x:\textsf{unlinked}}] \rrbracket_{M,tid}
\end{aligned}\\
\label{phu6st}
(\sigma',O'_{2},U'_{2}, T'_{2}, F'_2) \in \mathcal{R}(\{m\})
\\
\label{ahusigst'}
\sigma'_1 \bullet \sigma'_2 = \sigma'
\\
\label{phu7st}
O'_{1} \bullet_{O} O'_{2} = O'
\\
\label{phu8st}
U'_{1} \cup U'_{2} = U'
\\
\label{phu9st}
T'_{1} \cup T'_{2} = T'
\\
\\
\label{phustF}
F'_1 \uplus F'_2 = F'
\\
\label{phu10st}
\textsf{WellFormed}(\sigma',O'_{1},U'_{1},T'_{1},F'_1) \\
\label{phu11st}
\textsf{WellFormed}(\sigma',O'_{2},U'_{2},T'_{2}, F'_2)
\end{gather}
We also know from operational semantics that \lstinline|SyncStart| changes
\begin{gather}\label{ahuB}
  \sigma_1'.B = \sigma_1.B[\emptyset \mapsto R]
\end{gather}
Then \lstinline|SyncStop| changes it to $\emptyset$ so there exists no change in $B$ after \lstinline|SyncStart|;\lstinline|SyncStop|. So there is no change in machine state.
\begin{gather}\label{ahusst}
\sigma_1' =  \sigma_1
\end{gather}

There exists no change in the observation of heap locations
\begin{gather}\label{ahus1st}
  O'_1 =  O_1(\forall_{x \in \Gamma} \ldotp s(x,tid))[\textsf{unlinked} \mapsto \textsf{freeable}]
\end{gather}
and we pick free list to be
\begin{gather}\label{ahusF}
 F'_1 = F_1(\forall_{x:\textsf{unlinked} \in \Gamma, T\subseteq R} \ldotp s(x,tid)[ T \mapsto \{\emptyset\} ])
\end{gather}

\ref{ahus2st} follows from \ref{ahu1st}
\begin{gather}\label{ahus2st}
  T_1 = \{tid\} \text{ and } tid = \sigma.l
\end{gather}

Let $T'_1$ be $T_1$ and $\sigma'_1$(not changed) be determined by operational semantics. The undefined map need not change so we can pick $U'_1$ as $U_1$. Assuming \ref{ahu11st} and choices on maps makes $(\sigma_1',O'_{1},U'_{1}, T'_{1},F'_1)$ in denotation
\[ \llbracket \Gamma[\overline{x:\textsf{freeable}/x:\textsf{unlinked}}] \rrbracket_{M,tid}\]

In the rest of the proof, we prove \ref{phu10st} and \ref{phu11st} and show the composition of $(\sigma'_1, O'_1, U'_1,T'_1,F'_1)$ and $(\sigma'_2, O'_2, U'_2,T'_2,F'_2)$. To prove \ref{phu10st}, we need to show that each of the memory axioms in Section \ref{sec:memaxioms} holds for the state $(\sigma',O'_1,U_1',T_1',F'_1)$ which is trivial by assuming \ref{ahu16st}. We also know \ref{phu5st}(as we showed the support of state to the denotation).

To prove \ref{phu6st}, we need to show interference relation
\[(\sigma, O_2, U_2, T_2,F_2) \mathcal{R} (\sigma', O'_2, U'_2, T'_2,F_2')  \]
which by definition means that we must show 
\begin{gather}\label{phu17st}
  \sigma_2.l  \in  T_2 \rightarrow (\sigma_2.h =\sigma'_2.h \land \sigma_2.l=\sigma'_2.l)\\
  \label{phu18st}
  l\in T_2\rightarrow F_2=F_2'\\
  \label{phu20st}
  \forall tid,o\ldotp\textsf{iterator} \, tid \in O_2(o) \rightarrow o \in dom(\sigma_2.h) \\
  \label{phu21st}
  \forall tid,o\ldotp\textsf{iterator} \, tid \in O_2(o) \rightarrow o \in dom(\sigma'_2.h) \\
  \label{phu22st}
  O_2 = O_2' \land U_2 = U_2' \land T_2 = T_2'\land \sigma_2.R = \sigma'_2.R \land \sigma_2.rt = \sigma'_2.rt \\
  \label{phu23st}
  \forall x, t \in T_2 \ldotp \sigma_2.s(x,t) = \sigma'_2.s(x,t)\\
    \label{phustrt1}
  \forall tid,o\ldotp\textsf{root} \, tid \in O(o) \rightarrow o \in dom(h) \\
  \label{phustrt2}
  \forall tid,o\ldotp\textsf{root} \, tid \in O(o) \rightarrow o \in dom(h') 
\end{gather}
To prove all relations (\ref{phu17st}-\ref{phu23st}) we assume \ref{ahus2st} which is to assume $T_2$ as subset of reader threads and \ref{ahu17st}. Let $\sigma'_2$ be $\sigma_2$. $O_2$ need not change so we pick $O'_2$  as $O_2$. Since $T_2$ is subset of reader threads, we pick $T_2$ as  $T'_2$. By assuming \ref{ahu17st} we show \ref{phu11st}. \ref{phu20st} and \ref{phu21st} follows trivially. \ref{phu23st} follows from choice of $\sigma'_2$ and \ref{ahusst}(determined by operational semantics). 

\ref{phu17st} and \ref{phu18st} follow from \ref{ahus2st}. \ref{phu22st} are determined by choice of $\sigma'_2$ and operational semantics and choices made on maps related to the assertions.

\ref{phu8st}-\ref{phustF} follow from \ref{ahu13st}-\ref{ahustF} trivially by choices on maps of logical state and semantics of composition operators. \ref{ahusigst'} follow from \ref{ahusigst}, \ref{ahusst}-\ref{ahus2st} and choice of $\sigma'_2$. All compositions shown let us to derive conclusion for $(\sigma'_1, O'_1, U'_1, T'_1,F'_1) \bullet (\sigma'_2, O'_2, U'_2, T'_2,F'_2) $ .
 \end{proof}
 
  \begin{lemma}[\textsc{Alloc}]
   \label{lemma:alloc}
\begin{align*}
  \llbracket x:=new \rrbracket (\lfloor \llbracket \Gamma\, , x:\textsf{undef} \, \rrbracket_{M,tid} * \{m\}\rfloor)  \subseteq \\
                                                              \lfloor \llbracket \Gamma\,,  x:\textsf{rcuFresh} \, \N_{\emptyset}  \rrbracket  * \mathcal{R}(\{m\})\rfloor
\end{align*}
 \end{lemma}
 \begin{proof}
We assume
\begin{gather}\label{ahu1alc}
  \begin{aligned}
    (\sigma, O, U, T,F) \, \in &  \llbracket \Gamma\, , x:\textsf{undef} \, \rrbracket_{M,tid} * \{m\}\rfloor) 
    \end{aligned} \\
\textsf{WellFormed}(\sigma,O,U,T,F)
\label{ahu2alc}
\end{gather}

We split the composition in  \ref{ahu1alc} as 
\begin{gather} \label{ahu11alc}
  \begin{aligned}
    (\sigma, O_{1}, U_{1}, T_{1},F_1 ) \in & \llbracket \Gamma\, , x:\textsf{undef} \, \rrbracket_{M,tid}  \end{aligned}\\
  \label{ahu12alc}
(\sigma, O_{2}, U_{2}, T_{2},F_2) = m
  \\
  \label{ahualcsig}
  \sigma_1 \bullet \sigma_2 = \sigma
  \\
\label{ahu13alc}
O_{1} \bullet_{O} O_{2} = O
\\
\label{ahu14alc}
U_{1} \cup U_{2} = U
\\
\label{ahu15alc}
T_{1} \cup T_{2} = T
\\
\label{ahualcF}
F_1 \uplus F_2 = F
\\
\label{ahu16alc}
\textsf{WellFormed}(\sigma,O_{1},U_{1},T_{1},F_1)
\\
\label{ahu17alc}
\textsf{WellFormed}(\sigma,O_{2},U_{2},T_{2},F_2)
\end{gather}
We must show $\exists_{O'_{1}, O'_{2}, U'_{1}, U'_{2}, T'_{1}, T'_{2},F'_1, F'_2}$ such that
\begin{gather}\label{phu5alc}
\begin{aligned}
(\sigma',O'_{1},U'_{1}, T'_{1},F'_1)  \in \llbracket \Gamma\,,  x:\textsf{rcuFresh} \, \N_{\emptyset}  \rrbracket 
\end{aligned}\\
\label{phu6alc}
(\sigma',O'_{2},U'_{2}, T'_{2},F'_2) \in \mathcal{R}(\{m\})
\\
\label{phualcsig}
\sigma'_1 \bullet \sigma'_2 = \sigma'
\\
\label{phu7alc}
O'_{1} \bullet_{O} O'_{2} = O'
\\
\label{phu8alc}
U'_{1} \cup U'_{2} = U'
\\
\label{phu9alc}
T'_{1} \cup T'_{2} = T'
\\
\label{phualcF}
F'_1 \uplus F'_2 = F'
\\
\label{phu10alc}
\textsf{WellFormed}(\sigma',O'_{1},U'_{1},T'_{1}) \\
\label{phu11alc}
\textsf{WellFormed}(\sigma',O'_{2},U'_{2},T'_{2})
\end{gather}

From operational semantics we know that $s(y,tid)$ is $\ell$. We also know from operational semantics that the machine state has changed as
\begin{gather}\label{ahusalc}
\sigma' =  \sigma[h(\ell) \mapsto \textsf{nullmap} ]
\end{gather}

There exists no change in the observation of heap locations
\begin{gather}\label{ahus1alc}
  O'_1 =  O_1(\ell)[\textsf{undef}\mapsto \textsf{fresh}]
\end{gather}

\ref{ahus2alc} follows from \ref{ahu1alc}
\begin{gather}\label{ahus2alc}
  T_1 = \{tid\} \text{ and } tid = \sigma.l
\end{gather}

Let $T'_1$ to be $T_1$. Undefined map and free list need not change so we can pick $U'_1$ as $U_1$ and $F'_1$ as $F_1$ and show(\ref{phu5alc}) that $(\sigma',O'_1,U_1',T_1',F'_1)$ is in denotation of
\[\llbracket \Gamma\,,  x:\textsf{rcuFresh} \, \N_{\emptyset}  \rrbracket \]

In the rest of the proof, we prove \ref{phu10alc}, \ref{phu11alc} and $(\sigma'_1, O'_1, U'_1,T'_1,F'_1)$ and  $(\sigma'_2, O'_2, U'_2,T'_2,F'_2)$. To prove \ref{phu10alc}, we need to show that each of the memory axioms in Section \ref{sec:memaxioms} holds for the state $(\sigma',O'_1,U_1',T_1',F'_1)$.

\begin{case} - \textbf{UNQR} Determined by \ref{phu5alc} and operational semantics($\ell$ is fresh-unique).
\end{case}
\begin{case} - \textbf{RWOW}, \textbf{OW} By \ref{phu5alc}
\end{case}
\begin{case} - \textbf{AWRT} Determined by operational semantics($\ell$ is fresh-unique). 
\end{case}
\begin{case} - \textbf{IFL}, \textbf{ULKR}, \textbf{WULK}, \textbf{RINFL}, \textbf{UNQRT} Trivial.
\end{case}
\begin{case} - \textbf{FLR} determined by operational semantics and \ref{phu5alc}.
\end{case}
\begin{case} - \textbf{WF} By \ref{ahus2alc}, \ref{phu5alc} and \ref{ahus1alc}. 
\end{case}
\begin{case} - \textbf{FR} Determined by operational semantics($\ell$ is fresh-unique).
\end{case}
\begin{case} - \textbf{FNR} By \ref{phu5alc} and operational semantics($\ell$ is fresh-unique).
\end{case}
\begin{case} - \textbf{FPI} By \ref{phu5alc} and $\N_{f,\emptyset}$.
\end{case} 
\begin{case} - \textbf{HD} 
\end{case}
\begin{case} - \textbf{WNR} By \ref{ahus2alc}.
\end{case}

To prove \ref{phu6alc}, we need to show interference relation
\[(\sigma, O_2, U_2, T_2,F_2) \mathcal{R} (\sigma', O'_2, U'_2, T'_2,F'_2)  \]
which by definition means that we must show 
\begin{gather}\label{phu17alc}
  \sigma_2.l  \in  T_2 \rightarrow (\sigma_2.h =\sigma'_2.h \land \sigma_2.l=\sigma'_2.l)\\
  \label{phu18alc}
  l\in T_2\rightarrow F_2=F_2'\\
  \label{phu20alc}
  \forall tid,o\ldotp\textsf{iterator} \, tid \in O_2(o) \rightarrow o \in dom(\sigma_2.h) \\
  \label{phu21alc}
  \forall tid,o\ldotp\textsf{iterator} \, tid \in O_2(o) \rightarrow o \in dom(\sigma'_2.h) \\
  \label{phu22alc}
  O_2 = O_2' \land U_2 = U_2' \land T_2 = T_2'\land \sigma_2.R = \sigma'_2.R \land \sigma_2.rt = \sigma'_2.rt \\
  \label{phu23alc}
  \forall x, t \in T \ldotp \sigma_2.s(x,t) = \sigma'_2.s(x,t) \\
    \label{phualcrt1}
  \forall tid,o\ldotp\textsf{root} \, tid \in O(o) \rightarrow o \in dom(h) \\
  \label{phualcrt2}
  \forall tid,o\ldotp\textsf{root} \, tid \in O(o) \rightarrow o \in dom(h') 
\end{gather}
To prove all relations (\ref{phu17alc}-\ref{phualcrt2}) we assume \ref{ahus2alc} which is to assume $T_2$ as subset of reader threads and \ref{ahu17alc}. Let $\sigma'_2$ be $\sigma_2$. $F_2$ and $O_2$ need not change so we pick $O'_2$  as $O_2$ and $F'_2$ as $F_2$. Since $T_2$ is subset of reader threads, we pick $T_2$ as  $T'_2$. By assuming \ref{ahu17alc} and choices on maps we show \ref{phu11alc}. \ref{phu20alc} and \ref{phu21alc} follow trivially. \ref{phu23alc} follows from choice of $\sigma'_2$ and \ref{ahusalc}(determined by operational semantics). \ref{phu7alc}-\ref{phualcF} follow from \ref{ahu13alc}-\ref{ahualcF}, semantics of compositions operators and choices made for maps of the logical state.

\ref{phu17alc} and \ref{phu18alc} follow from \ref{ahus2alc} and choice on $F'_2$. \ref{phu22alc} are determined by operational semantics, operational semantics and choices made on maps related to the assertion.

$\sigma'_1.h \cap \sigma'_2.h = \emptyset$ is determined by operational semantics($\ell$ is unique and fresh). So, \ref{phualcsig} follows from \ref{ahualcsig} and choice of $\sigma'_2$. All compositions shown let us to derive conclusion for  $(\sigma'_1, O'_1, U'_1, T'_1,F'_1) \bullet (\sigma'_2, O'_2, U'_2, T'_2,F'_2) $.
 \end{proof}
 \begin{lemma}[\textsc{Free}]
   \label{lemma:free}
\begin{align*}
  \llbracket Free(x) \rrbracket (\lfloor \llbracket x:\textsf{freeable} \rrbracket_{M,tid} * \{m\}\rfloor)  \subseteq \\
                                                              \lfloor \llbracket x:\textsf{undef} \rrbracket  * \mathcal{R}(\{m\})\rfloor
\end{align*}
 \end{lemma}
 \begin{proof}
We assume
\begin{gather}\label{ahu1free}
  \begin{aligned}
    (\sigma, O, U, T,F) \, \in &  \llbracket x:\textsf{freeable} \, \rrbracket_{M,tid} * \{m\}\rfloor) 
    \end{aligned} \\
\textsf{WellFormed}(\sigma,O,U,T,F)
\label{ahu2free}
\end{gather}

We split the composition in  \ref{ahu1free} as 
\begin{gather} \label{ahu11free}
  \begin{aligned}
    (\sigma, O_{1}, U_{1}, T_{1},F_1 ) \in & \llbracket  x:\textsf{freeable}  \rrbracket_{M,tid}  \end{aligned}\\
  \label{ahu12free}
(\sigma, O_{2}, U_{2}, T_{2},F_2) = m
  \\
  \label{ahufreesig}
  \sigma_1 \bullet \sigma_2 = \sigma
  \\
\label{ahu13free}
O_{1} \bullet_{O} O_{2} = O
\\
\label{ahu14free}
U_{1} \cup U_{2} = U
\\
\label{ahu15free}
T_{1} \cup T_{2} = T
\\
\label{ahufreeF}
F_1 \uplus F_2 = F
\\
\label{ahu16free}
\textsf{WellFormed}(\sigma,O_{1},U_{1},T_{1},F_1)
\\
\label{ahu17free}
\textsf{WellFormed}(\sigma,O_{2},U_{2},T_{2},F_2)
\end{gather}

We must show $\exists_{O'_{1}, O'_{2}, U'_{1}, U'_{2}, T'_{1}, T'_{2},F'_1, F'_2}$ such that
\begin{gather}\label{phu5free}
\begin{aligned}
(\sigma',O'_{1},U'_{1}, T'_{1},F'_1)  \in \llbracket  x:\textsf{undef}  \rrbracket 
\end{aligned}\\
\label{phu6free}
(\sigma',O'_{2},U'_{2}, T'_{2},F'_2) \in \mathcal{R}(\{m\})
\\
\label{phufreesig}
\sigma'_1 \bullet \sigma'_2 = \sigma'
\\
\label{phu7free}
O'_{1} \bullet_{O} O'_{2} = O'
\\
\label{phu8free}
U'_{1} \cup U'_{2} = U'
\\
\label{phu9free}
T'_{1} \cup T'_{2} = T'
\\
\label{phufreeF}
F'_1 \uplus F'_2 = F'
\\
\label{phu10free}
\textsf{WellFormed}(\sigma',O'_{1},U'_{1},T'_{1},F'_1) \\
\label{phu11free}
\textsf{WellFormed}(\sigma',O'_{2},U'_{2},T'_{2},F'_2)
\end{gather}

From operational semantics we know that
\begin{gather}\label{ahusfree}
  \forall_{f,o'}\ldotp rt \neq s(x,tid) \land  o' \neq s(x,tid) \implies h(o',f) = h'(o',f) \land \forall_{f}\ldotp h'(o,f)=\textsf{undef}
\end{gather}

\begin{gather}\label{ahus1free}
  O'_1 =  O_1(s(x,tid))[\textsf{freeable}\mapsto \textsf{undef}]
\end{gather}

\begin{gather}\label{ahus3free}
  F'_1 = F_1 \setminus \{s(x,tid)\mapsto \{\emptyset\}\}
  \end{gather}

\begin{gather}\label{ahus4free}
  U'_1 = U_1 \cup \{(x,tid)\}
  \end{gather}

\ref{ahus2free} follows from \ref{ahu1free}
\begin{gather}\label{ahus2free}
  T_1 = \{tid\} \text{ and } tid = \sigma.l
\end{gather}

Let $T'_1$ to be $T_1$. All \ref{ahusfree}-\ref{ahus4free} show(\ref{phu5free}) that $(\sigma',O'_1,U_1',T_1',F'_1)$ is in denotation of  
\[\llbracket  x:\textsf{undef}  \rrbracket \]

In the rest of the proof, we prove \ref{phu10free}, \ref{phu11free} and show the composition of $(\sigma'_1, O'_1, U'_1,T'_1,F'_1)$ and  $(\sigma'_2, O'_2, U'_2,T'_2,F'_2)$.. To prove \ref{phu10free}, we need to show that each of the memory axioms in Section \ref{sec:memaxioms} holds for the state $(\sigma',O'_1,U_1',T_1',F'_1)$ and it it trivial by \ref{ahusfree}-\ref{ahus4free} and \ref{phu5free}.

To prove \ref{phu6free}, we need to show interference relation
\[(\sigma, O_2, U_2, T_2,F_2) \mathcal{R} (\sigma', O'_2, U'_2, T'_2,F'_2)  \]
which by definition means that we must show 
\begin{gather}\label{phu17free}
  \sigma_2.l  \in  T_2 \rightarrow (\sigma_2.h =\sigma'_2.h \land \sigma_2.l=\sigma'_2.l)\\
  \label{phu18free}
  l\in T_2\rightarrow F_2=F_2'\\
  \label{phu20free}
  \forall tid,o\ldotp\textsf{iterator} \, tid \in O_2(o) \rightarrow o \in dom(\sigma_2.h) \\
  \label{phu21free}
  \forall tid,o\ldotp\textsf{iterator} \, tid \in O_2(o) \rightarrow o \in dom(\sigma'_2.h) \\
  \label{phu22free}
  O_2 = O_2' \land U_2 = U_2' \land T_2 = T_2'\land \sigma_2.R = \sigma'_2.R \land \sigma_2.rt = \sigma'_2.rt \\
  \label{phu23free}
  \forall x, t \in T \ldotp \sigma_2.s(x,t) = \sigma'_2.s(x,t) \\
    \label{phufreert1}
  \forall tid,o\ldotp\textsf{root} \, tid \in O(o) \rightarrow o \in dom(h) \\
  \label{phufreert2}
  \forall tid,o\ldotp\textsf{root} \, tid \in O(o) \rightarrow o \in dom(h') 
\end{gather}

To prove all relations (\ref{phu17free}-\ref{phufreert2}) we assume \ref{ahus2free} which is to assume $T_2$ as subset of reader threads and \ref{ahu17alc}. Let $\sigma'_2$ be $\sigma_2$. $F_2$ and $O_2$ need not change so we pick $O'_2$  as $O_2$ and $F'_2$ as $F_2$. Since $T_2$ is subset of reader threads, we pick $T_2$ as  $T'_2$. By assuming \ref{ahu17alc} and choices on maps we show \ref{phu11free}. \ref{phu20free} and \ref{phu21free} follow trivially. \ref{phu23free} follows from choice of $\sigma'_2$ and \ref{ahusfree}(determined by operational semantics). \ref{phu7free}-\ref{phufreeF} follow from \ref{ahu14free}-\ref{ahufreeF}, semantics of composition operators and choices on related maps.

\ref{phu17free} and \ref{phu18free} follow from \ref{ahus2free} and choice on $F'_2$. \ref{phu22free} are determined by operational semantics, choice of $\sigma'_2$ and choices made on maps related to the assertion.

Composition for heap for case $\sigma'_1.h \cap \sigma'_2.h = \emptyset$ is trivial. $\sigma'_1.h \cap \sigma'_2.h \neq \emptyset$ is determined by semantics of heap composition operator $\bullet_h$( $v$ has precedence over \textsf{undef}) and this makes showing \ref{phu7free} straightforward. Since other machine components do not change(determined by operational semantics), \ref{phufreesig} follows from \ref{ahufreesig}, \ref{ahusfree} and choice of $\sigma'_2$. All compositions shown let us to derive conclusion for  $(\sigma'_1, O'_1, U'_1, T'_1,F'_1) \bullet (\sigma'_2, O'_2, U'_2, T'_2,F'_2) $.
 \end{proof} 
  \begin{lemma}[\textsc{RReadStack}]
   \label{lemma:rreadstack}
\begin{align*}
  \llbracket z:=x \rrbracket (\lfloor \llbracket \Gamma\,, z:\textsf{rcuItr}\, , x:\textsf{rcuItr} \rrbracket_{R,tid} * \{m\}\rfloor)  \subseteq \\
                                                              \lfloor \llbracket \Gamma\,, x:\textsf{rcuItr} \, , z:\textsf{rcuItr}  \rrbracket  * \mathcal{R}(\{m\})\rfloor
\end{align*}
 \end{lemma}
  \begin{proof}
  We assume
\begin{gather}\label{ahu1srr}
  \begin{aligned}
    (\sigma, O, U, T,F) \, \in & \llbracket \Gamma\,, \Gamma\,, z:\textsf{rcuItr}\, , x:\textsf{rcuItr}   \rrbracket_{R,tid} * \{m\}
    \end{aligned} \\
\textsf{WellFormed}(\sigma,O,U,T,F)
\label{ahu2srr}
\end{gather}

We split the composition in  \ref{ahu1srr} as 
\begin{gather} \label{ahu11srr}
  \begin{aligned}
    (\sigma_1, O_{1}, U_{1}, T_{1},F_1 ) \in & \llbracket \Gamma\,, \Gamma\,, z:\textsf{rcuItr}\, , x:\textsf{rcuItr} \rrbracket_{R,tid} \end{aligned}\\
  \label{ahu12srr}
(\sigma, O_{2}, U_{2}, T_{2},F_2) = m
\\
\label{ahu13srr}
O_{1} \bullet_{O} O_{2} = O
\\
\label{ahusigsrr}
\sigma_1 \bullet \sigma_2 = \sigma
\\
\label{ahu14srr}
U_{1} \cup U_{2} = U
\\
\label{ahu15srr}
T_{1} \cup T_{2} = T
\\
\label{ahusrrf}
F_1 \uplus F_2 = F
\\
\label{ahu16srr}
\textsf{WellFormed}(\sigma,O_{1},U_{1},T_{1},F_1)
\\
\label{ahu17srr}
\textsf{WellFormed}(\sigma,O_{2},U_{2},T_{2},F_2)
\end{gather}
We must show $\exists_{O'_{1}, O'_{2}, U'_{1}, U'_{2}, T'_{1}, T'_{2},F'_1,F'_2}$ such that
\begin{gather}\label{phu5srr}
\begin{aligned}
(\sigma',O'_{1},U'_{1}, T'_{1},F'_1)  \in\llbracket \Gamma\,, x:\textsf{rcuItr} \, , z:\textsf{rcuItr}   \rrbracket_{R,tid}
\end{aligned}\\
\label{phu6srr}
(\sigma',O'_{2},U'_{2}, T'_{2},F'_2) \in \mathcal{R}(\{m\})
\\
\label{phu7srr}
O'_{1} \bullet_{O} O'_{2} = O'
\\
\label{ahusigsrr'}
\sigma'_1 \bullet \sigma'_2 = \sigma'
\\
\label{phu8srr}
U'_{1} \cup U'_{2} = U'
\\
\label{phu9srr}
T'_{1} \cup T'_{2} = T'
\\
\label{phusrrf}
F'_1 \uplus F'_2 = F'
\\
\label{phu10srr}
\textsf{WellFormed}(\sigma',O'_{1},U'_{1},T'_{1},F'_1) \\
\label{phu11srr}
\textsf{WellFormed}(\sigma',O'_{2},U'_{2},T'_{2},F'_2)
\end{gather}

We also know from operational semantics that the machine state has changed as
\begin{gather}\label{ahussrr}
\sigma_1' =  \sigma_1
\end{gather}

There exists no change in the observation of heap locations
\begin{gather}\label{ahus1srr}
  O'_1 =  O_1
\end{gather}

\ref{ahus2srr} follows from \ref{ahu1srr}
\begin{gather}\label{ahus2srr}
  T_1 \subseteq R
\end{gather}

Let $T'_1$ be $T_1$ and $\sigma'_1$ be determined by operational semantics as $\sigma_1$. The undefined map and free list need not change so we can pick $U'_1$ as $U_1$ and $F'_1$ as $F_1$. Assuming \ref{ahu11srr} and choices on maps makes $(\sigma_1',O'_{1},U'_{1}, T'_{1},F'_1)$ in denotation
\[ \llbracket \Gamma\,, x:\textsf{rcuItr} \, , z:\textsf{rcuItr}  \rrbracket_{R,tid}\]

In the rest of the proof, we prove \ref{phu10srr}, \ref{phu11srr}\ref{phu10}, \ref{phu11} and show the composition of $(\sigma'_1, O'_1, U'_1,T'_1,F'_1)$ and  $(\sigma'_2, O'_2, U'_2,T'_2,F'_2)$. To prove \ref{phu10srr}, we need to show that each of the memory axioms in Section \ref{sec:memaxioms} holds for the state $(\sigma',O'_1,U_1',T_1',F'_1)$ which is trivial by assuming \ref{ahu16srr} and knowing \ref{ahus2srr}, \ref{ahus1srr} and components of the state determined by operational semantics.

To prove \ref{phu11srr}, we need to show that \textsf{WellFormed}ness is preserved under interference relation
\[(\sigma, O_2, U_2, T_2,F_2) \mathcal{R} (\sigma', O'_2, U'_2, T'_2,F'_2)  \]
which by definition means that we must show 
\begin{gather}\label{phu17srr}
  \sigma_2.l  \in  T_2 \rightarrow (\sigma_2.h =\sigma'_2.h \land \sigma_2.l=\sigma'_2.l)\\
  \label{phu18srr}
  l\in T_2\rightarrow F_2=F_2'\\
  \label{phu20srr}
  \forall tid,o\ldotp\textsf{iterator} \, tid \in O_2(o) \rightarrow o \in dom(\sigma_2.h) \\
  \label{phu21srr}
  \forall tid,o\ldotp\textsf{iterator} \, tid \in O_2(o) \rightarrow o \in dom(\sigma'_2.h) \\
  \label{phu22srr}
  O_2 = O_2' \land U_2 = U_2' \land T_2 = T_2'\land \sigma_2.B = \sigma'_2.B \land \sigma_2.rt = \sigma'_2.rt \\
  \label{phu23srr}
  \forall x, t \in T_2 \ldotp \sigma_2.s(x,t) = \sigma'_2.s(x,t) \\
    \label{phusrrrt1}
  \forall tid,o\ldotp\textsf{root} \, tid \in O(o) \rightarrow o \in dom(h) \\
  \label{phusrrrt2}
  \forall tid,o\ldotp\textsf{root} \, tid \in O(o) \rightarrow o \in dom(h') 
\end{gather}
$\sigma_2$, $O_2$, $U_2$ and $T_2$ need not change so that we choose $\sigma'_2$ to be $\sigma'_2$, $O'_2$ to be $O_2$, $U'_2$ to $U_2$ and $T'_2$ to be $T_2$. Let $F'_2$ be $F_2$. These choices make proving \ref{phu17srr}-\ref{phusrrrt2} trivial and  \ref{phu7srr}-\ref{phu9srr} follow from assumptions \ref{ahu13srr}-\ref{ahusrrf}, choices made for related maps and semantics of composition operations. All compositions shown let us derive conclusion for $(\sigma'_1,O'_1,U'_1,T'_1,F'_1) \bullet (\sigma'_2,O'_2,U'_2,T'_2,F'_2)$.
  \end{proof}

    \begin{lemma}[\textsc{RReadHeap}]
   \label{lemma:rreadheap}
\begin{align*}
  \llbracket z:=x.f \rrbracket (\lfloor \llbracket \Gamma\, , z:\textsf{rcuItr}\, ,  x:\textsf{rcuItr} \rrbracket_{R,tid} * \{m\}\rfloor)  \subseteq \\
                                                              \lfloor \llbracket \Gamma\,,  x:\textsf{rcuItr} \, ,z:\textsf{rcuItr}   \rrbracket  * \mathcal{R}(\{m\})\rfloor
\end{align*}
 \end{lemma}
    \begin{proof}
        We assume
\begin{gather}\label{ahu1hrr}
  \begin{aligned}
    (\sigma, O, U, T,F) \, \in &  \llbracket \Gamma\, , z:\textsf{rcuItr}\, ,  x:\textsf{rcuItr} \rrbracket_{R,tid} * \{m\}\rfloor)
    \end{aligned} \\
\textsf{WellFormed}(\sigma,O,U,T,F)
\label{ahu2hrr}
\end{gather}

We split the composition in  \ref{ahu1hrr} as 
\begin{gather} \label{ahu11hrr}
  \begin{aligned}
    (\sigma_1, O_{1}, U_{1}, T_{1},F_1 ) \in &  \llbracket \Gamma\, , z:\textsf{rcuItr}\, ,  x:\textsf{rcuItr}  \rrbracket_{R,tid} \end{aligned}\\
  \label{ahu12hrr}
(\sigma_2, O_{2}, U_{2}, T_{2},F_2) = m
  \\
  \label{ahusighrr}
  \sigma_1 \bullet \sigma_2 = \sigma
  \\
\label{ahu13hrr}
O_{1} \bullet_{O} O_{2} = O
\\
\label{ahu14hrr}
U_{1} \cup U_{2} = U
\\
\label{ahu15hrr}
T_{1} \cup T_{2} = T
\\
\label{ahuhrrf}
F_1 \uplus F_2 = F
\\
\label{ahu16hrr}
\textsf{WellFormed}(\sigma_1,O_{1},U_{1},T_{1},F_1)
\\
\label{ahu17hrr}
\textsf{WellFormed}(\sigma_2,O_{2},U_{2},T_{2},F_2)
\end{gather}
We must show  $\exists_{\sigma'_1,\sigma'_2,O'_{1}, O'_{2}, U'_{1}, U'_{2}, T'_{1}, T'_{2},F'_1,F'_2}$ such that
\begin{gather}\label{phu5fhrr}
\begin{aligned}
(\sigma_1',O'_{1},U'_{1}, T'_{1},F'_1)  \in   \lfloor \llbracket \Gamma\,,  x:\textsf{rcuItr} \, ,z:\textsf{rcuItr}  \rrbracket  
\end{aligned}\\
\label{phu6hrr}
(\sigma_2',O'_{2},U'_{2}, T'_{2}, F'_2) \in \mathcal{R}(\{m\})
\\
  \label{ahusighrr'}
  \sigma'_1 \bullet \sigma'_2 = \sigma'
\\
\label{phu7hrr}
O'_{1} \bullet_{O} O'_{2} = O'
\\
\label{phu8hrr}
U'_{1} \cup U'_{2} = U'
\\
\label{phu9hrr}
T'_{1} \cup T'_{2} = T'
\\
\label{phu10hrr}
\textsf{WellFormed}(\sigma_1',O'_{1},U'_{1},T'_{1},F'_1) \\
\label{phu11hrr}
\textsf{WellFormed}(\sigma_2',O'_{2},U'_{2},T'_{2},F'_2)
\end{gather}

Let $h(s(x,tid),f)$ be $o_x$. We also know from operational semantics that the machine state has changed as
\begin{gather}\label{ahushrr}
\sigma_1' =  \sigma_1[s(z,tid)\mapsto o_x]
\end{gather}

There exists no change in the observation of heap locations
\begin{gather}\label{ahus1hrr}
  O'_1 =  O_1
\end{gather}

\ref{ahus2hrr} follows from \ref{ahu1hrr}
\begin{gather}\label{ahus2hrr}
  T_1 \subseteq R
\end{gather}
Proof is similar to Lemma \ref{lemma:rreadstack}.
      \end{proof}
\subsection{Soundness Proof of Structural Program Actions}
\label{lem:lemstructural}
In this section, we introduce soundness Theorem \ref{thm:snd} for structural rules of the type system. We consider the cases of the induction on derivation of  $\Gamma \vdash C \dashv \Gamma$ for all type systems,$R,M$.

Although we have proofs for read-side structural rules, we only present proofs for write-side structural type rules in this section as read-side rules are simple versions of write-side rules and proofs for them are trivial and already captured by proofs for write-side structural rools.
\begin{theorem}[Type System Soundness]
  \label{thm:snd}
\[
\forall_{\Gamma,\Gamma',C}\ldotp \Gamma \vdash C \dashv \Gamma' \implies \llbracket \Gamma \vdash C \dashv \Gamma' \rrbracket 
\]
\end{theorem}

\begin{proof}
  Induction on derivation of $\Gamma \vdash_{M} C \dashv \Gamma$.

  \begin{case}-\textbf{M}: consequence where $C$ has the form $\Gamma \vdash_{M} C \dashv \Gamma'''$.
    We know
    \begin{gather}\label{s'l6a1}
      \Gamma' \vdash_{M}  C \dashv \Gamma'' \\
      \label{s'l6a2}
      \Gamma \subt \Gamma'\\
      \label{s'l6a3}
      \Gamma'' \subt \Gamma''' \\
      \label{s'l6a4}
      \{\llbracket \Gamma' \rrbracket_{M,tid} \}C\{\llbracket \Gamma'' \rrbracket_{M,tid} \}
    \end{gather}

    We need to show
    \begin{gather}\label{s'l6p1}
\{ \llbracket \Gamma \rrbracket_{M,tid} \}C\{\llbracket \Gamma''' \rrbracket_{M,tid}\}
    \end{gather}

    The $\subt$ relation translated to entailment relation in Views Logic. The relation is established over the action judgement for identity label/transition
    
From \ref{s'l6a2} and Lemma \ref{lem:cntxsubt-m} we know 
\begin{gather}\label{s'l6a6}
\llbracket \Gamma \rrbracket_{M,tid} \sqsubseteq \llbracket \Gamma' \rrbracket_{M,tid}
\end{gather}

From \ref{s'l6a3} and \ref{lem:cntxsubt-m} we know 
\begin{gather}\label{s'l6a7}
\llbracket \Gamma'' \rrbracket_{M,tid} \sqsubseteq \llbracket \Gamma''' \rrbracket_{M,tid}
\end{gather}

By using \ref{s'l6a6}, \ref{s'l6a7} and \ref{s'l6a4} as antecedentes of Views Logic's consequence rule, we conclude \ref{s'l6p1}.
  \end{case}
  
%%HERE sequence
  \begin{case}-\textbf{M}: where $C$ is sequence statement. $C$ has the form $C_1;C_2$. Our goal is to prove
  \begin{gather}
    \label{sl1p1}
   \{ \llbracket \Gamma \rrbracket_{M,tid} \} \vdash_{M} C_1;C_2 \dashv \{ \llbracket \Gamma'' \rrbracket_{M,tid} \}
 \end{gather}
We know 
  \begin{gather}\label{sl1a1}
    \Gamma \vdash_{M} C_1 \dashv \Gamma' \\
    \label{sl1a2}
    \Gamma' \vdash_{M} C_2 \dashv \Gamma''\\
    \label{sl1a3}
    \{\llbracket \Gamma \rrbracket_{M,tid}\}   C_1  \{ \llbracket \Gamma' \rrbracket_{M,tid}\} \\
    \label{sl1a4}
    \{ \llbracket \Gamma' \rrbracket_{M,tid} \}  C_2  \{\llbracket \Gamma'' \rrbracket_{M,tid}\} 
  \end{gather} 
  
  By using \ref{sl1a3} and \ref{sl1a4} as the antecedents for the Views sequencing rule, we can derive the conclusion for \ref{sl1p1}.
  \end{case}
  
  \begin{case}-\textbf{M}: where $C$ is loop statement. $C$ has the form $while\left(x\right)\{C\}$.

     \begin{gather}\label{sl2a1}
       \Gamma \vdash_{M} C \dashv \Gamma \\
       \label{sl2a2}
       \Gamma(x) = \textsf{bool} \\ 
       \label{sl2a3}
       \{\llbracket \Gamma \rrbracket_{M,tid} \}  C  \{\llbracket \Gamma \rrbracket_{M,tid}\} 
     \end{gather}
    Our goal is to prove
    \begin{gather}\label{sl2p1}
      \{\llbracket \Gamma \rrbracket_{M,tid} \} \left(assume\left(x\right);C\right)^{*};assume(\lnot x) \{\llbracket \Gamma \rrbracket_{M,tid} \}
    \end{gather}
 We prove \ref{sl2p1} by from the consequence rule, based on the proofs of the following \ref{sl2p2} and \ref{sl2p3}
    \begin{gather}\label{sl2p2}
      \{ \llbracket \Gamma \rrbracket_{M,tid} \}\left(assume\left(x\right);C\right)^{*} \{ \llbracket \Gamma \rrbracket_{M,tid} \}
    \end{gather}
    \begin{gather}\label{sl2p3}
      \{ \llbracket \Gamma \rrbracket_{M,tid} \}assume\left(\lnot x\right)\{ \llbracket \Gamma \rrbracket_{M,tid} \}
    \end{gather}
    
   The poof of \ref{sl2p2} follows from Views Logic's proof rule for assume construct by using
      \[\{\llbracket\Gamma \rrbracket_{M,tid}\} assume\left(x\right) \{\llbracket \Gamma \rrbracket_{M,tid} \}\]
      as antecedent. We can use this antecedent together with the antecedent we know from \ref{sl2a3}
      \[\{\llbracket \Gamma \rrbracket_{M,tid} \}C\{\llbracket \Gamma \rrbracket_{M,tid}\}\]
      as antecedents to the Views Logic's proof rule for sequencing. Then we use the antecedent
      \[ \{\llbracket \Gamma \rrbracket_{M,tid} \}assume\left(x\right);C\{\llbracket \Gamma \rrbracket_{M,tid} \}\] to the proof rule for nondeterministic looping.
  
    The proof of \ref{sl2p3} follows from Views Logic's proof rule for assume construct by using the
    \[ \{ \llbracket \Gamma \rrbracket_{M,tid} \}assume\left(\lnot x\right)\{ \llbracket \Gamma \rrbracket_{M,tid} \}\] as the antecedent.
  \end{case}
  \begin{case}-\textbf{M}: where $C$ is a loop statement. $C$ has the form $while(x.f\neq \texttt{null})\{C\}$
    Proof is similar to the one for \textsc{T-Loop1}. 
  \end{case}
  \begin{case}-\textbf{M}: case where $C$ is branch statement. $C$ has the form $if\left(e\right)then\{C_1\}else\{C_2\}$.
      \begin{gather}\label{sl4a1}
       \Gamma, x:\textsf{rcuItr}\;\rho\;\N([f_1 \rightharpoonup z]) \vdash_{M} C_1 \dashv \Gamma' \\
       \label{sl4a2}
       \Gamma, x:\textsf{rcuItr}\;\rho\;\N([f_2 \rightharpoonup z]) \vdash_{M} C_2 \dashv \Gamma' \\
       \label{sl4a4}
       \{\llbracket \Gamma, \,  x:\textsf{rcuItr}\;\rho\;\N([f_1 \rightharpoonup z]) \rrbracket_{M,tid} \}  C_1  \{\llbracket \Gamma' \rrbracket_{M,tid}\}\\
        \label{sl4a5}
       \{\llbracket \Gamma , \, x:\textsf{rcuItr}\;\rho\;\N([f_2 \rightharpoonup z]) \rrbracket_{M,tid} \}  C_2  \{\llbracket \Gamma' \rrbracket_{M,tid}\}
      \end{gather}
      
    Our goal is to prove
    \begin{gather}\label{sl4p1}
      \begin{array}{l}
      \{\llbracket \Gamma,x:\textsf{rcuItr}\;\rho\;\N([f_1 | f_2 \rightharpoonup z]) \rrbracket_{M,tid} \} \\
      y = x.f_1;\left(assume\left(z = y\right);C_1\right)+\left(assume\left(y \neq z\right);C_2 \right)\\
      \{\llbracket \Gamma' \rrbracket_{M,tid} \}
      \end{array}
    \end{gather}
    where the desugared form includes a fresh variable y. We use fresh variables just for desugaring and they are not included in any type context.
    We prove\ref{sl4p1} from the consequence rule  of Views Logic based on the proofs of the following \ref{sl4p2} and \ref{sl4p3}
    \begin{gather}\label{sl4p2}
      \begin{array}{l}
      \{\llbracket\Gamma,   x:\textsf{rcuItr}\;\rho\;\N([f_1 | f_2 \rightharpoonup z])\rrbracket_{M,tid}\} \\
      \left(assume\left(z = y\right);C_1\right)+\left(assume\left(y \neq z\right);C_2 \right) \\
      \{\llbracket \Gamma' \rrbracket_{M,tid}\}
      \end{array}
    \end{gather}
    and
    \begin{gather}\label{sl4p3}\begin{array}{l}
      \{\llbracket \Gamma,   x:\textsf{rcuItr}\;\rho\;\N[f_1 | f_2 \rightharpoonup z] \rrbracket_{M,tid}\} \\
      y = x.f_1\\
      \{\llbracket \Gamma,   x:\textsf{rcuItr}\;\rho\;\N([f_1 | f_2 \rightharpoonup z]) \llbracket_{M,tid} \cap \llbracket x:\textsf{rcuItr}\;\rho\;\N([f_1 \rightharpoonup y]) \rrbracket_{M,tid} \}
      \end{array}
      \end{gather}

    \ref{sl4p3} is trivial from the fact that y is a fresh variable and it is not included in any type context and just used for desugaring.

    We prove \ref{sl4p2} from the branch rule of Views Logic based on the proofs of the following \ref{sl4p4} and \ref{sl4p5}
    \begin{gather}\label{sl4p4}
      \begin{array}{l}
        \{\llbracket \Gamma, x:\textsf{rcuItr}\;\rho\;\N([f_1 | f_2 \rightharpoonup z]) \rrbracket_{M,tid} \cap \\
        \llbracket x:\textsf{rcuItr}\;\rho\;\N([f_1 \rightharpoonup y]) \rrbracket_{M,tid} \}\\
      \left(assume\left(z = y\right);C_1\right)\\
      \{\llbracket \Gamma' \rrbracket_{M,tid}\}
      \end{array}
    \end{gather}
    and
    \begin{gather}\label{sl4p5}
      \begin{array}{l}
        \{\llbracket \Gamma, x:\textsf{rcuItr}\;\rho\;\N([f_1 | f_2 \rightharpoonup z]) \rrbracket_{M,tid} \cap\\
        \llbracket x:\textsf{rcuItr}\;\rho\;\N([f_1 \rightharpoonup y]) \rrbracket_{M,tid} \}\\
        \left(assume\left(z \neq y\right);C_2\right)\\
      \llbracket \Gamma' \rrbracket_{M,tid}\}
      \end{array}
    \end{gather}
    We show \ref{sl4p4} from Views Logic's proof rule for the assume construct by using 
      \[
      \begin{array}{l}
        \{\llbracket \Gamma, x:\textsf{rcuItr}\;\rho\;\N([f_1 | f_2 \rightharpoonup z]) \rrbracket_{M,tid} \cap \\
        \llbracket x:\textsf{rcuItr}\;\rho\;\N([f_1 \rightharpoonup y]) \rrbracket_{M,tid} \} \\
        assume\left(y=z\right)\\
        \{\llbracket \Gamma, x:\textsf{rcuItr}\;\rho\;\N([f_1 \rightharpoonup z]) \rrbracket_{M,tid} \}
      \end{array}
      \]
      as the antecedent. We can use this antecedent together with 
      \[ \{\llbracket \Gamma, x:\textsf{rcuItr}\rho\N([f_1 \rightharpoonup z]) \rrbracket_{M,tid}\} C_1 \{\llbracket\Gamma' \rrbracket_{M,tid}\} \] as antecedents to the View's Logic's proof rule for sequencing.
    
We show \ref{sl4p5} from Views Logic's proof rule for the assume construct by using 
      \[
      \begin{array}{l}
      \{\llbracket \Gamma, x:\textsf{rcuItr}\;\rho\;\N([f_1 | f_2 \rightharpoonup z]) \cap\\
      x:\textsf{rcuItr}\;\rho\;\N([f_1 \rightharpoonup y]) \rrbracket_{M,tid} \}\\
      assume(x\neq y)\\
      \{\llbracket  \Gamma,x:\textsf{rcuItr}\;\rho\;\N([f_2 \rightharpoonup z]) \rrbracket_{M,tid} \}
      \end{array}
      \]
      as the antecedent. We can use this antecedent together with 
      \[ \{\llbracket\Gamma,   x:\textsf{rcuItr}\rho\N([f_2  \rightharpoonup z])\rrbracket_{M,tid}\} C_2 \{\llbracket\Gamma' \rrbracket_{M,tid}\} \]
      as antecedents to the Views Logic's proof rule for sequencing.
  \end{case}
  
  \begin{case}-\textbf{M}: case where $C$ is branch statement. $C$ has the form $if(x.f == \texttt{null})then\{C_1\}else\{C_2\}$.
    Proof is similar to one for \textsc{T-Branch1}.
   \end{case}

\begin{case}-\textbf{O}: parallel where $C$ has the form $\Gamma_1,\Gamma_2\vdash_{O} C_1 || C_2 \dashv \Gamma'_1,\Gamma'_2$
We know 
\begin{gather} \label{sl7a1}
\Gamma_1 \vdash C_1 \dashv \Gamma'_1\\
\label{sl7a2}
\Gamma_2 \vdash C_2 \dashv \Gamma'_2\\
\label{sl7a3}
\{ \llbracket \Gamma_1 \rrbracket \} C_1 \{ \llbracket \Gamma'_1 \rrbracket\}\\
\label{sl7a4}
\{ \llbracket \Gamma_2 \rrbracket \} C_2  \{ \llbracket \Gamma'_2 \rrbracket \}
\end{gather}

We need to show 
\begin{gather}\label{sl7p1}
\{\llbracket \Gamma_1, \Gamma_2 \rrbracket \} C_1 || C_2 \{\llbracket \Gamma'_1, \Gamma'_2 \rrbracket \}
\end{gather}

By using \ref{sl7a3} and \ref{sl7a4} as antecedents to Views Logic's parallel rule, we can draw conclusion for \ref{sl7p2}
\begin{gather}\label{sl7p2}
\{\llbracket \Gamma_1 \rrbracket * \llbracket \Gamma_2 \rrbracket\} C_1 || C_2\{\llbracket \Gamma'_1 \rrbracket * \llbracket \Gamma'_2 \rrbracket\}
\end{gather}

Showing \ref{sl7p1} requires showing 
\begin{gather}\label{sl7a5}
\llbracket \Gamma_1,\Gamma_2 \rrbracket \sqsubseteq \llbracket \Gamma_1\rrbracket * \llbracket  \Gamma_2 \rrbracket
\end{gather}

\begin{gather}\label{sl7a6}
\llbracket \Gamma'_1 \rrbracket * \llbracket \Gamma'_2 \rrbracket \sqsubseteq \llbracket \Gamma'_1, \Gamma'_2 \rrbracket
\end{gather}

By using \ref{sl7a5} and \ref{sl7a6}(trivial to show as "," and "*" for denotation of type contexts are both semantically equivalent to $\cap$) as antecedents to Views Logic's consequence rule, we can conclude \ref{sl7p1}.
\end{case}
\begin{case}-\textbf{M} where C has form $\textsf{RCUWrite}\, x.f \text{ as } y \text{ in } C$
  which desugars into
  \[\textsf{WriteBegin}; x.f:= y ; C ;\textsf{WriteEnd}\]

  We assume from the rule \textsc{ToRCUWrite}
 \begin{gather}\label{dsuga1}
       \Gamma, y:\textsf{rcuItr}\;\_ \vdash_{M} C \dashv \Gamma' \\
       \label{dsuga2}
       \textsf{FType}(f) = \textsf{RCU} \\
       \label{dsuga3}
       \textsf{NoFresh}(\Gamma') \\
       \label{dsugaes}
       \textsf{NoFreeable}(\Gamma')\\
       \label{dsuga4}
       \textsf{NoUnlinked}(\Gamma')\\
        \label{dsuga5}
       \{\llbracket \Gamma\,, y:\textsf{rcuItr}\;\_ \rrbracket_{M,tid} \}  C \{\llbracket \Gamma' \rrbracket_{M,tid}\}
      \end{gather}
      
    Our goal is to prove
    \begin{gather}\label{dsugp1}
      \begin{array}{l}
      \{\llbracket \Gamma \rrbracket_{M,tid} \} 
      \textsf{WriteBegin}; C ;\textsf{WriteEnd}
      \{\llbracket \Gamma' \rrbracket_{M,tid} \}
      \end{array}
    \end{gather}
Any case of $C$ does not change the state(no heap update) by assumptions \ref{dsuga3}-\ref{dsuga4} therefore \ref{dsugp1} follows from assumptions \ref{dsuga1}-\ref{dsuga5} trivially.
\end{case}
\end{proof}
%  \end{itemize}
\begin{lemma}[Context-SubTyping-M]\label{lem:cntxsubt-m}
\[ \Gamma \subt \Gamma'  \implies \llbracket \Gamma \rrbracket_{M,tid} \sqsubseteq \llbracket  \Gamma' \rrbracket_{M,tid} \]
\end{lemma}
\begin{proof}
  
Induction on the subtyping derivation. Then inducting on the first entry in the non-empty context(empty case is trivial) which follows from \ref{lem:cntxsubt-m-s}.
\end{proof}
\begin{lemma}[Context-SubTyping-R]\label{lem:cntxsubt-r}
\[ \Gamma \subt \Gamma'  \implies \llbracket \Gamma \rrbracket_{R,tid} \sqsubseteq \llbracket  \Gamma' \rrbracket_{R,tid} \]
\end{lemma}
\begin{proof}
Induction on the subtyping derivation. Then inducting on the first entry in the non-empty context(empty case is trivial) which follows from \ref{lem:cntxsubt-r-s}.
\end{proof}
\begin{lemma}[Singleton-SubTyping-M]\label{lem:cntxsubt-m-s}
  \[ x:T \subt x:T'  \implies \llbracket x:T \rrbracket_{M,tid} \sqsubseteq \llbracket  x:T' \rrbracket_{R,tid} \]
\end{lemma}
\begin{proof}
  Proof by case analysis on structure of $T'$ and $T$. Important case includes the subtyping relation is defined over components of \textsf{rcuItr} type. $T'$ including approximation on the path component
  \[\rho.f_1 \subt \rho.f_1|f_2\]
  together with the approximation on the field map
  \[\N([f_1 \rightharpoonup \_ ]) \subt \N([f_1|f_2 \rightharpoonup \_ ])\]
  lead to subset inclusion in between a set of states defined by denotation of the $x:T'$ the set of states defined by denotation of the $x:T$(which is also obvious for \textsc{T-Sub}). Reflexive relations and relations capturing base cases in subtyping are trivial to show.
  \end{proof}
\begin{lemma}[Singleton-SubTyping-R]\label{lem:cntxsubt-r-s}
  \[ x:T \subt x:T'  \implies \llbracket x:T \rrbracket_{M,tid} \sqsubseteq \llbracket  x:T' \rrbracket_{M,tid} \]
\end{lemma}
\begin{proof}
  Proof is similar to \ref{lem:cntxsubt-m-s} with a single trivial reflexive derivation relation (\textsc{T-TSub2})
\[\textsf{rcuItr} \subt \textsf{rcuItr}\]
  \end{proof}

\newpage
\makeatletter
\newcommand{\srcsize}{\@setfontsize{\srcsize}{2pt}{2pt}}
\makeatother
\section{RCU BST Delete}
\label{appendix:bst_del}
\[\small
\begin{array}{@{}l@{}}
void\;delete(\;\INT\;data)\;\{\\
 \;\;WriteBegin;\\
   \;\;\text{// Find data in the tree} \\
   \;\;\text{// Root is never empty and its value is unique id} \\
    \;\;BinaryTreeNode \;current,\;parent = root;  \\
    \;\;\specline{parent:rcuItr\; \epsilon \; \{ \} }\\
    \;\;current = parent.Right;\\
    \;\;\specline{parent:rcuItr\; \epsilon \; \{ Right \mapsto current\} }\\
    \;\;\specline{current:rcuItr\; Right \; \{ \} }\\  
    \;\;while\;(current!=null \&\& current.data != data)\\
    \;\;\{\\
        \;\;\specline{parent:rcuItr\; (Left|Right)^{k} \; \{ (Left|Right) \mapsto current\} }\\
        \;\;\specline{current:rcuItr\; (Left|Right)^{k}.(Left|Right) \; \{ \} }\\
        \;\;\;\;if\;(current.data > data)\\
        \;\;\;\;\{\\
            \;\;\;\;\;\;\text{//if data exists it's in the left subtree}\\
            \;\;\;\;\;\;parent = current;\\
            \;\;\;\;\;\;\specline{parent:rcuItr\; (Left|Right)^{k} \; \{ \} }\\
            \;\;\;\;\;\;\specline{current:rcuItr\; (Left|Right)^{k} \; \{ \} }\\
            \;\;\;\;\;\;current = parent.Left;\\
            \;\;\;\;\;\;\specline{parent:rcuItr\; (Left|Right)^{k} \; \{ Left\mapsto current\} }\\
            \;\;\;\;\;\;\specline{current:rcuItr\; (Left|Right)^{k}.Left \; \{ \} }\\
        \;\;\;\;\}\\
        \;\;\;\;else \;if\; (current.data < data)\\
        \;\;\;\;\{\\
            \;\;\;\;\;\;\text{//if data exists it's in the right subtree}\\
            \;\;\;\;\;\;parent = current; \\
            \;\;\;\;\;\;\specline{parent:rcuItr\; (Left|Right)^{k} \; \{ \} }\\
            \;\;\;\;\;\;\specline{current:rcuItr\; (Left|right)^{k} \; \{ \} }\\
            \;\;\;\;\;\;current = current.Right; \\
            \;\;\;\;\;\;\specline{parent:rcuItr\; (Left|Right)^{k} \; \{ Right\mapsto current\} }\\
            \;\;\;\;\;\;\specline{current:rcuItr\; (Left|Right)^{k}.Right \; \{ \} }\\
        \;\;\;\;\}\\
    \;\;\} \\
    \;\;\specline{parent:rcuItr\; (Left|Right)^{k} \; \{(Left|Right)\mapsto current\} }\\
    \;\;\specline{current:rcuItr\; (Left|Right)^{k}.(Left|Right) \; \{ \} }\\
    \;\;\text{// At this point, we've found the node to remove}\\
    \;\;BinaryTreeNode \;lmParent = current.Right;\\
    \;\;BinaryTreeNode \;currentL = current.Left;\\
    \;\;\specline{current:rcuItr (Left|Right)^{k}.(Left|Right) \{ Left \mapsto currentL, Right \mapsto lmParent\} }\\
    \;\;\specline{currentL:rcuItr\; (Left|Right)^{k}.(Left|Right).Left \; \{\} }\\
     \;\;\specline{lmParent:rcuItr\; Left|Right)^{k}.(Left|Right).Right \; \{ \} }\\
    
\end{array}
\]
\[\small
\begin{array}{@{}l@{}}
    \;\;\text{// We now need to "rethread" the tree}\\
    \;\;\text{// CASE 1: If current has no right child, then current's left child becomes}\\
    \;\;\text{//         the node pointed to by the parent}\\
    \;\;if\; (current.Right == null)\\
    \;\;\{\\
        \;\;\;\;\specline{parent:rcuItr\; (Left|Right)^{k} \; \{ (Left|Right) \mapsto current\} }\\
        \;\;\;\;\specline{current:rcuItr\; (Left|Right)^{k}.(Left|Right) \; \{Left\mapsto currentL , Right \mapsto null\} }\\
        \;\;\;\;\specline{currentL:rcuItr\; (Left|Right)^{k}.(Left|Right).Left \; \{\} }\\
        
            \;\;\;\;if\; (parent.Left == current)\\
                \;\;\;\;\;\;\text{// parent.Value is greater than current.Value}\\
                \;\;\;\;\;\;\text{// so make current's left child a left child of parent}\\
                \;\;\;\;\;\;\specline{parent:rcuItr\; (Left|Right)^{k} \; \{ Left \mapsto current\} }\\
                \;\;\;\;\;\;\specline{current:rcuItr\; (Left|Right)^{k}.Left \; \{Left\mapsto currentL, Right\mapsto null\} }\\
                \;\;\;\;\;\;\specline{currentL:rcuItr\; (Left|Right)^{k}.Left.Left \; \{\} }\\
                \;\;\;\;\;\;parent.Left = currentL;\\
                \;\;\;\;\;\;\specline{parent:rcuItr\; (Left|Right)^{k} \; \{ Left \mapsto current\} }\\
                \;\;\;\;\;\;\specline{current:unlinked}\\
                \;\;\;\;\;\;\specline{currentL:rcuItr\; (Left|Right)^{k}.Left \; \{\} }\\
                
            \;\;\;\;else \\

                \;\;\;\;\;\;\text{// parent.Value is less than current.Value}\\
                \;\;\;\;\;\;\text{// so make current's left child a right child of parent}\\
                \;\;\;\;\;\;\specline{parent:rcuItr\; (Left|Right)^{k} \; \{ Right \mapsto current\} }\\
                \;\;\;\;\;\;\specline{current:rcuItr\; (Left|Right)^{k}.Right \; \{Left\mapsto currentL, Right \mapsto null\} }\\
                \;\;\;\;\;\;\specline{currentL:rcuItr\; (Left|Right)^{k}.Right.Left \; \{\} }\\
                \;\;\;\;\;\;parent.Right = currentL;\\
                \;\;\;\;\;\;\specline{parent:rcuItr\; (Left|Right)^{k} \; \{ Right \mapsto current\} }\\
                \;\;\;\;\;\;\specline{currentL:rcuItr\; (Left|Right)^{k}.Right \; \{\} }\\
                \;\;\;\;\;\;\specline{current:unlinked }\\
                \;\;\;\;\;\;SyncStart;\\
                \;\;\;\;\;\;SyncStop;\\
                \;\;\;\;\;\;\specline{current:freeable}\\
                \;\;\;\;\;\;Free(current);\\
                \;\;\;\;\;\;\specline{current:undef}\\
    \;\;\}\\
\end{array}
\]

\[\small
\begin{array}{@{}l@{}}
 \;\;\text{// CASE 2: If current's right child has no left child, then current's right child}\\
   \;\;\text{ //         replaces current in the tree}\\
   \;\; else\; if\; (current.Left == null)\\
    \;\;\{\\
        \;\;\;\;\specline{parent:rcuItr\; (Left|Right)^{k} \; \{(Left|Right)\mapsto current\} }\\
        \;\;\;\;\specline{current:rcuItr\; (Left|Right)^{k}.(Left|Right) \; \{ Left \mapsto null, Right \mapsto lmParent\} }\\
        \;\;\;\;\specline{currentL:rcuItr\; (Left|Right)^{k}.(Left|Right).Left \; \{\} }\\
        \;\;\;\;\specline{lmParent:rcuItr\; (Left|Right)^{k}.(Left|Right).Right \; \{ \} }\\
             
            \;\;\;\;if\; (parent.Left == current)\\
                \;\;\;\;\;\;\specline{parent:rcuItr\; (Left|Right)^{k} \; \{ Left \mapsto current\} }\\
                \;\;\;\;\;\;\specline{current:rcuItr\; (Left|Right)^{k}.Left \; \{Left\mapsto null, Right\mapsto lmParent\} }\\
                \;\;\;\;\;\;\specline{lmParent:rcuItr\; (Left|Right)^{k}.Left.Right \; \{\} }\\
                \;\;\;\;\;\;\text{// parent.Value is greater than current.Value}\\
                \;\;\;\;\;\;\text{// so make current's right child a left child of parent}\\
                \;\;\;\;\;\;parent.Left = lmParent;\\
                \;\;\;\;\;\;\specline{parent:rcuItr\; (Left|Right)^{k} \; \{ Left \mapsto lmParent\} }\\
                \;\;\;\;\;\;\specline{current:unlinked }\\
                \;\;\;\;\;\;\specline{lmParent:rcuItr\; (Left|Right)^{k}.Left \; \{\} }\\                 
            \;\;\;\;else\\
                \;\;\;\;\;\;\specline{parent:rcuItr\; (Left|Right)^{k} \; \{ Right \mapsto current\} }\\
                \;\;\;\;\;\;\specline{current:rcuItr\; (Left|Right)^{k}.Right \; \{Left\mapsto null, Right\mapsto lmParent\} }\\
                \;\;\;\;\;\;\specline{lmParent:rcuItr\; (Left|Right)^{k}.Right.Right \; \{\} }\\
                \;\;\;\;\;\;\text{// parent.Value is less than current.Value}\\
                \;\;\;\;\;\;\text{// so make current's right child a right child of parent}\\
                \;\;\;\;\;\;parent.Right = lmParent;\\
                \;\;\;\;\;\;\specline{parent:rcuItr\; (Left|Right)^{k} \; \{ Right \mapsto lmParent\} }\\
                \;\;\;\;\;\;\specline{lmParent:rcuItr\; (Left|Right)^{k}.Right \; \{\} }\\
                \;\;\;\;\;\;\specline{current:unlinked  }\\
                \;\;\;\;\;\;SyncStart;\\
                \;\;\;\;\;\;SyncStop;\\
                \;\;\;\;\;\;\specline{current:freeable}\\
                \;\;\;\;\;\;Free(current);\\
                \;\;\;\;\;\;\specline{current:undef}\\
       
    \;\;\}\\
\end{array}
\]
\[\small
\begin{array}{@{}l@{}}
    \;\;\text{// CASE 3: If current's right child has a left child, replace current with current's}\\
    \;\;\text{//          right child's left-most descendent}\\
    \;\;else\\
    \;\;\{\\
        \;\;\;\;\specline{parent:rcuItr\; (Left|Right)^{k} \; \{ (Left|Right)\mapsto current\} }\\
        \;\;\;\;\specline{current:rcuItr\; (Left|Right)^{k}.(Left|Right) \; \{Right\mapsto lmParent, Left \mapsto currentL \} }\\
        \;\;\;\;\specline{lmParent:rcuItr\; (Left|Right)^{k}.(Left|Right).Right \; \{ \} }\\
        \;\;\;\;\specline{currentL:rcuItr\; (Left|Right)^{k}.(Left|Right).Left \; \{ \} }\\ 
        \;\;\;\;\text{// We first need to find the right node's left-most child}\\    
        \;\;\;\;BinaryTreeNode \;currentF = new;\\
        \;\;\;\;\specline{currentF:rcuFresh }\\
        \;\;\;\;currentF.Right = lmParent;\\
        \;\;\;\;\specline{currentF:rcuFresh \; \{Right\mapsto lmParent\}}\\
        \;\;\;\;currentF.Left = currentL;\\
        \;\;\;\;\specline{currentF:rcuFresh \; \{Right\mapsto lmParent, Left\mapsto currentL\}}\\
        \;\;\;\;BinaryTreeNode \; leftmost = lmParent.Left;\\
        \;\;\;\;\specline{lmParent:rcuItr\; (Left|Right)^{k}.(Left|Right).Right \; \{ Left \mapsto leftmost\} }\\
        \;\;\;\;\specline{leftmost:rcuItr\; (Left|Right)^{k}.(Left|Right).Right.Left \; \{\} }\\
        \;\;\;\;if\;(lmParent.Left == null)\{\\
        \;\;\;\;\specline{lmParent:rcuItr\; (Left|Right)^{k}.(Left|Right).Right \; \{ Left \mapsto null\} }\\
         \;\;\;\;\;\;currentF.data = lmParent.data;\\
         \;\;\;\;\;\;if\; (parent.Left == current)\{\\
                \;\;\;\;\;\;\;\;\specline{parent:rcuItr\; (Left|Right)^{k} \; \{ Left\mapsto current\} }\\
                \;\;\;\;\;\;\;\;\specline{current:rcuItr\; (Left|Right)^{k}.Left \; \{Right\mapsto lmParent, Left \mapsto currentL \} }\\
                \;\;\;\;\;\;\;\;\specline{currentF:rcuFresh \; \{Right\mapsto lmParent, Left\mapsto currentL\}}\\
                \;\;\;\;\;\;\;\;\text{//current's right child a left child of parent}\\
                \;\;\;\;\;\;\;\;parent.Left = currentF; \\
                \;\;\;\;\;\;\;\;\specline{parent:rcuItr\; (Left|Right)^{k} \; \{ Left\mapsto currentF\} }\\
                \;\;\;\;\;\;\;\;\specline{current:unlinked }\\
                \;\;\;\;\;\;\;\;\specline{currentF:rcuItr \; (Left|Right)^{k}.Left \; \{Right\mapsto lmParent, Left\mapsto currentL\}}\\
                \;\;\;\;\;\;\;\;SyncStart;\\
                \;\;\;\;\;\;\;\;SyncStop;\\
                \;\;\;\;\;\;\;\;\specline{current:freeable}\\
                \;\;\;\;\;\;\;\;Free(current);\\
                \;\;\;\;\;\;\;\;\specline{current:undef}\\
            \;\;\;\;\;\;\} \\
            \;\;\;\;\;\;else\{ \\
                \;\;\;\;\;\;\;\;\specline{parent:rcuItr\; (Left|Right)^{k} \; \{ Right\mapsto current\}}\\
                \;\;\;\;\;\;\;\;\specline{current:rcuItr\; (Left|Right)^{k}.Right \; \{Right\mapsto lmParent, Left \mapsto currentL \}}\\
                \;\;\;\;\;\;\;\;\specline{currentF:rcuFresh \; \{Right\mapsto lmParent, Left\mapsto currentL\}}\\
                \;\;\;\;\;\;\;\;\text{//current's right child a right child of parent}\\
                \;\;\;\;\;\;\;\;parent.Right = currentF; \\
                \;\;\;\;\;\;\;\;\specline{parent:rcuItr\; (Left|Right)^{k} \; \{ Right\mapsto currentF\} }\\
                \;\;\;\;\;\;\;\;\specline{current:unlinked }\\
                \;\;\;\;\;\;\;\;\specline{currentF:rcuItr \; (Left|Right)^{k}.Right \; \{Right\mapsto lmParent, Left\mapsto currentL\}}\\
                \;\;\;\;\;\;\;\;SyncStart;\\
                \;\;\;\;\;\;\;\;SyncStop;\\
                \;\;\;\;\;\;\;\;\specline{current:freeable}\\
                \;\;\;\;\;\;\;\;Free(current);\\
                \;\;\;\;\;\;\;\;\specline{current:undef}\\
          \;\;\;\;\;\;\}\\
        \;\;\;\;\}
        \end{array}
\]

\[\small
\begin{array}{@{}l@{}}
        \;\;\;\;else\{\\
        %%else -- leftmost is not null -- so find it
        \;\;\;\;\;\;\specline{lmParent:rcuItr\; (Left|Right)^{k}.(Left|Right).Right \; \{ Left \mapsto leftmost\} }\\
        \;\;\;\;\;\;\specline{leftmost:rcuItr\; (Left|Right)^{k}.(Left|Right).Right.Left \; \{\} }\\
       \;\;\;\;\;\;while \;(leftmost.Left != null)\\
        \;\;\;\;\;\;\{\\
             \;\;\;\;\;\;\;\;\specline{lmParent:rcuItr\; (Left|Right)^{k}.(Left|Right).Right.Left(Left)^{l} \; \{Left \mapsto leftmost\} }\\
             \;\;\;\;\;\;\;\;\specline{leftmost:rcuItr\; (Left|Right)^{k}.(Left|Right).Right.Left(Left)^{l}.Left \; \{\} }\\
             \;\;\;\;\;\;\;\;lmParent = leftmost;\\
             \;\;\;\;\;\;\;\;\specline{lmParent:rcuItr\; (Left|Right)^{k}.(Left|Right).Right.Left(Left)^{l}.Left \; \{\} }\\
             \;\;\;\;\;\;\;\;\specline{leftmost:rcuItr\; (Left|Right)^{k}.(Left|Right).Right.Left(Left)^{l}.Left \; \{\} }\\
             \;\;\;\;\;\;\;\;leftmost = lmParent.Left;\\
             \;\;\;\;\;\;\;\;\specline{lmParent:rcuItr\; (Left|Right)^{k}.(Left|Right).Right.Left(Left)^{l}.Left \; \{Left \mapsto leftmost\} }\\
            \;\;\;\;\;\;\;\;\specline{leftmost:rcuItr\; (Left|Right)^{k}.(Left|Right).Right.Left(Left)^{l}.Left.Left \; \{\} }\\
        \;\;\;\;\;\;\}\\
        \;\;\;\;\;\;currentF.data = leftmost.data;\\
         \;\;\;\;\;\;if\; (parent.Left == current)\{\\
                \;\;\;\;\;\;\;\;\specline{parent:rcuItr\; (Left|Right)^{k} \; \{ Left\mapsto current\} }\\
                \;\;\;\;\;\;\;\;\specline{current:rcuItr\; (Left|Right)^{k}.Left \; \{Right\mapsto lmParent, Left \mapsto currentL \} }\\
                \;\;\;\;\;\;\;\;\specline{currentF:rcuFresh \; \{Right\mapsto lmParent, Left\mapsto currentL\}}\\
                \;\;\;\;\;\;\;\;\text{//current's right child a left child of parent}\\
                \;\;\;\;\;\;\;\;parent.Left = currentF; \\
                \;\;\;\;\;\;\;\;\specline{parent:rcuItr\; (Left|Right)^{k} \; \{ Left\mapsto currentF\} }\\
                \;\;\;\;\;\;\;\;\specline{current:unlinked }\\
                \;\;\;\;\;\;\;\;\specline{currentF:rcuItr \; (Left|Right)^{k}.Left \; \{Right\mapsto lmParent, Left\mapsto currentL\}}\\
                \;\;\;\;\;\;\;\;SyncStart;\\
                \;\;\;\;\;\;\;\;SyncStop;\\
                \;\;\;\;\;\;\;\;\specline{current:freeable}\\
                \;\;\;\;\;\;\;\;Free(current);\\
                \;\;\;\;\;\;\;\;\specline{current:undef}\\
            \;\;\;\;\;\;\} \\
            \;\;\;\;\;\;else\{ \\
                \;\;\;\;\;\;\;\;\specline{parent:rcuItr\; (Left|Right)^{k} \; \{ Right\mapsto current\}}\\
                \;\;\;\;\;\;\;\;\specline{current:rcuItr\; (Left|Right)^{k}.Right \; \{Right\mapsto lmParent, Left \mapsto currentL \}}\\
                \;\;\;\;\;\;\;\;\specline{currentF:rcuFresh \; \{Right\mapsto lmParent, Left\mapsto currentL\}}\\
                \;\;\;\;\;\;\;\;\text{//current's right child a right child of parent}\\
                \;\;\;\;\;\;\;\;parent.Right = currentF; \\
                \;\;\;\;\;\;\;\;\specline{parent:rcuItr\; (Left|Right)^{k} \; \{ Right\mapsto currentF\} }\\
                \;\;\;\;\;\;\;\;\specline{current:unlinked }\\
                \;\;\;\;\;\;\;\;\specline{currentF:rcuItr \; (Left|Right)^{k}.Right \; \{Right\mapsto lmParent, Left\mapsto currentL\}}\\
                \;\;\;\;\;\;\;\;SyncStart;\\
                \;\;\;\;\;\;\;\;SyncStop;\\
                \;\;\;\;\;\;\;\;\specline{current:freeable}\\
                \;\;\;\;\;\;\;\;Free(current);\\
                \;\;\;\;\;\;\;\;\specline{current:undef}\\
          \;\;\;\;\;\;\}\\
 \end{array}
\]
\[\small
\begin{array}{@{}l@{}}
        \;\;\;\;\;\;\specline{lmParent:rcuItr\; (Left|Right)^{k}.(Left|Right).Right.Left(Left)^{l} \; \{Left \mapsto leftmost\} }\\
        \;\;\;\;\;\;\specline{leftmost:rcuItr\; (Left|Right)^{k}.(Left|Right).Right.Left(Left)^{l}.Left \; \{Left \mapsto null\} }\\
        \;\;\;\;\;\;BinaryTreeNode\; leftmostR = leftmost.Right;\\
        \;\;\;\;\;\;\specline{leftmost:rcuItr\; (Left|Right)^{k}.(Left|Right).Right.Left(Left)^{l}.Left \;  \left\{ \begin{array}{ll} Left \mapsto null, \\ Right \mapsto leftmostR \end{array} \right\} }\\
        \;\;\;\;\;\;\specline{lmParent:rcuItr\; (Left|Right)^{k}.(Left|Right).Right.Left(Left)^{l} \; \{Left \mapsto leftmost\} }\\
        \;\;\;\;\;\;\specline{leftmostR:rcuItr\; (Left|Right)^{k}.(Left|Right).Right.Left(Left)^{l}.Left.Right \; \{\} }\\
        \;\;\;\;\;\;\text{// the parent's left subtree becomes the leftmost's right subtree}\\
        \;\;\;\;\;\;lmParent.Left = leftmostR;\\
        \;\;\;\;\;\;\specline{leftmost:unlinked  }\\
        \;\;\;\;\;\;\specline{lmParent:rcuItr\; (Left|Right)^{k}.(Left|Right).Right.Left(Left)^{l} \; \{Left \mapsto leftmostR\} }\\
        \;\;\;\;\;\;\specline{leftmostR:rcuItr\; (Left|Right)^{k}.(Left|Right).Right.Left(Left)^{l}.Left \; \{\} }\\
        \;\;\;\;\;\;SyncStart;\\
        \;\;\;\;\;\;SyncStop;\\
		\;\;\;\;\;\;\specline{leftmost:freeable}\\
        \;\;\;\;\;\;Free(leftmost);\\
        \;\;\;\;\;\;\specline{leftmost:undef}\\
        \;\;\;\;\}\\
    \;\;\}\\
     \;\;WriteEnd;\\
\}

\end{array}
\]

\newpage
\makeatletter
\makeatother
\section{RCU Bag with Linked-List}
\label{appendix:bag_paul}
\[\small
\begin{array}{@{}l@{}}
  BagNode\; head;\\
  int \;member\;(int\;toRead)\;\{\\
   \;\;ReadBegin; \\
   \;\;int\;result =0; \\
   \;\;\specline{parent:undef,\, head:rcuRoot}\\
   \;\;BagNode \;parent = head; \\
   \;\;\specline{parent:rcuItr}\\
\;\;\specline{current:\_}\\
   \;\;current = parent.Next; \\
   \;\;\specline{current:rcuItr,\, parent:rcuItr}\\
      \;\;\specline{current:rcuItr}\\
   \;\;while(current.data\; != \;toRead \&\& current.Next \neq null)\{\\
     \;\;\;\;\specline{parent:rcuItr}\\
      \;\;\;\;\specline{current:rcuItr}\\
     \;\;\;\;parent= current;\\
     \;\;\;\;current = parent.Next ;\\
  \;\;\;\;\specline{parent:rcuItr}\\
      \;\;\;\;\specline{current:rcuItr}\\
     \;\;\}\\
   \;\;\specline{parent:rcuItr}\\
   \;\;\specline{current:rcuItr}\\
   \;\;result = current.data;\\
   \;\;ReadEnd;\\
   \;\;return \; result;\\
  \}\\
\end{array}
\]
\[\small
\begin{array}{@{}l@{}}
   void\; remove\;(int \;toDel\;)\;\{\\
   \;\;WriteBegin;\\
   \;\;BagNode\; current,\;parent\;=\;head;\\
   \;\;current\;=\; parent.Next;\\
   \;\;\specline{current:rcuItr\; Next \; \{\}}\\
   \;\;\specline{parent:rcuItr\; \epsilon\; \{Next \mapsto current\}}\\
   \;\;while\;(current.Next != null \&\& current.data \neq toDel)\;\{\\
   \;\;\;\;\specline{parent:rcuItr\; (Next)^{k} \; \{Next \mapsto current \}}\\
   \;\;\;\;\specline{current:rcuItr\; Next.(Next)^{k}.Next \; \{ \}}\\
   \;\;\;\;parent= current;\\
   \;\;\;\;\specline{current:rcuItr\; Next.(Next)^{k}.Next \; \{ \}}\\
   \;\;\;\;\specline{parent:rcuItr\; Next.(Next)^{k}.Next \; \{ \}}\\
   \;\;\;\;current = parent.Next;\\
   \;\;\;\;\specline{parent:rcuItr\; Next.(Next)^{k}.Next \; \{Next \mapsto current \}}\\
   \;\;\;\;\specline{current:rcuItr\; Next.(Next)^{k}.Next.Next \; \{ \}}\\
   \;\;\}\\
   \;\;\text{//We don't need to be precise on whether next of current is null or not} \\
   \;\;\specline{parent:rcuItr\; Next.(Next)^{k}.Next \; \{Next \mapsto current \}}\\
   \;\;\specline{current:rcuItr\; Next.(Next)^{k}.Next.Next.Next \; \{Next \mapsto null \} }\\
   \;\;BagNode \;currentL= current.Next;\\
   \;\;\specline{parent:rcuItr\; Next.(Next)^{k}.Next \; \{Next \mapsto itr \}}\\
   \;\;\specline{currentL:rcuItr\; Next.(Next)^{k}.Next.Next.Next \; \{\} }\\
   \;\;\specline{current:rcuItr \; Next.(Next)^{k}.Next.Next \;\{Next \mapsto currentL\} }\\
   \;\;current.Next=currentL;\\
   \;\;\specline{parent:rcuItr\; Next.(Next)^{k}.Next \; \{Next \mapsto itrN \}}\\
   \;\;\specline{currentL:rcuItr\; Next.(Next)^{k}.Next.Next \; \{\} }\\
   \;\;\specline{current:unlnked }\\
   \;\;SyncStart;\\
   \;\;SyncStop;\\
   \;\;\specline{current:freeable}\\
   \;\;Free(current);\\
   \;\;\specline{current:undef}\\
   \;\;WriteEnd;\\
   \}\\
   \end{array}
\]
\[\small
\begin{array}{@{}l@{}}
  void \; add(int toAdd)\{\\
    \;\;WriteBegin;\\
    \;\;BagNode \;nw = new;\\
    \;\; nw.data  =  toAdd;\\
    \;\;\specline{nw:rcuFresh \, \{ \} }\\
    \;\;BagNode\; current,\;parent\;=\;head;\\
    \;\;parent.Next\;=\;current;\\
    \;\;\specline{current:rcuItr\; Next \; \{\}}\\
    \;\;\specline{parent:rcuItr\; \epsilon\; \{Next \mapsto current\}}\\
    \;\;while\;(current.Next != null )\;\{\\
    \;\;\;\;\specline{parent:rcuItr\; (Next)^{k} \;\{Next \mapsto current \}}\\
    \;\;\;\;\specline{current:rcuItr\; Next.(Next)^{k}.Next\; \{ \}}\\
    \;\;\;\;parent = current;\\
    \;\;\;\;current = parent.Next;\\
    \;\;\;\;\specline{parent:rcuItr\; (Next)^{k}.Next \;\{Next \mapsto current \}}\\
    \;\;\;\;\specline{current:rcuItr\; Next.(Next)^{k}.Next.Next\; \{ \}}\\
    \;\;\}\\
    \;\;\specline{parent:rcuItr\; (Next)^{k}.Next \;\{Next \mapsto current \}}\\
    \;\;\specline{current:rcuItr\; Next.(Next)^{k}.Next.Next\; \{ Next \mapsto null \}}\\
    \;\;nw.next= null;\\
    \;\;\specline{nw:rcuFresh\; \{Next \mapsto null\}}\\
    \;\;current.Next=nw\\
    \;\;\specline{parent:rcuItr\; (Next)^{k}.Next \;\{Next \mapsto nw \}}\\
    \;\;\specline{current:rcuItr\; (Next)^{k}.Next.Next \;\{Next \mapsto nw \}}\\
    \;\;\specline{nw:rcuItr\; Next.(Next)^{k}.Next.Next.Next\; \{Next \mapsto null \}}\\
    \;\;WriteEnd;\\
  \}\\  
\end{array}
\]

\newpage
\section{Safe Unlinking}
\label{appendix:rcuunlink}
\begin{figure}[H]
 \centering
 \begin{subfigure}[b]{.4\linewidth}
\centering
 \begin{tikzpicture}[>=stealth',node distance=1.3cm,semithick,auto]
 \tikzstyle{hollow node}=[circle,draw,inner sep=1]
 \tikzstyle{sub node}=[triangle,draw,inner sep=1]
 \tikzstyle{solid node}=[rectangle,draw,inner sep=1.5]
 \tikzstyle{solids node}=[rectangle,draw=red,inner sep=1.5]
\tikzstyle{null node}=[circle,draw=red,fill=red]
 \tikzset{
 	red node/.style={rectangle,draw=black,fill=red,inner sep=1},
 	blue node/.style={rectangle,draw=black,inner sep=1},
 	reader node/.style={circle,draw=black,inner sep=1},
        redreader node/.style={circle,draw=red,inner sep=1},
        readerr node/.style={dashed,circle,draw=black,inner sep=1},
 	writer node/.style={circle,draw=black,inner sep=1}
 }

       \node[solid node] (R) {$R$};
       \node[solid node] (1) [right of=R] {$H_0$};
       \node[solids node] (2) [right of=1] {$H_1$};
       \node[solids node] (3) [right of=2] {$H_2$};
       \node[solids node] (4) [right of=3] {$H_3$};
       \node[solid node] (5) [above  of=3] {$H_4$};
       \node[null node] (nl) [above right of = 3]{};
       \node[solid node] (6) [above  of=4] {$H_5$};
       
       \node[redreader node] (r0) [below  of= 2]  {$pr$};
       \node[redreader node] (r1) [below  of= 3]  {$cr$};
       \node[redreader node] (r2)  [below  of= 4] {$crl$};

       \node[readerr node] (r0a) [above  of= 1]  {$a_1$};
       \node[readerr node] (r1a) [above   of= 2]  {$a_2$};
       \node[readerr node] (r2a)  [below right of= 4] {$a_3$};

       \node[reader node] (r1f) [above  of= 5]  {$a_4$};
       \node[reader node] (r2f)  [above  of= 6] {$a_5$};

     \path[->]  (R) edge node[below] {$l$} (1);
     \path[->]  (1) edge node[below] {$l$} (2);
     \path[draw=red,->]  (2) edge node[below] {$l$} (3);
     \path[draw=red,->]  (3) edge node[below] {$l$} (4);
     \path[dashed,->]  (5) edge node {$l$} (3);
     \path[dashed,->]  (6) edge node {$l$} (4);

     \path[draw=red,->]  (r0) edge node {} (2);
     \path[draw=red,->]  (r1) edge node  {}  (3);
     \path[draw=red,->]  (r2) edge  node  {}   (4);

     \path[dashed,->]  (r0a) edge node {} (2);
     \path[dashed,->]  (r1a) edge node  {}  (3);
     \path[dashed,->]  (r2a) edge  node  {}   (4);

     \path[->]  (r1f) edge node  {}  (5);
     \path[->]  (r2f) edge  node  {}   (6);
    \path[draw=red,->] (3) edge node[below] {$r$} (nl);
%\path[dotted,->] (1) edge[bend left] node  {$f^{*}$} (5)
 ;
 
 \end{tikzpicture}
 \caption{\textsf{Framing} before unlinking the heap node pointed by \texttt{current}-$cr$.}
 \label{fig:bframeout}
 \end{subfigure}\quad \quad \quad
\begin{subfigure}[b]{.4\linewidth}
 \centering
 \begin{tikzpicture}[>=stealth',node distance=1.3cm,semithick,auto]
 \tikzstyle{hollow node}=[circle,draw,inner sep=1]
 \tikzstyle{sub node}=[triangle,draw,inner sep=1]
 \tikzstyle{solid node}=[rectangle,draw,inner sep=1.5]
 \tikzstyle{solids node}=[rectangle,draw=red,inner sep=1.5]
  \tikzstyle{solidss node}=[dashed,rectangle,draw=red,inner sep=1.5]
 \tikzstyle{null node}=[circle,draw=red,fill=red]
 \tikzset{
 	red node/.style={rectangle,draw=black,fill=red,inner sep=1.5},
 	blue node/.style={rectangle,draw=black,inner sep=1.5},
 	reader node/.style={circle,draw=black,inner sep=1},
        redreader node/.style={circle,draw=red,inner sep=1},
                rredreader node/.style={dashed,circle,draw=red,inner sep=1},
        readerr node/.style={dashed,circle,draw=black,inner sep=1},
 	writer node/.style={circle,draw=black,inner sep=1}
 }

       \node[solid node] (R) {$R$};
       \node[solid node] (1) [right of=R] {$H_0$};
       \node[solids node] (2) [right of=1] {$H_1$};
       \node[solidss node] (3) [right of=2] {$H_2$};
       \node[solids node] (4) [right of=3] {$H_3$};
       \node[solid node] (5) [above  of=3] {$H_4$};
       \node[solid node] (6) [above  of=4] {$H_5$};
       \node[null node] (nl) [above right of = 3]{};

       \node[redreader node] (r0) [below  of= 2]  {$pr$};
       \node[rredreader node] (r1) [below  of= 3]  {$cr$};
       \node[redreader node] (r2)  [below  of= 4] {$crl$};

       \node[readerr node] (r0a) [above  of= 1]  {$a_1$};
       \node[readerr node] (r1a) [above   of= 2]  {$a_2$};
       \node[readerr node] (r2a)  [below right of= 4] {$a_3$};

       \node[reader node] (r1f) [above  of= 5]  {$a_4$};
       \node[reader node] (r2f)  [above  of= 6] {$a_5$};

     \path[->]  (R) edge node[below] {$l$} (1);
     \path[->]  (1) edge node[below] {$l$} (2);
     \path[draw=red,->]  (2) edge [bend right] node[below left] {$l$} (4);
     \path[dashed,draw=red,->]  (3) edge node[below] {$l$} (4);
     \path[dashed,->]  (5) edge node {$l$} (3);
     \path[dashed,->]  (6) edge node {$l$} (4);

     \path[draw=red,->]  (r0) edge node {} (2);
     \path[dashed,draw=red,->]  (r1) edge node  {}  (3);
     \path[draw=red,->]  (r2) edge  node  {}   (4);

     \path[dashed,->]  (r0a) edge node {} (2);
     \path[dashed,->]  (r1a) edge node  {}  (3);
     \path[dashed,->]  (r2a) edge  node  {}   (4);

     \path[->]  (r1f) edge node  {}  (5);
     \path[->]  (r2f) edge  node  {}   (6);
     \path[dashed,draw=red, ->] (3) edge node[below] {$r$} (nl);
 ;
 \end{tikzpicture}

 \caption{Safe unlinking of the heap node pointed by \texttt{current}-$cr$ via \textsf{Framing}}
 \label{fig:bunlinkframeout}
\end{subfigure}
\caption{Safe unlinking of a heap node from a \textsf{BST}}
 \label{fig:unlkappndx}\vspace{-2mm}
 \end{figure}
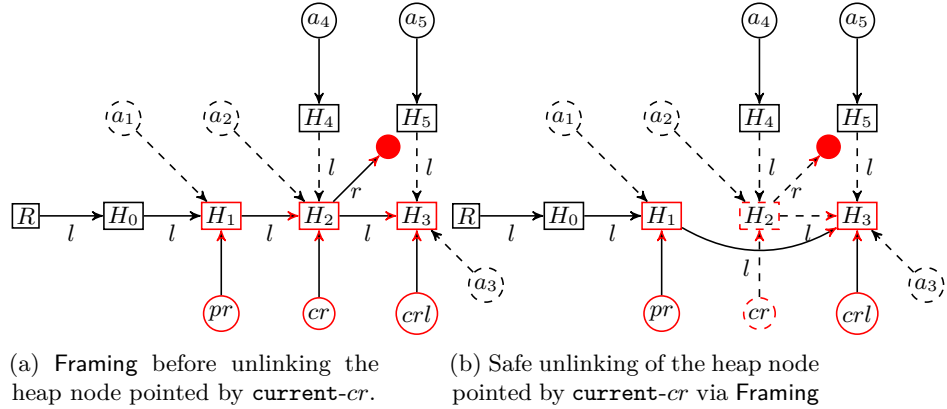

Preserving invariants of a data structure against possible mutations under \textsf{RCU} semantics is challenging. Unlinking a heap node is one way of mutating the heap. To understand the importance of the locality on the possible effects of the mutation, we illustrate a setting for unlinking a heap in Figures  \ref{fig:bframeout} and \ref{fig:bunlinkframeout}. The square nodes filled with $R$ -- a root node -- and $H$ -- a heap node -- are heap nodes. The hollow nodes are stack pointers to the square heap nodes. All resources in red form the memory foot print of unlinking. The hollow red nodes -- $pr$, $cr$ and $crl$ -- point to the red square heap nodes which are involved in unlinking of the heap node pointed by \texttt{cr}. We have $a_1$, $a_2$ and $a_3$ which are aliases with \texttt{parent}-$pr$, \texttt{current}-$cr$ and \texttt{currenL}-$crl$ respectively. We call them the \textit{path-aliases} as they share the same path from root to the node that they reference. The red filled circle depicts \texttt{null}, $l$ field which depicts $Left$ and $r$ depicts $Right$ field.

The type rule for unlinking must assert the "proper linkage" in between the heap nodes involved in the action of unlinking. We see the proper linkage relation between in Figure \ref{fig:bframeout} as red $l$ links between $H_1$, $H_2$ and $H_3$ which are referenced by $pr$, $cr$ and $crl$ respectively. Our type rule for unlinking(\textsc{T-UnlinkH}) asserts that $x$ (\texttt{parent}), $y$ (\texttt{current}) and $z$ (\texttt{currentL}) pointers are linked with field mappings $\N([f_1\rightharpoonup z])$ ($Left \mapsto current$) of $x$, $\N_1([f_2\rightharpoonup r])$ ($Left \mapsto currentL$) of $y$. In accordance with the field mappings, the type rule also asserts that $x$ has the path $\rho$ ($(Left)^{k}$), $y$ has the path $\rho.f_1$ ($(Left)^{k}.Left$) and $z$ has the path $\rho.f_1.f_2$ ($(Left)^{k}.Left.Left$).

Being able to localize the effects of the mutation is important in a sense that it prevents unexpected side effects of the mutation. So, sharing through aliases to the resources under mutation, e.g. aliasing to \texttt{parent}, \texttt{current} and \texttt{currentL}, needs to be handled carefully. Aliasing can occur via either through  object fields -- via field mappings -- or stack pointers -- via path components. We see path aliases, $a_1$, $a_2$ and $a_3$, illustrated with dashed nodes and arrows to the heap nodes in Figures \ref{fig:bframeout} and \ref{fig:bunlinkframeout}. They are depicted as dashed because they are not safe resources to use when unlinking so they are \textit{framed-out} by the type system via
\[
\begin{array}{l}
(\neg\mathsf{MayAlias}(\rho_3,\{\rho,\rho_1,\rho_2\})  ) 
\end{array}
\]
which ensures the non-existence of the \textit{path-aliases} to any of $x$, $z$ and $r$ in the rule which corresponds to $pr$, $cr$ and $crl$ respectively.

Any heap node reached from root by following a path($\rho_3$) deeper than the path reaching to the last heap node($crl$) in the footprint cannot be pointed by any of the heap nodes($pr$, $cr$ and $crl$) in the footprint. We require this restriction to prevent inconsistency on path components of references, $\rho_3$, referring to heap nodes deeper than memory footprint
\[
 (\forall_{\rho_4\neq \epsilon} \ldotp \neg\mathsf{MayAlias}(\rho_3, \rho_2.\rho_4) )
\]
The reason for framing-out these dashed path aliases is obvious when we look at the changes from the Figure \ref{fig:bframeout} to Figure \ref{fig:bunlinkframeout}. For example, $a_1$ points to $H_1$ which has object field $Left$-$l$ pointing to $H_2$ which is also pointed by \texttt{current} as depicted in the Figure \ref{fig:bframeout}. When we look at Figure \ref{fig:bunlinkframeout}, we see that $l$ of $H_1$ is pointing to $H_3$ but $a_1$ still points to $H_1$. This change invalidates the field mapping $Left \mapsto current$ of $a_1$ in the \textsf{rcuItr} type.

One another safety achieved with framing shows up in a setting where \texttt{current} and $a_2$ are aliases. In the Figure \ref{fig:bframeout}, both \texttt{current} and $a_2$ are in the \textsf{rcuItr} type and point to $H_2$. After the unlinking occurs, the type of \texttt{current} becomes \texttt{unlinked} although $a_2$ is still in the \texttt{rcuItr} type. Framing out $a_2$ prevents the inconsistency in its type under the unlinking operation.

One interesting and not obvious inconsistency issue shows up due to the aliasing between $a_3$ and \texttt{currentL}-$crl$. Before the unlinking occurs, both \texttt{currentL} and $a_3$ have the same path components. After the unlinking, the path of \texttt{currentL}-$crl$ gets shortened as the path to heap node it points, $H_3$, changes  to $(Left)^{k}.Left$ . However, the path component of $a_3$  would not change so the path component of $a_3$ in the \textsf{rcuItr} would become inconsistent with the actual path reaching to $H_3$.

In addition to \textit{path-aliasing}, there can also be aliasing via \textit{field-mappings} which we call \textit{field-aliasing}. We see field aliasing examples in Figures \ref{fig:bframeout} and \ref{fig:bunlinkframeout}: $pr$ and $a_1$ are field aliases with $Left-l$ from $H_0$ points to $H_1$, $cr$ and $a_2$ are field aliases with $Left-l$ from $H_4$ points to $H_2$  and $crl$ and $a_3$ are field aliases with $Left-l$ from $H_5$ points to $H_3$. We do not discuss the problems that can occur due to the \textit{field-aliasing} as they are same with the ones due to \textit{path-aliasing}. What we focus on is how the type rule prevents \textit{field-aliases}. The type rule asserts $ \land (m\not\in\{z,r\} )$ to make sure that there exists no object field from any other context pointing either to the variable points the heap node that is mutation(unlinking) -- \texttt{current}-$cr$ -- or to the variable which points to the new $Left$ of \texttt{parent} after unlinking -- \texttt{currentL}-$crl$. We should also note that it is expected to have object fields in other contexts to point to $pr$ as they are not in the effect zone of unlinking. For example, we see the object field $l$ points from $H_0$ to  $H_1$ in Figures \ref{fig:bframeout} and \ref{fig:bunlinkframeout}.

Once we unlink the heap node, it cannot be accessed by the new coming reader threads the ones that are currently reading this node cannot access to the rest of the heap. We illustrate this with dashed red $cr$, $H_2$ and object fields in Figure \ref{fig:bunlinkframeout}.

Being aware of how much of the heap is under mutation is important, e.g. a whole subtree or a single node. Our type system ensures that there can be only just one heap node unlinked at a time by atomic field update action. To be able to ensure this, in addition to the proper linkage enforcement, the rule also asserts that all other object fields which are not under mutation must either not exists or point to \texttt{null} via
\[\forall_{f\in dom(\N_1)} \ldotp f\neq f_2 \implies (\N_1(f) = \textsf{null})\] 

\newpage
\section{Types Rules for \textsf{RCU} Read Section}
\label{appendix:readtypes}
\begin{figure}[!htb]
%\begin{subfigure}[b]{.4\linewidth}
\begin{mathpar}
  \fbox{$\Gamma \vdash_{R} \alpha \dashv \Gamma'$} \;\;
  \inferrule[\scriptsize(T-ReadS)]
{
z \notin  \textsf{FV}(\Gamma)
}
{
\Gamma\,, z:\_ \, , x:\textsf{rcuItr}\vdash z=x \dashv  x:\textsf{rcuItr} \, , z:\textsf{rcuItr}\, , \Gamma
}
\and
\inferrule[\scriptsize(T-Root)]
{ y \notin \textsf{FV}(\Gamma) }
{\Gamma\,,r:\textsf{rcuRoot}\,,y:\textsf{undef} \vdash y = r \dashv y: \textsf{rcuItr}\,,r:\textsf{rcuRoot}\,,\Gamma}
\and
\inferrule[\scriptsize(T-ReadH)]
{
 z \notin \textsf{FV}(\Gamma)
}
{
 \Gamma\, , z: \_ \, ,  x:\textsf{rcuItr} \, \N   \vdash
 	z=x.f
 \dashv  x:\textsf{rcuItr} \, ,z:\textsf{rcuItr}  \, , \Gamma
}
\and
\fbox{$\Gamma \vdash_{R} C \dashv \Gamma'$} \;\;
\inferrule[\scriptsize(ToRCURead)]
{
 \Gamma\, , y : \textsf{rcuItr} \vdash_R \bar{s} \dashv \Gamma' \quad \textsf{FType}(f)=\textsf{RCU}
}
{
\Gamma \vdash \textsf{RCURead}\, x.f \textsf{ as }y\textsf{ in } \{ \bar{s} \}
}
\and
\fbox{$\Gamma \vdash_{M,R} C \dashv \Gamma'$} \;\;
\inferrule[\scriptsize(T-Branch2)]
{
\Gamma(x)= \textsf{bool} \\
 \Gamma \vdash C_1 \dashv\Gamma' \and \Gamma \vdash C_2 \dashv \Gamma'
}
{
\Gamma \vdash \textsf{if}(x) \textsf{ then } C_1  \textsf{ else } C_2 \dashv \Gamma'
}
\and
\inferrule[\scriptsize(T-Seq)]
{
\Gamma_1 \vdash C_1 \dashv   \Gamma_2  \qquad   \Gamma_2 \vdash C_2 \dashv   \Gamma_3
}
{
\Gamma_1  \vdash C_1 \; ; \; C_2 \dashv  \Gamma_3
}
\and
\inferrule[\scriptsize(T-Par)]
{
 \Gamma_1 \vdash_{R} C_1 \dashv   \Gamma'_1  \qquad   \Gamma_2 \vdash_{M,R} C_2 \dashv   \Gamma'_2
}
{
 \Gamma_1, \; \Gamma_2  \vdash C_1 || C_2 \dashv  \Gamma'_1 \; , \Gamma'_2
}
\and
\inferrule[\scriptsize(T-Exchange)]
{
 \Gamma, \, y:T' , \, x:T, \, \Gamma' \vdash C \dashv   \Gamma''
}
{
\Gamma, \, x:T , \, y:T' , \, \Gamma' \vdash C \dashv   \Gamma''
}
\and
\inferrule[\scriptsize(T-Conseq)]
{
 \Gamma \subt \Gamma'  \and \Gamma' \vdash C \dashv \Gamma'' \and \Gamma'' \subt \Gamma'''
}
{
 \Gamma \vdash C \dashv \Gamma'''
}
\and
\inferrule[\scriptsize(T-Skip)]
{
}
{
 \Gamma \vdash \textsf{skip} \dashv   \Gamma
}
\end{mathpar}
\caption{Type Rules for \textsf{Read} critical section for \textsf{RCU} Programming}
\label{fig:tssr}
\end{figure}

\end{document}